\documentclass[a4paper]{amsart}
\usepackage{amssymb,stmaryrd}
\usepackage{amsfonts}
\usepackage{amstext}
\usepackage{algorithmic}
\usepackage{algorithm}
\usepackage{xcolor}
\usepackage{graphicx}
\usepackage{epstopdf}
\usepackage[all]{xy}
\usepackage{MnSymbol}
\numberwithin{equation}{section}
\newtheorem{theorem}{Theorem}[section]
\newtheorem{lemma}[theorem]{Lemma}
\newtheorem{proposition}[theorem]{Proposition}
\newtheorem{corollary}[theorem]{Corollary}

\theoremstyle{definition}

\newtheorem{example}[theorem]{Example}

\theoremstyle{remark}

\newcommand{\R}{{\mathbb{R}}}

\newcommand{\CA}{{\mathcal{A}}}
\newcommand{\C}{{\mathbb{C}}}
\newcommand{\Z}{{\mathbb{Z}}}

\newcommand{\<}{{\langle}}

\renewcommand{\>}{{\rangle}}

\newcommand{\CH}{{\mathcal{H}}}
\newcommand{\CE}{{\mathcal{E}}}

\newcommand{\CL}{{\mathcal{L}}}

\newcommand{\wedgeq}{{\wedge\kern-5pt\cdot}}
\newcommand*{\doublenabla}{\nabla\mkern-12mu\nabla}

\newcommand{\cu}{{\mathfrak u}}
\newcommand{\ch}{{\mathfrak{h}}}
\newcommand{\cX}{{\mathfrak{X}}}

\newcommand{\tens}{\otimes}

\newcommand{\id}{{\rm id}}
\newcommand{\bo}{{}^{(1)}}
\newcommand{\bt}{{}^{(2)}}

\newcommand{\extd}{{\rm d}}
\newcommand{\del}{{\partial}}
\newcommand{\eps}{\epsilon}
\newcommand{\ev}{{\rm ev}}

\renewcommand{\imath}{\mathrm{i}}

\allowdisplaybreaks
\begin{document}

\title{Quantum geodesics in quantum mechanics}


\author{Edwin Beggs and Shahn Majid}
\address{ Department of Mathematics, Bay Campus, Swansea University, SA1 8EN, UK; Queen Mary University of London, 
School of Mathematical Sciences, Mile End Rd, London E1 4NS, UK}

\email{e.j.beggs@swansea.ac.uk,   s.majid@qmul.ac.uk}


\begin{abstract}  We show that the standard Heisenberg algebra of quantum mechanics admits a noncommutative differential calculus $\Omega^1$ depending on the Hamiltonian $p^2/2m + V(x)$, and  a flat quantum connection $\nabla$ with torsion such that a previous quantum-geometric formulation of flow along autoparallel curves (or `geodesics') is exactly Schr\"odinger's equation. The connection $\nabla$ preserves a generalised `skew metric' given by the canonical symplectic structure lifted to a certain rank (0,2) tensor on the extended phase space where we adjoin a time variable. We also apply the same approach to the Klein Gordon equation on Minkowski spacetime with a background electromagnetic field, formulating quantum `geodesics' on the relativistic Heisenberg algebra with proper time for the external geodesic parameter. Examples include a   relativistic free particle wave packet and a hydrogen-like atom.   \end{abstract}
\maketitle 

\section{Introduction}

Quantum Riemannian geometry, in the sense of quantum metrics and connections on possibly noncommutative `coordinate algebras', has been extensively developed since the 1980s and now has an accepted role as a plausibly better description of spacetime (i.e. `quantum spacetime') that includes Planck scale effects.  Many authors have written on this topic and we refer to our book
\cite{BegMa} for a bibliography as well as an introduction to a particular constructive approach developed in \cite{BegMa1,BegMa2,BegMa3}, among other works, which we will adopt. This approach uses bimodule connections\cite{DV1,Mou} and quantum metrics to build up the quantum Riemannian geometry on a chosen differential algebra, making it complementary to  the well-known Connes'  approach where the noncommutative geometry is encoded in a spectral triple\cite{Con} or abstract Dirac operator as starting point. The two approaches can be compatible and with  interesting results where they meet\cite{BegMa3,LirMa}. Recent applications  to models of quantum gravity itself are in \cite{Ma:squ,ArgMa,LirMa1}. 

In the present paper, we apply this powerful machinery of quantum Riemannian geometry to the more obvious context
of ordinary quantum mechanics and quantum theory. Here the noncommutativity parameter will not be the Planck scale but just the usual $\hbar$. The noncommutativity inherent in quantum theory has long been one of the motivations for results in operator algebras in general and noncommutative geometry in particular, and the role of latter in actual quantum systems has already been noted in Connes' approach, for example to understand the quantum Hall effect\cite{Bel}. The role of the  quantum Riemannian geometry formalism\cite{BegMa}, however, has not been explored so far in this context but makes sense once we note that the `quantum metric' need not be symmetric. Rather, we will be led to an antisymmetric `generalised quantum metric' and on an extended phase space where we adjoin time. The quantum metric will be degenerate and $\nabla$, although compatible with it, will have a small amount of torsion, both features relating to the extra time direction. Thus, there are some differences but in broad terms we will effectively  formulate  ordinary quantum mechanics somewhat more in the spirit of gravity, rather than the more well-studied idea of formulating gravity in a quantum manner.   

We make use of the notion of `quantum geodesics' with respect to any $\nabla$, as recently introduced in \cite{Beg:geo}. The preliminary Section~\ref{seci} provides the algebraic definition of a `geodesic bimodule' $E$ and its application to one classical geodesic in a manifold, and Section~\ref{secii} its further application for another choice of $E$ to a dust of particles with density $\rho$, where each particle moves along a classical geodesic. The tangent vector to all these particles will be a vector field $X$ obeying an autoparallel `geodesic velocity equation'. The actual particle flows are then given classically by exponentiating the vector field $X$ to a diffeomorphism of the manifold, while the natural way to do this in our algebraic formulation  turns out to be a corresponding flow equation not for $\rho$ but for an amplitude $\psi$, where $\rho=|\psi|^2$. This algebraic formalism then makes sense when the coordinate algebra of the manifold is replaced by a noncommutative algebra $A$, i.e. in noncommutative geometry. For our purposes now, we need to go further and Section~\ref{seciii} introduces a new choice of $E$ in which the `geodesic flow'  takes place more generally on a representation space of an algebra $A$ rather than on $A$ itself. We can then apply this  in Section~\ref{secheis} to $A$ the  Heisenberg algebra in the Schr\"odinger representation, allowing us to express the standard Schr\"odinger equation for a Hamiltonian $h= p^2/2m + V(x)$ as a quantum geodesic flow. The new result here is not the flow equation, which is just Schr\"odinger's equation, but the noncommutative geometric structures that we find behind it: quantum differential forms, the quantum vector field $X$, the antisymmetric quantum metric and the metric compatible connection $\nabla$. 

Specifically, the algebra $A$ in Section~\ref{secheis} will be the standard $\R^{2n}$ Heisenberg algebra 
\begin{equation} [x^i,p_j]=\imath\hbar\delta^i{}_j,
\quad [x^i,x^j]=[p_i,p_j]=0,\quad 1\le i,j\le n,\end{equation}
but equipped now with a certain differential calculus $\Omega^1_A$ defined by the Hamiltonian $h$. This idea to use the freedom of the noncommutative differential structure to encode the physical dynamics is in the spirit of \cite{Ma:newt} where one of the authors showed how Newtonian gravity can be encoded in the choice of differential structure on quantum spacetime, but now applied to quantum mechanics. Specifically, the exterior algebra in Proposition~\ref{heiscalc} has the commutation relations
\begin{align}
[\extd p_i,p_j]= -\mathrm{i}\hbar\frac{\partial^2 V}{\partial x^i \,\partial x^j}\,\theta',\quad 
[\extd p_i,x^j]=[\extd x^i,p_j]= 0,\quad
[\extd x^i,x^j]= -\frac{\mathrm{i}\hbar}{m}\,\delta_{ij}\,\theta'
\end{align}
between functions and differentials, where $\theta'$ is a graded central extra direction initially with no classical analogue but is dictated by the algebra. Here 
\begin{equation*} \theta'={ m\over \imath\hbar}[x^i,\extd x^i]\end{equation*}
for any fixed $i$ makes clear that this has its origins in the noncommutativity of the quantum geometry. The need for an extra direction $\theta'$ in the cotangent bundle has emerged in recent years as a somewhat common phenomenon in noncommutative model building \cite{Sit,Ma:spo,Ma:alm,Ma:rec}. Its associated partial derivative in the expansion of the exterior derivative $\extd$ is typically a second order `Laplacian' of some kind and that will be our case as well. Mathematically, it means that the calculus we use is a central extension of a commutative differential calculus on the Heisenberg algebra, which is recovered by projection via $\theta'=0$. All of our results become empty if we set $\theta'=0$, which means that our entire point of view is purely quantum and not visible at the classical level. Rather, we find that the natural interpretation of this emergent 1-form is $\theta'=\extd t$ on an extended heisenberg algebra $\tilde A$ where we adjoin a central variable $t$. This fits in with the idea that highly noncommutative systems tend to generate their own evolution\cite{Ma:spo}. (This is different from but reminiscent of the observation that von Neumann algebras have an associated modular automorphism group.) In our specific case, and using this extra cotangent dimension, we will arrive at a rather unusual geometric picture in which the central 1-forms 
\begin{equation} \omega_i:=\extd p_i+{\del V\over \del x^i} \theta', \quad \eta^i:=\extd x^i- {p_i\over m}\theta'\end{equation}
are covariantly constant under $\nabla$ and killed by the geodesic velocity field so that $X(\omega_i)=X(\eta^i)=0$. If we identify $\theta'=\extd t$ for an external time variable $t$, then setting $\omega_i=\eta^i=0$ exactly reproduces a quantum version $\extd p_i=-{\del V\over \del x^i} \extd t, \extd x^i= {p_i\over m}\extd t$ of the Hamilton-Jacobi equations of motion in our approach. The antisymmetric quantum metric in Proposition~\ref{Gcov}  is 
\begin{equation*} G=\omega_i\tens\eta^i-\eta^i\tens\omega_i=\extd p_i\tens\extd x^i-\extd x^i\tens\extd p_i+O(\theta')\end{equation*}
where the second expression shows its origin in a lift of the classical symplectic 2-form $\omega=\extd p_i\wedge\extd x^i$  and the first expression shows that $G$  vanishes on solutions of the Hamilton-Jacobi equation of motion. The classical interior product $i_{X_h}(\omega)=\extd h$ saying that $X_h$ is the Hamiltonian vector field for the Hamiltonian function appears differently now in our extended phase space geometry as $(X\tens\id)(G)=(\id \tens X)(G)=0$, i.e. as the kernel of the antisymmetric quantum metric. 

Section~\ref{secKG} does the same as Section~\ref{secheis} but for $A$ now the Heisenberg algebra with $x^a,p_b$  Minkowski (eg 4-vectors and covectors) and with quantum differential calculus defined now by an external $U(1)$ gauge  potential. In this section, $t=x^0/c$ is a coordinate variable (with metric -1 in this direction and $c$ the speed of light) and the quantum geodesic time parameter is instead denoted $s$. Moreover, $[p_a,p_b]$ are no longer zero when there is electromagnetic curvature. This time,  we propose a novel  quantum `geodesic' evolution based on the Klein-Gordon operator, which was  not previously studied but is suggested by our formalism as a relativistic version of Section~\ref{secheis}. The geodesic parameter $s$ now plays a role more like proper time. Moreover, the differential algebra has a central 1-form 
\[\zeta=\extd t+{p_0\over m c} \theta'\]
and in the natural quotient of $\Omega^1_A$ where this is set to zero,  $\theta'$ has the same role as the relativistic proper time interval in relation to the Minkowski coordinate time interval $\extd t$. The geodesic time element $\extd s$ plays a similar role to $\theta'$ but as before  it is external to the calculus on $A$; identifying the two now imposes the time-dilation relation in a similar spirit to the way that we imposed the Hamilton-Jacobi equations in Section~\ref{secheis}. Our approach here is very different from previous discussions  of proper time in the Klein Gordon context,  such as \cite{KalVig} where the proper time and rest mass come from a canonically conjugate pair of observables. We illustrate our proposal with the easy case of a free particle in 1+1 Minkowski space, where we analyse a proper time wave packet centred around an on-shell Klein Gordon field (Example~\ref{freeflow}), and we also outline a proper time atomic model similar to a hydrogen atom. 

Section~\ref{secsym} ends the paper with a self-contained Poisson-level extended  phase space formalism as suggested by our results of Section~\ref{secheis} at the semiclassical level. This helps to clarify the geometric content of our constructions and also provides the physical meaning of $\nabla$ as infinitesimal data for the quantisation of the differential structure in the same way as a Poisson bracket is usually regarded as the data for the quantisation of the algebra. Some concluding remarks in Section~\ref{seccon} provide directions for further work.

\section{Preliminaries: Algebraic formulation of geodesics}\label{secpre}

Here, we give a minimal but self-contained account of the algebraic set up of differentials and connections and the formulation of quantum geodesics introduced in \cite{Beg:geo} in terms of $A$-$B$ bimodule connections\cite{BegMa}. A possibly noncommutative unital algebra $A$ equipped with an exterior algebra $(\Omega_A,\extd)$ will play the role of a manifold, and $B=\C^\infty(\R)$ expresses a geodesic time parameter $t$ with its classical differential $\extd t$. The formalism also allows for more general and possibly noncommutative $B$ and $(\Omega_B,\extd)$, but we will only need the classical choice in the present paper.  Proposition~\ref{propi} is a general version of the classical case treated in \cite{Beg:geo}, Proposition~\ref{propii} is essentially in \cite{Beg:geo} but reworked for right connections (which is needed to mesh later with conventions in quantum mechanics), while Corollary~\ref{corii} is new. 

\subsection{Algebraic set up and the case of a single geodesic}\label{seci}   
We recall that geodesics on a smooth Riemannian or pseudo-Riemannian manifold $M$ can be expressed
as the autoparallel condition $\nabla_{\dot \gamma}\dot\gamma=0$
for a curve $\gamma$ in $M$, parametrised appropriately. Explicitly, this is 
\begin{equation}\label{classgeo} \ddot\gamma^\mu + \Gamma^\mu{}_{\alpha\beta}\dot\gamma^\alpha\dot\gamma^\beta=0\end{equation} 
and as such makes sense for any linear connection on a manifold (it does not have to be the Levi-Civita connection for a  metric if we are not seeking to obey a variational principle). In quantum geometry, there is not yet a convincing calculus of variations and instead, by `geodesic', we mean this autoparallel sense with respect to any linear connection (albeit one of geometric interest). Also note that $\dot\gamma$ is not actually a vector field, being defined only along a particular curve. Fortunately $\nabla$ is only being taken along the same curve, but this does suggest that there there is a more geometric point of view. To explain it, we will need a fair bit of algebra. 

Our first task for an algebraic version  is the differential structure. If $A$ is any unital algebra, we define a  `differential structure' formally by fixing a bimodule $\Omega^1_A$ over $A$ of 1-forms. This means a vector space where we can associatively multiply by elements of $A$ from either side and a map $\extd: A\to \Omega^1_A$ sending a `function' to a `differential form' obeying the Leibniz rule $\extd(aa')=\extd a.a'+a.\extd a'$). In the $*$-algebra case over $\C$, we require  $\Omega^1_A$ to also have a $*$-operation for which $(a.\extd a')^*=(\extd a'{}^*).a^*$ for all $a,a'\in A$. One normally demands that $\Omega^1_A$ is spanned by elements of the form $a\extd a'$ for $a,a'\in A$, otherwise one has a generalised differential calculus.  Any $\Omega^1_A$ then extends to an exterior algebra $\Omega_A$ with exterior derivative increasing degree by 1 and obeying a graded-Leibniz rule and $\extd^2=0$. There is a canonical `maximal prolongation' of any $\Omega^1_A$ which will actually be sufficient in our examples, but one can also consider quotients of it for $(\Omega,\extd)$. In fact the choice of higher degrees does not directly impact the geodesic theory but is relevant to the torsion and curvature of a connection. We also define  left and right vector fields  as respectively left and right module maps $X:\Omega^1_A\to A$ (i.e. maps which are tensorial in the sense of commuting with the left and right multiplication by $A$). 

Next, let $A,B$ be unital algebras with differential structure and $E$ an $A$-$B$-bimodule (so we can associatively multiply elements of $E$ by elements of $A$ from the left and of $B$ from the right). We define a right $A$-$B$-connection\cite{BegMa}  on $E$ as a map $\nabla_E:E\to E\tens_B\Omega^1_B$ subject to two Leibniz rules. On the right, 
\begin{equation}\label{rightleib}  \nabla_E(e.b)=(\nabla_E e).b+e\tens\extd b,\quad \forall e\in E, b\in B\end{equation} 
as usual for a right connection in noncommutative geometry. From the other side, 
\begin{equation}\label{sigleib} \nabla_E(a.e)= a.\nabla_E e+\sigma_E(\extd a\tens e),\quad\forall e\in E, a\in A; \quad \sigma_E: \Omega^1_A\tens_A E\to E\tens_B \Omega^1_B\end{equation} 
for a certain bimodule map $\sigma_E$ as shown, called the `generalised braiding'. Being a bimodule map means that it is fully tensorial in the sense of commuting with the algebra actions from either side.  This map, if it exists, is uniquely determined by $\nabla_E$ and the bimodule structure and we say in this case that $\nabla_E$ is a (right) $A$-$B$-bimodule connection. 
If $X$ is a vector field $\Omega^1_B\to B$ then we have an associated covariant derivative $D_X=(\id\tens X)\nabla_E:E\to E$. The collection of categories ${}_A\CE_B$ of such $A$-$B$-bimodule connections itself forms a coloured bicategory\cite{BegMa} with a tensor product ${}_A\CE_B\times {}_B\CE_C\to {}_A\CE_C$ defined by 
\[ \nabla_{E\tens F}=(\id\tens \sigma_F)(\nabla_E\tens\id)+\id\tens\nabla_F\]
for all $(E,\nabla_E)\in {}_A\CE_B$ and $(F,\nabla_F)\in {}_B\CE_C$. We defer discussion of the $*$-operation to where we need it Section~\ref{secflow}. 

The above generalises  the monoidal category ${}_A\CE_A$ of right $A$-$A$-bimodule connections for any fixed differential algebra $A$. This diagonal case, in a  left-handed version, is more familiar in noncommutative geometry\cite{DV1,DV2,Mou}. 
By a linear connection on $A$, we mean an $A$-$A$-bimodule connection  $\nabla_{\Omega^1_A}$ with associated braiding $\sigma_{\Omega^1_A}$ (or just $\nabla$ with associated braiding $\sigma$ when the context is clear). In this case,  the covariant derivative associated to a left vector field $X$ will be denoted $\nabla_X:\Omega^1_A\to \Omega^1_A$. We will also adopt an explicit notation $\nabla\xi=\xi\bo\tens_A\xi\bt$ (summation understood), so that $\nabla_X(\xi)=\xi\bo X(\xi\bt)$ for all $\xi\in \Omega^1$.

We now express dependence on a time variable $t$ by values in an algebra $B=C^\infty(\R)$ with its usual $\extd t$ and the usual (commutative) bimodule structure on $\Omega^1_B$.  We take $\nabla\extd t=0$ as defining a trivial classical linear connection acting on this. Here $\nabla(b\extd t)=\extd b\tens_B\extd t=\dot b\extd t\tens\extd t$ and $\sigma(\extd t\tens\extd t)=\extd t\tens\extd t$. Now consider a linear connection $\nabla$ on $A$ and an $A$-$B$-bimodule connection $E$ with the domain and codomian of  $\sigma_E$ in (\ref{sigleib}). Each of the factors has a connection and hence we have two tensor product  $A$-$B$-bimodule connections
\begin{align}\label{nabtena}\nabla_{\Omega^1_A\tens_A E}&=(\id\tens \sigma_E)(\nabla \tens\id) + \id\tens \nabla_E:\Omega^1_A\tens_A E\to \Omega^1_A\tens_A E\tens_B \Omega^1_B\\
\label{nabtenb}\nabla_{E\tens_B\Omega^1_B}&=(\id\tens \sigma_{\Omega^1_B})(\nabla_E\tens\id) + \id\tens \nabla_B:E\tens_B\Omega^1_B\to E\tens_B\Omega^1_B\tens_B\Omega^1_B\end{align}
(albeit in our case $\sigma_{\Omega^1_B}$ acts as the identity map so one does not need to include it). Next, any bimodule map between $A$-$B$-bimodules with connection has covariant derivative $\doublenabla$ which measures the extent to which the map fails to intertwine the connections (classically, in the familiar diagonal case, this would be the induced covariant derivative of the map viewed as a tensor). With this machinery, \cite{Beg:geo} proposed 
\begin{align}\label{covsig}
\doublenabla(\sigma_E) &:= \nabla_{E\tens_B\Omega^1_B} \sigma_E-(\sigma_E\tens\id) \nabla_{\Omega^1_A\tens_A E}=0
\end{align}
as a kind of universal `geodesic equation' including and generalising (\ref{classgeo}), depending on the choice of $E$. 

To describe a single geodesic, note that a smooth curve in a manifold $M$ defines an algebra map $\gamma: A\to B$ compatible with the differential structures. Here $A=C^\infty(M)$ for the classical setting, but we can proceed at the algebraic level more generally. We say that $\gamma$ is `differentiable'  if it extends to an $A$-bimodule map $\gamma_*: \Omega^1_A\to \Omega^1_B$  for the pull-back action on $\Omega^1_B$ by $\gamma_*(a\extd a')=\gamma(a)\extd \gamma(a')$ for $a,a'\in A$, see \cite{BegMa}. In our case, since $\Omega^1_B$ has basis $\extd t$, we can also write $\gamma_*$ explicitly as
\begin{equation}\label{gammasquare} \gamma_*(\xi):=\gamma_*[\xi]\extd t,\quad \gamma_*[a\extd a']= \gamma(a)\dot\gamma(a')\end{equation}
where $\gamma_*[\xi]\in B$.  Also note that $E=B$ is an an $A$-$B$-bimodule by 
\[ a\cdot e= \gamma(a)e,\quad e.b=eb,
\quad \forall a\in A, e\in E, b\in B\]
and in this case $\nabla_E$, $\sigma_E$ are maps
\[ \nabla_E: E\to E\tens_B\Omega^1_B=\Omega^1_B,\quad \sigma_E:\Omega^1_A\tens_A E\to E\tens_B\Omega^1_B=\Omega^1_B.\] 
 with the proviso that $A$ acts from the left on $\Omega^1_B$ via $\gamma$. The trivial choice is $\nabla_E=\extd$. As above, we also fix the trivial linear connection with $\nabla\extd t=0$ on $B$. 
 
\begin{proposition}\label{propi} Let $A$ be a differential algebra with linear connection $\nabla\xi:=\xi\bo\tens_A\xi\bt$ and  let $\gamma:A\to B$ a differentiable algebra map  and $E=B$ an $A$-$B$-bimodule as above. Then  the trivial connection $\nabla_E e=\extd e = \dot e\tens \extd t$ is an $A$-$B$-bimodule connection with
\[ \sigma_E(\xi\tens_A e)=\gamma_*(\xi)e\]
and $\doublenabla(\sigma_E)=0$ reduces to 
\[ {\extd\over\extd t}\gamma_*[\xi]=\gamma_*[\xi\bo]\gamma_*[\xi\bt],\quad\forall \xi\in \Omega^1_A\]
where $\gamma_*[\xi]$ is defined by (\ref{gammasquare}).
\end{proposition}
\proof Here $\nabla_E e=\dot e\tens \extd t$ (the same as $\nabla$ on $B$). We have a right connection as this is the same as classically. Using this, the left action and that $B$ is commutative, we have \begin{align*}\nabla_E(a.e)&=\nabla_E(\gamma(a)e)=\nabla_E(e\gamma(a))=\nabla_E(e)\gamma(a)+e\dot\gamma(a)\tens\extd t\\
&=\dot e\gamma(a)\tens\extd t+ e\dot \gamma(a)\tens\extd t=\gamma(a)\dot  e\tens\extd t+ \dot \gamma(a)e\tens\extd t  \end{align*} and for the left Leibniz rule, this should equal
\[\sigma_E(\extd a\tens e)+a.\nabla_E(e)=\sigma_E(\extd a\tens e)+\gamma(a)\dot e\tens\extd t.\] Comparing these
and extending $\sigma_E$ as a left module map gives the formula stated, which is well-defined by the assumption that $\gamma$ is differentiable. Hence we have a bimodule connection. There is therefore a well-defined equation $\doublenabla(\sigma_E)=0$. Here $E\tens_B\Omega^1_B=\Omega^1_B$ has the trivial linear connection $\nabla\extd t=0$ but just viewed as an $A$-$B$-bimodule connection with $A$ acting by pull back along $\gamma$. Then
\[ \nabla_{\Omega^1_A\tens_A E}(\xi\tens e)= \xi\bo\tens_A\gamma_*(\xi\bt)e+\xi\tens_A\extd e\in \Omega^1_A\tens_A E\tens_B\Omega^1_B=\Omega^1_A\tens_A\Omega^1_B\]
with the above proviso for the left action of $A$ on $\Omega^1_B$. It follows that $\doublenabla(\sigma_E)=0$ appears as
\[ \nabla(\gamma_*(\xi)e)=( (\gamma_*\tens_B\gamma_*)\nabla\xi)e+\gamma_*(\xi)\tens_B\extd e\in \Omega^1_B\tens_B\Omega^1_B.\]
Expanding the linear connection on $B$ on the left by the right Leibniz rule we cancel the $\extd e=\dot e\tens\extd t$ term from both sides. This can be written more explicitly in terms of $B$ as
\[( {\extd\over\extd t}\gamma_*[\xi])e=\gamma_*[\xi\bo]\gamma_*[\xi\bt]e\]
and requiring this for all $e\in E$ is the condition stated. \endproof

This equation makes sense for any differential algebra. We could also have defined $\nabla_E e=(\dot e+e\kappa_t)\tens\extd t$ slightly more generally with the same $\sigma_E$, albeit this generalisation is of no particular interest at this level. If $A=\C^\infty(M)$ and $\gamma:\R\to M$ is a smooth curve, then in local coordinates, $\gamma_*[\xi_i\extd x^i]=\gamma^*(\xi_i)\dot\gamma^i\in B$, where $\gamma^*(\xi_i)(t)=\xi_i(\gamma(t))$ pulls back the coefficients of a 1-form $\xi$. If we also write $\nabla\extd x^i=-\Gamma^i{}_{jk}\extd x^j\tens\extd x^k$ for Christoffel symbols $\Gamma^i{}_{jk}$, then the algebraic geodesic equation
reduces to (\ref{classgeo}), as analysed in \cite{Beg:geo}.  Thus, we have introduced $\doublenabla(\sigma_E)=0$ in generality as the notion of a `geodesic bimodule' and shown that the simplest choice $E=B$ as noncommutative bimodule where the left action is defined by a curve $\gamma$ reduces to a single classical geodesic in the classical case. 
 
\subsection{Algebraic setting of geodesic velocity fields}\label{secii} In noncommutative geometry there can often not be enough algebra maps and one indeed needs a more general concept such as a correspondence or, in our case, an $A$-$B$-bimodule $E$. The root of this is that if there are not enough points, one should not expect enough curves either if these are defined pointwise. The natural thing to do here from a physical point of view is to consider not one geodesic but a density distribution of them with every point moving on a geodesic. Their collective tangents define  a time-dependent velocity field $X_t$ (a path in the space of vector fields)  subject to 
the {\em velocity equation}
\begin{equation}\label{veleq} \dot X_t+ \nabla_{X_t}X_t=0.\end{equation}
The idea is to take $X_t$ as the starting point and first solve this equation with some initial value $X_0\in {\rm Vect}(M)$. Any one geodesic is then recovered as a curve $\gamma(t)$ such that
\begin{equation}\label{geoX}\dot\gamma(t)= X_t(\gamma(t)),\quad \dot\gamma(0)=X_0(\gamma(0)).\end{equation}
This is different from the notion of `geodesic spray', being more directly tied to the manifold itself. 

In algebraic terms, this is just our universal $\doublenabla(\sigma_E)=0$ equation for a different choice of bimodule, namely now $E=A\tens B$, where $A$ is a differential algebra in the role of the manifold equipped with a linear connection and $B=C^\infty(\R)$ as above with its classical calculus and trivial linear connection $\nabla\extd t=0$.  More precisely, for a topological algebra, we can take $E=C^\infty(\R,A)$ with the left action by $A$ and right action by $B$ when viewed as subalgebras in the obvious way, but to keep things simple we will give formulae for $A\tens B$. 

\begin{proposition} \label{propii} cf\cite{Beg:geo} A right $A$-$B$-bimodule connection $\nabla_E$ on $E=A\tens B$ has the form
\[
\sigma_E(\xi\tens e)=X_t(\xi.e)\tens\extd t,\quad 
\nabla_E e=(\dot e+e\kappa_t+X_t(\extd e))\tens\extd t,
\]
where $X_t$ is a left vector field on  $A$ and  $\kappa_t\in A$. Let  
$\nabla:\Omega^1_A\to \Omega^1_A\tens_A \Omega^1_A$ be a right bimodule connection. Then
\begin{enumerate}\item    $\doublenabla(\sigma_E)$ is a bimodule map if and only if   \[
(\sigma_E\tens\id)(\id\tens\sigma_E)\big((\sigma-\id)\tens\id\big)=0, \]
which is equivalent to
\[
X_t(\id\tens X_t)(\sigma-\id)=0;
\]
\item  $\doublenabla(\sigma_E)=0$ if and only if in addition, for all $\xi\in\Omega^1_A$, 
\[
\dot X_t(\xi) + [X_t, \kappa_t ] (\xi) +X_t(\extd X_t(\xi)  )  
 - X_t (\id\tens X_t) \nabla(\xi)=0\ .
\]
\end{enumerate}
\end{proposition}
\proof This is a right-handed version of a result for left connections in \cite{Beg:geo}, but we include a brief proof for completeness. Since $\Omega^1_A\tens E=\Omega^1_A\tens B$ and $E\tens_B\Omega^1_B= A\tens \Omega^1_B$ in the obvious way, and since $\extd t$ is a basis of $\Omega^1_B$, the content of the bimodule map $\sigma_E$ is a bimodule map $\Omega^1_A\tens B\to A\tens B$ which when restricted to $\Omega^1_A\tens 1$ implies it is given by a time dependent left vector field $X_t$ on $A$ as stated. 

Next, writing $e=a\tens f(t)$ we have $\nabla_E(e)=\nabla_E(a.1\tens f)=a.\nabla_E(1\tens 1. f) +\sigma_E(\extd a\tens f)= a.\nabla_E(1\tens1).f+ a\tens \extd f +\sigma_E(\extd a\tens 1).f$ gives the formula for $\nabla_E$, for some undetermined $\nabla_E(1\tens 1)=\kappa_t\tens\extd t$ and some $\sigma_E(\extd a\tens 1)=X_t(\extd a)\tens \extd t$.

Next, by similar arguments to those at the start of the proof of \cite[Lemma 4.13]{BegMa}, we see that $\doublenabla(\sigma_E)$ is a bimodule map if and only if  
\[
(\sigma_E\tens\id)\sigma_{\Omega^1_A\tens E} = \sigma_{E\tens \Omega^1_B} (\id\tens\sigma_E), 
\]
where $\sigma_{\Omega^1_A\tens E} = ( \id\tens\sigma_E  )(\sigma\tens\id)$ and 
 $\sigma_{E\tens \Omega^1_B}  = \sigma_E\tens\id$ as $\sigma_{\Omega^1_B}$ is the identity, which is (1).  Since $\sigma_E$ is given by $X_t$, we obtain the second form as a map $\Omega^1_A\tens_A\Omega^1_A\to A$.  

For (2),  as  $\doublenabla(\sigma_E)$ is a right module map, we only have to calculate, for $\xi\in\Omega^1_A$,
\begin{align*}
\doublenabla(\sigma_E)(\xi\tens 1) &=  \big(\id\tens\nabla_{\Omega^1_B}+(\id\tens\sigma_{\Omega^1_B})(\nabla_E\tens\id)\big)(X_t(\xi)\tens\extd t) \cr
&\quad -  (\sigma_E\tens\id) 
 \big(\id\tens\nabla_E + (\id\tens\sigma_E)(\nabla\tens\id)\big)(\xi\tens 1)   \cr
 &= \nabla_E(X_t(\xi)) \tens\extd t -  (\sigma_E\tens\id) 
 \big(\xi \tens\nabla_E(1) + (\id\tens\sigma_E)(\nabla(\xi)\tens 1)\big) \cr
 &= \big(\dot X_t(\xi) + X_t(\xi)  \kappa_t  +X_t(\extd X_t(\xi)  )  - X_t(\xi  \kappa_t ) 
 - X_t (\id\tens X_t) \nabla(\xi) \big)\tens\extd t   \tens\extd t     
\end{align*}
and the vanishing of this is  (2). As $X_t$ is a left vector field, we wrote $(X_t.\kappa_t)(\xi)=X_t(\xi)  \kappa_t $ and 
$(\kappa_t.X_t)(\xi)=X_t(\xi  \kappa_t) $.  \endproof

It is also possible to restate the conditions in Proposition~\ref{propii} in terms of a connection on the vector fields, bypassing the 1-forms entirely, but we have to be careful of the sides of the connections. 
To do this, we first assume that $\nabla$ above has $\sigma$  invertible. In this case,  \cite[Lemma~3.70]{BegMa} tells us that $\nabla^L=\sigma^{-1}\nabla$ is a left connection on $\Omega^1_A$. If we further assume that $\Omega^1_A$ is left finitely generated projective as a left module, which classically reduces to saying that the cotangent space is locally trivial, then by  \cite[Prop.~3.80]{BegMa} we can dualise a left bimodule connection$\nabla^L$  on $\Omega^1_A$ to a right one $\nabla_\cX$ on $\cX$. 
In terms of the evaluation map $\ev:\Omega^1_A\tens_A \cX\to A$, we have
\begin{align*}
\extd\, \ev(\xi\tens X) &= \big( (\id\tens\ev)(\nabla^L\tens\id) +(\ev\tens\id)(\id\tens\nabla_\cX)\big)(\xi\tens X),\\
(\id\tens\ev)(\sigma^L\tens\id) &= (\ev\tens\id)(\id\tens\sigma_\cX): \Omega^1_A \tens_A \Omega^1_A \tens_A \cX\to \Omega^1_A\ .
\end{align*}
We can also define $\sigma_{\cX\cX}  :   \cX \tens_A \cX \to   \cX \tens_A \cX$ such that
\[
(\ev\tens\id)(\id\tens\sigma_{\cX\cX})= (\id\tens\ev)(\sigma_\cX\tens\id):
\Omega^1_A \tens_A \cX \tens_A \cX\to \cX\ .
\]

\begin{corollary}\label{corii} In Proposition~\ref{propii}, let $\nabla$ have $\sigma$  invertible,  $\Omega^1_A$ be left finitely generated projective as a left module and $\nabla_\cX$ be the associated right connection on $\cX$. In these terms, the corresponding conditions are
\begin{enumerate}\item $\sigma_{\cX\cX}(X_t\tens X_t)=X_t\tens X_t$; 
\item $ \dot X_t +[X_t,\kappa_t]+(\id\tens X_t)\nabla_\cX(X_t)=0.$ 
\end{enumerate}
\end{corollary}
\proof 
From Proposition~\ref{propii}  (1), we have
$X_t(\id\tens X_t)\sigma=X_t(\id\tens X_t)$, so (2) can be rewritten as
\[
\dot X_t(\xi) + [X_t, \kappa_t ] (\xi) +X_t(\extd X_t(\xi)  )  
 - X_t (\id\tens X_t)\sigma^{-1} \nabla(\xi)=0\ ,
\]
which by duality is the displayed equation (2). The other part is given by
\[
\ev(\id\tens\ev\tens\id)(\sigma^L\tens X_t\tens X_t)=\ev(\id\tens\ev\tens\id)(\id\tens\id\tens\sigma_{\cX\cX}(X_t\tens X_t))
\]
as a function $:\Omega^1_A \tens_A\Omega^1_A \to A$,  where 
$\sigma_L=\sigma^{-1}$ is inverse to $\sigma$ in  Proposition~\ref{propii} (1).
\endproof

\subsection{Probabilistic geodesic flow and the $\nabla_E e=0$ equation}\label{secflow} Next, returning to our classical model at the start of Section~\ref{secii},   if we have a perfect fluid with an evolving density $\rho(t)$ on a manifold $M$, where each particle moves according to a velocity field $X_t$, then conservation of mass (the continuity equation in fluid mechanics\cite{fluid}) dictates
\[ \dot\rho + X_t(\extd \rho)+ \rho\, {\rm div} (X_t)=0.\]
In our algebraic formulation, we were led to $e\in E=C^\infty(\R,C^\infty(M))$ or $e(t)\in C^\infty(M)$ at each $t$, with complex values, and we now identify $\rho(t)=\overline{e(t)}e(t)$ as playing the role of the probability density. Its evolution then corresponds to 
\begin{equation}\label{classe} \dot e + X_t(\extd e)+ e \kappa_t=0,\quad \kappa_t +\overline\kappa_t=  {\rm div}(X_t)  \end{equation}
for the amplitude $e(t)$, which is exactly $\nabla_E e=0$ in the classical limit of Proposition~\ref{propii}. For an actual probabilistic interpretation, we need a measure and to maintain the total probability with respect to it.  For example, in the Riemannian case with the Levi-Civita connection, we want to maintain
\[ \phi_0(\rho(t)):=\int_M\extd x \sqrt{|g|} \rho(t)=1\]
as the probability density $\rho(t)$ evolves. Here $|g|$ is the determinant of $g_{\mu\nu}$. There is an associated inner product 
\[  \<f(t) | e(t)\>= \phi_0(\bar f(t) e(t))\]
where $e(t),f(t)\in L^2(M)$ with respect to the Riemannian measure as above, and we used the usual bra-ket notation. From this point of view, $\nabla_E e=0$ ensures that $\<e(t)|e(t)\>=1$ as $e(t)$ evolves. Thus, our approach to geodesics in Section~\ref{secii} leads us naturally into a quantum mechanics-like interpretation, even though we are doing classical geodesics with $A=C^\infty(M)$.

For the algebraic formalism, we take $A$ and $B$ $*$-algebras with $*$-differential structures as in Section~\ref{seci}. Any  $A$-$B$-bimodule $E$ has a conjugate  
$\overline{E}$, which is a $B$-$A$-bimodule with elements $\overline{e}$ for $e\in E$ and vector space structure $\overline{e}+\overline{f}=\overline{e+f}$ and $\overline{\lambda\,e}=\bar\lambda\,\overline{e}$ for $\lambda\in \C$ and $e,f\in E$, see \cite{BegMa}. The algebra actions are
$\overline{e}.a=\overline{a^*.e}$ and $b.\overline{e}=\overline{e.b^*}$ for $a\in A$ and $b\in B$. Now suppose that $E$ is equipped with a $B$ valued inner product $\<\,,\>:\overline{E}\tens_A E\to B$ which is bilinear and hermitian in the sense $\<\overline{e},f\>^*=\<\overline{f},e\>$, where $\<\overline{e},a.f\>=\<\overline{e}.a,f\>$ for all $a\in A$. When $B$ is a dense subalgebra of a $C^*$-algebra, we call the inner product positive if 
$\<\overline{e},e\>>0$ for all $e\in E$.  In this context, a right $A$-$B$-bimodule connection $\nabla_E$ is said to preserve the inner product if for all $e,f\in E$ we have\cite{BegMa}
 \begin{equation}\label{invariantinner}
 \extd \<\overline{e},f\>=(\id\tens\<\,,\>)(\nabla_{\overline E}(\overline e)\tens f) + (\<\,,\>\tens\id)(\overline e\tens \nabla_E(f))\ .
 \end{equation}
Here the {\em left} connection $\nabla_{\overline E}:\overline{E}\to \Omega^1_B\tens_B \overline{E}$ is defined by
 $\nabla_{\overline E}(\overline e)=\xi^*\tens\overline p$ if  $\nabla_E(e)=p\tens \xi$ (sum of such terms implicit).  Both $A$ and $B$ could be noncommutative.  
 
 In our case of interest,  $B=C^\infty(\R)$ and we consider the $B$-valued output to define a function of `time' $t\in\R$ with $t^*=t$. Then $\extd$ on the left is  derivative in the $\R$ coordinate. Hence, {\em if} $\nabla_E$ preserves $\<\ ,\ \>$ and $e$ obeys $\nabla_Ee=0$ as above for geodesic evolution then $\frac{\extd}{\extd t}\<\overline{e},e\>=0$. We adopted a more mathematical notation but this is equivalent to the usual bra-ket notion other than the values being in $B$.  
 
It remains to analyse the content of inner product preservation for our specific $E$ where  $E=A\tens B$ (or $E=C^\infty(\R,A)$). Since $A$ could be noncommutative, instead of a measure we fix a positive linear functional $\phi_0:A\to \C$ or `vacuum state' and define $\<\bar f , e\>=\phi_0(f^*e)$ as above, pointwise at each $t$ so that the result is $B$-valued. Equivalently, we can suppose we are given $\<\ ,\ \>$ and define $\phi_0(a)=\<\bar 1, a\>$, where $a$ is viewed in $E$ as constant in time. 

\begin{proposition}  \label{presip} \cite{Beg:geo}
The connection on $E=A\tens B$ in Proposition~\ref{propii} preserves the inner product on $E$ if and only if 
for all $a\in A$ and $\xi\in\Omega^1_A$, 
\[
 \< \overline{1}\ , \kappa_t{}^*a+ a\kappa_t+ X_t(\extd a) 
 \>=0, \quad \< \overline{1}, X_t(\xi^*)-X_t(\xi)^*\>=0.
\]
\end{proposition}
\proof  This is again from \cite{Beg:geo} but we provide a short explanation. The condition for preservation is, for $a,c\in A$,
\begin{eqnarray*}
0 &=& \<  \overline{   c\kappa_t+X_t(\extd c)  } ,  a \> + \<\overline{ c},  a\kappa_t+X_t(\extd a)  \>.\extd t \cr
&=&    \< \overline{1}\ ,\big(\kappa_t{}^*c^*a+X_t(\extd c)^*a+ c^*a\kappa_t+ c^*X_t(\extd a) 
\big) \>.\extd t     \cr
&=&   \< \overline{1}\ ,\big(\kappa_t{}^*c^*a+X_t(a^*\extd c)^*+ c^*a\kappa_t+ X_t(c^*\extd a) 
\big) \>.\extd t    
\end{eqnarray*}
and putting $c=1$ gives the first displayed equation. Using this with $c^*a$ instead of $c$ in the condition for preservation gives
\begin{eqnarray*}
0 =  \< \overline{1}\ ,X_t(a^*\extd c)^* - X_t(\extd(c^*a))+ X_t(c^*\extd a)  \> =  \< \overline{1}\ , X_t(a^*\extd c)^* - X_t(\extd c^*.a) 
 \> \ .\quad\square
\end{eqnarray*}

\medskip We call the first of the displayed equations in Proposition~\ref{presip} the \textit{divergence condition} for $\kappa_t$ and the second the \textit{reality condition} for $X_t$. The first generalises the second half of (\ref{classe}) to potentially noncommutative $A$ and the second would be automatic on a real manifold.  When $A$ is a noncommutative, one cannot think of $\rho(t)=e(t)^*e(t)$ as a time-dependent probability density, but rather we adopt the usual formalism of quantum theory where any $e$ implies an associated  positive linear functional $\phi:A\to B$ or `state' given by 
\begin{equation}\label{phia} \phi(a)=\<e|a|e\>=\<\bar e, a.e\>=\<\overline{a^*.e},e\>=\phi_0(e^*ae)\end{equation}
 in our two notations for the inner product. Here, positive means  $\phi(a^*a)\ge 0$ for all $a\in A$ and usually we normalise it so that  $\phi(1)=\<\bar e,e\>=\<e|e\>=1$ as we have assumed above.   If $A$ and $B$ were $C^*$-algebras then we would have the standard notion \cite{Lance} of a Hilbert $C^*$ bimodule upon completion with respect to the induced norm $|e|^2=\|\<\overline{e},e\>\|_B$. In our case of interest,  $B=C^\infty(\R)$ and for every $e\in E$ we have a possibly un-normalised state $\phi_t$ at each time defined by $\phi_t(a)=\<e(t)|a|e(t)\>$. 

\section{Quantum mechanical geodesics on Hilbert spaces}\label{seciii} 

We now go beyond the noncommutative differential geometric formulation of geodesics\cite{Beg:geo} covered in the preceding section, extending this to a $*$-algebra $A$ of observables represented on a Hilbert space $\CH$ as in quantum mechanics. We still employ the `universal equation' $\doublenabla(\sigma_E)=0$ but for a new choice of bimodule $E=\CH\tens B$ with $B=C^\infty(\R)$, or more precisely $E=\C^\infty(\R,\CH)$ with its canonical $A$-$B$-bimodule structure
\[(a.\psi)(t)=\rho_t(a)\psi(t),\quad (\psi.b)(t)=\psi(t)b(t).\]
Here $\psi\in E$ and $\psi(t)\in \CH$, while $\rho_t$ at each $t$ is a representation of $A$ on a vector space $\CH$ (we will only use the constant case where $\rho$ is fixed but the more general case costs little to include and will be needed to recover the case of a single geodesic). Here $\rho_t$ should not be confused with probability densities $|\psi|^2$ which we no longer consider separately. We let $L(\CH)$ be the (possibly unbounded) linear operators from $\CH$ to itself, and make this into an $A$-bimodule in the obvious way by $a.T=\rho_t(a)\circ T$ and $T. a=T\circ \rho_t(a)$ for all $T\in L(\CH)$. 

\begin{lemma}\label{yyuuoo} In this context, a bimodule map $\sigma_E:\Omega^1_A\tens_A E\to E\tens_{C^\infty(\R)} \Omega^1(\R)$ necessarily has the form
\[
\sigma_E(\xi\tens \psi) = \tilde X(\xi)(\psi)\tens\extd t
\]
for all $\xi\in \Omega^1_A$,  for some  bimodule map $\tilde X:\Omega^1_A\to L(\CH)\tens C^\infty(\R)$. 
\end{lemma}
\proof As $\sigma_E$ is a right $C^\infty(\R)$-module map and $\Omega^1(\R)$ has basis $\extd t$, we can write  $\sigma_E(\xi\tens \psi ) = \tilde X(\xi)(\psi)\tens\extd t$ for $\psi \in \CH$ and some linear map $\tilde X\in L(\CH,\CH\tens C^\infty(\R))$, which is more or less $L(\CH)\tens C^\infty(\R)$. Now
\begin{align*}
&\tilde X(\xi a)(\psi )=\sigma_E(\xi a\tens \psi )=\sigma_E(\xi \tens  \rho(a)\psi )=\tilde X(\xi)\rho(a)(\psi )\ ,\\
& \tilde X(a \,\xi )(\psi )=\sigma_E(a\xi \tens \psi )=a\,\sigma_E(\xi \tens \psi )=\rho(a)\tilde X( \xi )(\psi )\ .
\end{align*}
gives the bimodule map.
 \endproof 

If we think of being valued in $C^\infty(\R)$ as dependence on $t$,  the required data will be  an `operator-valued  time-dependent vector field' which for each $t$ is a bimodule map $\tilde X_t:\Omega^1_A\to L(\CH)$ in the sense  $\tilde X_t(a.\xi)=\rho_t(a) \tilde X_t(\xi)$ and $\tilde X_t(\xi.a)=\tilde X_t(\xi)\rho_t(a)$. 

\begin{lemma}\label{Xiii} For a time dependent operator $\ch_t\in C^\infty(\R, L(\CH))$ we define a right $C^\infty(\R)$ connection on $E=\C^\infty(\R,\CH)$ by 
\[  (\nabla_E\psi)(t)=\left(\dot\psi(t)+ \ch_t(\psi(t))\right)\tens\extd t\ .\]
This is a bimodule connection if and only of  $\tilde X_t(\extd a)=[\ch_t,\rho_t(a)]+\dot\rho_t(a)$ extends to a well defined bimodule map $\tilde X_t:\Omega^1_A\to L(\CH)$ at each $t$, in which case $\sigma_E(\xi\tens \psi)(t)=\tilde X_t(\xi)(\psi(t))\tens\extd t.$
\end{lemma}
\proof We prove the result in the given direction but, because of the complications of unbounded operators and topologies, we just give a motivation as to why this is a reasonably general case. 
First the linear map $\nabla_E: E\to E\tens_B \Omega^1_B$ when restricted to time independent $\psi\in \CH$
should be of the form $(\nabla_E\psi)(t)= \ch_t(\psi(t))\tens\extd t$ for some time dependent linear operator $\ch_t$. Then multiplying $\psi\in \CH$ by a function of time and using the Leibniz rule (\ref{rightleib}) gives the form for general
$\psi\in E$ in the statement. Now we assume that this assumption is true in what follows. 
\newline\indent
For a bimodule connection we can similarly justify a time dependent linear map $\tilde X_t:\Omega^1_A\to  L(\CH)$ 
so that $\sigma_E(\xi\tens \psi)(t)=\tilde X_t(\xi)(\psi(t))\tens\extd t$. Now $\tilde X_t$ is, by definition of $\sigma_E$, a left $A$-module map. Also since $\sigma_E$ is a map from $\Omega^1_A\tens_A \CH$ we see that
$\tilde X_t(\xi)\rho_t(a)=\tilde X_t(\xi.a)$, i.e.\ $\tilde X_t$ is a right $A$-module map. Now we use (\ref{sigleib}) to write
\begin{align*}
\sigma_E(\extd a\tens e) &=\nabla_E(a.\psi) - a.\nabla_E \psi =
\nabla_E(\rho_t(a)(\psi)) - \rho_t(a)\nabla_E \psi \\
&= \big(  \dot \rho_t(a)(\psi) + \rho_t(a)( \dot \psi) +    \ch_t \rho_t(a)(\psi)
-\rho_t(a)\dot\psi - \rho_t(a) \ch_t(\psi)
 \big)\tens\extd t \\
 &= ([\ch_t,\rho_t(a)]+\dot\rho_t(a))(\psi(t))\tens\extd t
\end{align*}
as required. We require this to be a well-defined bimodule map as analysed in Lemma~\ref{yyuuoo}. 
\endproof

There are two important comments to make about this result. The first is that the well definedness of $\tilde X_t$ could be read, starting with the calculus $\Omega^1_A$ as a condition on $\ch_t$. However, in this paper we prefer to start with $\ch_t$ and read the condition as a constraint on the calculus $\Omega^1_A$. The second is that in many applications the time dependent operator $\ch_t$ will be the action of a time dependent element of the algebra $H_t\in A$ via the representation, i.e.\ $\ch_t=\rho_t(H_t)$. On the assumption that the representation is faithful, the condition in Lemma~3.2 for a bimodule connection is then equivalent to whether there is a well defined bimodule map $X_t:\Omega^1_A\to A$ satisfying
\begin{equation}\label{kappat} \rho_t(X_t(\extd a))=\rho_t([H_t,a])+\dot\rho_t(a).\end{equation}
 This $X_t$ would then be a geometric time dependent vector field on $A$ and $\tilde X_t$ in the lemma is then its image.  This is typically the case as follows. Assuming that the left $A$-module $\CH$ has a cyclic vector (often a vacuum vector in physics), i.e.\ $x_0\in \CH$ so that $\{\rho_t(a)(x_0):a\in A\}$ is dense in $\CH$ at each $t$ and that the operators contain $x_0$ in their domain, we have
$\tilde X(\xi)(\rho_t(a)(x_0))=\tilde X(\xi a)(x_0)\in \CH$. This means that the left module map $\widetilde X:\Omega^1_A\to \CH\tens C^\infty(\R)$
given by $\widetilde X(\xi)=\tilde X(\xi)(x_0)$
 recovers the bimodule map $\tilde X$ on the dense subset $\rho_t(A)(x_0)$ by the formula
 $\tilde X(\xi)(\rho_t(a)(x_0))=\widetilde X(\xi a)$. If we  further suppose that $\rho_t$ satisfies $\rho_t(a)(x_0)=\rho_t(a')(x_0)$ only when $a=a'$, and that every $\tilde X_t(\xi)$ maps the cyclic vector $x_0$ into the dense subset $\rho_t(A)(x_0)$, then we obtain  a left module map $X:\Omega^1_A\to A\tens C^\infty(\R)$ characterised by
 \[
\rho_t\big( X_t(\xi)\big)(x_0)= \tilde X_t(\xi)(x_0).
 \]
 As $\tilde X$ is also a bimodule map, it follows that $\rho_t\big( X_t(\xi)\big)=\tilde X_t(\xi)$ on the dense subset $\rho_t(A)(x_0)$, and therefore that $\rho_t\big( X_t(\xi)\big)=\tilde X_t(\xi)$ as desired. 

We now proceed as in Section~\ref{secpre} to take the trivial linear connection $\nabla\extd t=0$ acting on on $\Omega^1_B$ and an arbitrary linear right connection $\nabla$ on $\Omega^1_A$, and the tensor product connections (\ref{nabtena})-(\ref{nabtenb}) on the domain and codomain of $\sigma_E$. 

\begin{proposition}\label{propiii}  For $E=C^\infty(\R,\CH)$ and $\nabla_E$ defined by $\ch_t$, $\doublenabla(\sigma_E)=0$ is equivalent to an auxiliary condition of the same form as (1) in Proposition~\ref{propii} and the further condition
\[ 
 [\ch_t, \tilde X_t(\xi)]+ \dot {\tilde X}_t(\xi)=\tilde X_t(\id\tens \tilde X_t)\nabla\xi,\quad \forall \xi\in \Omega^1_A\ .\]
 This can be read as a time evolution equation for $\tilde X_t$, and this is consistent with an initial $A$-bimodule map $\tilde X_0$ giving a bimodule map $\tilde X_t$ for $t\ge 0$. (In the absence of a uniqueness result for the differential equation we cannot make a stronger statement.) For such a bimodule map the evolution equation is determined by its value on $\xi=\extd a$, 
\begin{equation*}  [\ch_t,[\ch_t, \rho_t(a)]]+[\dot \ch_t, \rho_t(a)]+  2[\ch_t,\dot\rho_t(a)]+\ddot\rho_t(a)=\tilde X_t(\id\tens \tilde X_t)\nabla\extd a,\quad \forall a\in A. \end{equation*}
\end{proposition}  
\proof 
To calculate $\doublenabla( \sigma_E)$, we need
\begin{align*}
\nabla_{E\tens\Omega^1_\R}   \sigma_E(\xi\tens \psi) &= \nabla_{E\tens\Omega^1_\R}  ( \tilde X_t(\xi)(\psi(t))\tens\extd t) \\
&=\big(       \tilde X_t(\xi)( \dot\psi(t))  + \dot{\tilde X}_t(\xi)(\psi(t))  + \ch_t \tilde X_t(\xi)(\psi(t))    \big)\tens\extd t  \tens\extd t
\end{align*}
and also
\begin{align*}
( \sigma_E\tens\id)  & \nabla_{\Omega^1_A\tens E}(\xi\tens \psi) =( \sigma_E\tens\id)  ( \id\tens \sigma_E  )(\nabla\xi  
\tens\psi) +   \sigma_E( \xi \tens  (\dot\psi(t)+ \ch_t\psi(t)) )\tens\extd t \\
&= \big( (\tilde X_t(\id\tens \tilde X_t)\nabla\xi)\psi + \tilde X_t(\xi)  (\dot\psi(t)+ \ch_t\psi(t)) \big)  \tens\extd t  \tens\extd t
\end{align*}
so we get the equation
\begin{align*}
 (\tilde X_t(\id\tens \tilde X_t)\nabla\xi)\psi + \tilde X_t(\xi)  (\dot\psi(t)+ \ch_t\psi(t)) =  \tilde X_t(\xi)( \dot\psi(t))  + \dot{\tilde X}_t(\xi)(\psi(t))  + \ch_t \tilde X_t(\xi)(\psi(t)) 
\end{align*}
which gives the first displayed equation in the statement. Replacing $\xi$ with $\xi.a$ in the displayed equation and using $\tilde X_t$ being a right module map gives $\dot{\tilde X}_t(\xi.a)=\dot{\tilde X}_t(\xi)\rho_t(a)+ \tilde X_t(\xi)\dot \rho_t(a)$, showing that the time evolution equation for $\tilde X_t$ is consistent with $\tilde X_t$ being a right $A$-module map. Similarly replacing $\xi$ with $a.\xi$ shows that the evolution equation for $\tilde X_t$ is consistent with $\tilde X_t$ being a left $A$-module map, but this requires using (1) of Proposition~\ref{propii}. Now just use the formula for $\tilde X_t(\extd a)$ for the second displayed equation in the statement.  \endproof

The single geodesic case of Section~\ref{seci} is recovered by $\CH=\C$ and $\rho_t(a)=a(\gamma(t))$ i.e. the evaluation representation along the image of the curve in the classical case, or $\rho_t(a)=\gamma(a)(t)$ in the algebraic version. In this case, $\ch_t$ is some function of $t$ and does not enter, while 
\[ \tilde X:\Omega^1(M)\to C^\infty(\R),\quad \tilde X_t(a\extd x^i)=a(\gamma(t))\dot\gamma^i(t)\]
in the classical case or $\tilde X_t(\xi)=\gamma_*[\xi]$ for $\xi\in \Omega^1_A$ in the algebraic version.  Then Proposition~\ref{propiii} reduces to Proposition~\ref{propi}. 

Similarly, the geodesic flow of Section~\ref{secii} is recovered with $\CH=C^\infty(M)$ (or some completion thereof) in the classical case or $\CH=A$ with the left regular representation (or a completion thereof) in a potentially noncommutative version.  In the classical case, compatibility with the calculus $\Omega^1(M)$ requires the Hamiltonian to have the form $\ch_t= \tilde X_t+ \kappa_t$ for some time dependent vector field $\tilde X_t$ and function $\kappa_t$ acting by left multiplication on $M$ and the condition in Propostion~\ref{propiii} reduces to the velocity field equation (\ref{veleq}). In the noncommutative case with $\rho$ the left regular representation, we have discussed in (\ref{kappat})  how $\tilde X_t$ can arise as the image of a bimodule map $X_t$. This is not quite as general as Proposition~\ref{propii}, where we only assumed a left vector field.

It remains to extend Section~\ref{secflow} to our more general setting. We suppose that $\rho_t$ is unitary for each $t$.  $\phi_t(a)=\<e(t)|a|e(t)\>=\<e(t)|\rho_t e(t)\>$ or $\phi(a)=\<\bar e,a.e\>$ when viewed as a $B$-valued inner product. The main  difference is that now the inner product is assumed in the Hilbert space and not given by a vacuum state $\phi_0$ or a preferred element $1\in E$ as was possible before. 

\begin{corollary} In the setting of Proposition~\ref{propiii}, the inner product $\<\ ,\ \>:\overline{E}\tens_A E\to B$ is preserved by $\nabla_E$  if and only if $\ch_t$ is antihermitian.   \end{corollary}
\proof For $\psi,\zeta\in E$ we have
\[
(\id\tens\<,\>)(\nabla_{\overline{E}}(\overline{\zeta})\tens\psi)+ (\<,\>\tens\id)
(\zeta\tens \nabla_{E}(\psi))=(\<\overline{(\dot\zeta +\ch_t \zeta)} ,\psi \> + \<\overline{\zeta} ,\dot\psi+\ch_t \psi  \>  )\extd t
\]
and this is required to equal 
\[
\frac{\partial}{\partial t} \<\overline{\zeta},\psi\> \, \extd t = 
(\<\overline{\dot\zeta} ,\psi \> + \<\overline{\zeta} ,\dot\psi\>  )\extd t, 
\]
which is just the condition that $\ch_t$ is antihermitian. 
 \endproof

If we consider Proposition~\ref{propii} with $\tilde X_t$ a bimodule map as a special case with $\ch_t$ as in (\ref{kappat}), then $\ch_t$ anti-hermtian essentially reduces to the two conditions in Proposition~\ref{presip}.

We close with a couple of remarks in the case where the representation $\rho$ does not depend on $t$. The first is, by the same reasoning as in \cite{Beg:geo}, if $\psi$ obeys $\nabla_E\psi=0$  then 
\begin{equation}\label{dexpdt}
\frac{\extd}{\extd t} \<\psi | a | \psi \>= \<\psi  | \tilde X_t(\extd a) | \psi \>\ .
\end{equation}
This comes from $\nabla_E$ preserving the inner product and the definition of $\sigma$, so that $
\extd\<\psi|a\psi\>=(\<|\ |\>\tens\id)(\id\tens\sigma)(\bar\psi\tens\extd a\tens\psi)=\<\psi|\tilde X_t(\xi) |\psi\>\,\extd t.$ 
In terms of cochain complexes, this says that the expectation map $\mathbb{E}:A\to C^\infty(\mathbb{R})$ is a cochain map. 

Our second remark is that if we are in the setting of (\ref{kappat}) where $\tilde X_t$ is the image of a bimodule map  $X_t(\extd a)=[H_t,a]$ for all $a\in A$ (with $\rho$ time-independent and faithful) then we also have a solution of the geodesic velocity equation in Proposition~\ref{propii} with $\kappa_t=0$. This implies a quantum geodesic flow on  $E=A\tens C^\infty(\R)$ which, from $\nabla_E$ in Proposition~\ref{propii}, comes out as
\begin{equation}\label{antiheisflow} \dot a_t=-\tilde X_t(\extd a_t)=-[H_t,a_t] \end{equation}
for $a_t$ a time-dependent element of $A$ (denoted $e\in E=A\tens B$ there). This is {\em minus} the usual Heisenberg evolution for an actual quantum system with Hamiltonian $h_t$ and $H_t=\imath h_t/\hbar$, so we call it the `{\em anti-Heisenberg} flow'  underlying the `Schr\"odinger flow' studied above. We can also interpret this flow probabilistically as in Section~\ref{secflow} if we fix a hermitian inner product on the $*$-algebra $A$ by means of a positive linear functional $\phi_0: A\to \C$. For the unitarity conditions in Proposition~\ref{presip} to apply with $\kappa_t=0$, we need 
\begin{equation}\label{phi0trace} \phi_0(X_t(\extd a))=\phi_0([H_t,a])=0\end{equation}
for all $a\in A$, which happens automatically if $\phi_0$ is a trace, and 
\[ \phi_0(X_t((\extd a)^*)-(X_t(\extd a))^*)=\phi_0([H_t,a^*]-[H_t,a]^*)=\phi_0([H_t+H_t^*,a^*])=0\]
which is automatic as $H_t=\imath h_t/\hbar$ with $h_t$ Hermitian. This then applies to $\xi=a\extd b$ and hence to general $\xi\in\Omega^1_A$ since $X_t$ is a bimodule map. Note that $\phi_0$ does not have to be a trace, for example we can let $\phi_0(a)=\<\psi|\rho(a)|\psi\>$ where $|\psi\>$ is any eigenvector of the Hamiltonian (such as the ground state). For then,  $\phi_0([H_t,a])=\<\psi|   [\ch_t,\rho(a)]   |\psi\>=\<\ch_t^*\psi|  \rho(a)  |\psi\>- \<\psi|  \rho(a)  | \ch_t\psi\>=0$ if $\psi$ is an eigenvector. Similarly for $\phi_0$ any convex linear combination of pure states given by eigenvectors of the Hamiltonian. The reason for the opposite sign in (\ref{antiheisflow}) is that applying the  flow for the same $\tilde X_t$ to a time dependent $a_t$ has the same general flavour (but in a noncommutative algebra) as applying the flow to $\psi(t)\in \CH$ or to the density $|\psi(t)|^2$ if $\CH$ is $L^2$ of a configuration space as in usual quantum mechanics. It is also comparable to the von Neumann evolution for density operators which has opposite sign to the Heisenberg evolution.  By contrast,  the  usual equivalence between the Schr\"odinger and Heisenberg evolution equations is based on equating ${\extd \over\extd t}\<\psi|a|\psi\>=\<\psi|[H_t,a]|\psi\>$ if $\psi$ obeys the Schr\"odinger equation and $a$ is constant, to $\<\psi|\dot a|\psi\>$ for $a$ time-dependent and  $\psi$ constant from the Heisenberg point of view. This is a contravariant relationship in that $\<\psi|a|\psi\>$ is being interpreted from dual points of view, evolving due to $\psi$ or evolving due to $a$. This is very different from our more direct point of view. 

We also note in passing that given our assumption of a bimodule map $X_t$, the geodesic velocity equation in Proposition~\ref{propii} also holds with arbitrary $\kappa_t$ since then $[X_t,\kappa_t]=0$. This more general flow is then not connected with the Schr\"odinger flow above but we can still consider it. Moreover, choosing $\kappa_t=H_t$ means that the unitarity conditions in Proposition~\ref{presip} now hold automatically for any $\phi_0$ as then $\kappa_t^*a+a\kappa_t+ X_t(\extd a)=0$ for the first condition, while the second condition holds automatically as already noted. For this second choice of $\kappa_t$, the quantum geodesic flow on $A$ given by $\nabla_E=0$ becomes
\begin{equation}\label{nsflow} \dot a_t= - a_t \kappa_t-X_t(\extd a_t)= -H_t a_t\end{equation}
for the evolution of $a_t\in A$. The significance of this second `non-standard flow' is unclear as it  is not something we would normally consider in quantum mechanics.

\section{Quantum geodesics for the Heisenberg algebra}\label{secheis}

In this section, we consider what the quantum geodesic formalism of Section~\ref{seciii} amounts to for $A$ the standard  Heisenberg algebra  with generators $x^i$ and $p_i$ for $i=1,\dots,n$ and relations
\[[x^i,p_j]=\mathrm{i}\hbar\delta^i{}_j,\quad [x^i,x^j]=[p_i,p_j]=0\]
for a suitable choice of differential calculus. We fix our Hamiltonian in the standard form
\[ h= {p_1^2+\cdots+p_n^2\over 2m} +V(x^1,\cdots,x^n)\]
for some real potential $V$. We avoid any normal ordering problems due to the decoupled form. The algebra $B=C^\infty(\R)$ as usual, with its classical calculus and trivial linear connection with $\nabla_{\Omega^1_B} \extd t=0$ and $\sigma_{\Omega^1_B}$ the identity map. 

Now we choose $E= C^\infty(\R,L^2(\R^n))$ with $A$ acting in the standard Schr\"odinger representation. Here  $\psi\in E$ is a time dependent element $\psi(t)\in L^2(\R^n)$, where $\R^n$ has standard basis $x^1,\dots ,x^n$, and $x^i\in A$ act on $\psi$ by multiplication and $p_i$   by $-\mathrm{i}\hbar\frac{\partial}{\partial x^i}$. The geodesic flow will be given by  $\nabla_E:E\to E\tens_B\Omega^1_B=E\tens_B\extd t$  (we can also view the connection as an operator $\nabla_E:E\to E$ if we leave $\tens_B\extd t$ understood). We will take this to be 
\[
\nabla_E \psi := \left(\frac{\partial}{\partial t}\psi-\frac{1}{\mathrm{i}\hbar}h\psi\right) \tens \extd t\] 
so that $\nabla_E\psi=0$ lands on the {\em standard Schr\"odinger equation}. Our new result will be all the quantum geometry for this to work, i.e.\ for Schr\"odinger's equation to appear as a quantum geodesic flow:  the choice of calculus $\Omega^1_A$, a linear connection $\nabla$ on $\Omega^1_A$, the geodesic velocity field $X$, and an associated generalised metric that $\nabla$ is metric compatible with.  To achieve this we consider a class of `almost commutative' centrally extended differential structures\cite{Ma:alm} where the classical commutation relations of differentials on phase space acquire a multiple of a central 1-form $\theta'$.  

\begin{proposition}\label{heiscalc} There is a unique centrally extended differential calculus $\Omega^1_A$ on the Heisenberg algebra such that $\nabla_E $ is an $A$-$B$-bimodule connection with $\sigma_E(\theta'\tens \psi)=\psi \tens \extd t$, namely with the bimodule relations
\begin{align*}
[\extd p_i,p_j]= -\mathrm{i}\hbar\frac{\partial^2 V}{\partial x^i \,\partial x^j}\,\theta',\quad 
[\extd p_i,x^j]=[\extd x^i,p_j]= 0,\quad
[\extd x^i,x^j]= -\frac{\mathrm{i}\hbar}{m}\,\delta_{ij}\,\theta'.
\end{align*}
Moreover $\sigma_E (\xi\tens\psi)=
X(\xi)\psi\extd t$ for all $\xi\in\Omega^1_A$, for a bimodule map  
\[ X:\Omega^1_A\to A,\quad X(\theta')=1,\quad X(\extd p_i)=-\frac{\partial V}{\partial x^i},\quad X(\extd x^i)=\frac{p_i}{m}\]
acting on $\psi$ in the Schr\"odinger representation. 
\end{proposition}
\proof  Recall that a bimodule connection involves the existence of a bimodule map $ \sigma_E :\Omega^1_A\tens_A E \to E \tens_B \Omega^1_B=E\tens_B\extd t$, and that by definition of $ \sigma_E $ we have
\begin{align*}
 \sigma_E (\extd p_i\tens\psi)&=\nabla_E (p_i \psi)-p_i\,\nabla_E (\psi)=-\frac{\partial V}{\partial x^i}\,\psi\extd t,\cr
 \sigma_E (\extd x^i\tens\psi)&=\nabla_E (x^i \psi)-x^i\,\nabla_E (\psi)=\frac{1}{m}\,p_i\,\psi \extd t.
\end{align*}
For the central element we assume that $ \sigma_E (\theta'\tens\psi)=\psi \tens \extd t$. 
Now we find the commutation relations in the calculus as follows:
\begin{align*}
 \sigma_E ([\extd p_i,p_j]\tens\psi)&= \sigma_E (\extd p_i\tens p_j\psi)- p_j \,  \sigma_E (\extd p_i\tens\psi)= -\mathrm{i}\hbar\frac{\partial^2 V}{\partial x^i \,\partial x^j}\,\psi \tens \extd t,\cr
 \sigma_E ([\extd p_i,x^j]\tens\psi)&= \sigma_E (\extd p_i\tens x^j\psi)- x^j \,  \sigma_E (\extd p_i\tens\psi)=0,\cr
 \sigma_E ([\extd x^i,p_j]\tens\psi)&= \sigma_E (\extd x^i\tens p_j\psi)- p_j \,  \sigma_E (\extd x^i\tens\psi)= 0,\cr
 \sigma_E ([\extd x^i,x^j]\tens\psi)&= \sigma_E (\extd x^i\tens x^j\psi)- x^j \,  \sigma_E (\extd x^i\tens\psi)= -\frac{\mathrm{i}\hbar}{m}\,\delta_{ij}\psi \tens \extd t.
\end{align*}
From these we are led to the commutation relations as stated, which can be shown give a calculus. 

For the last part, the value of $X(\theta')$ is a definition and for the other values of $X$, we use the formula
$X(\extd b)= [\ch,\rho(b)]$ from Lemma~\ref{Xiii} in the time-independent case, with 
\begin{equation}\label{gothht}\ch=-\frac{1}{\mathrm{i}\hbar}\rho(h).\end{equation} In the lemma, $X$ is an  operator-valued map but we see that this operator factors through a map $X:\Omega^1_A\to A$ and the Schr\"odinger representation of $A$ (the $p_i,x^i$  in the formulae for $\sigma_E$ act on $\psi$).   We then check directly that this $X$ respects the commutation relations of the calculus so as to give a bimodule map, i.e. a left and right vector field. \endproof

This dictates both the differential calculus $\Omega^1_A$ and $\sigma_E$, which is uniquely determined from the pre-chosen $\nabla_E$ once $\Omega^1_A$ is fixed. The form of $\sigma_E$ then determined $X$ uniquely. Note that the exterior derivative on general elements is determined from the Leibniz rule and the stated commutation relations. For example,  if $f(x)$ is a function of the $x^i$ only, then 
\begin{equation}\label{dfx}
\extd f(x)=\frac{\partial f}{\partial x^i}\,\extd x^i-\frac{\mathrm{i}\hbar}{2m} (\del^2 f)\theta',
\end{equation}
where $\del^2f=\sum_i\frac{\partial^2 f}{\partial x^i\partial x^i}$ is the $\R^n$ Laplacian.  The structure of the calculus here is that of a central extension\cite{Ma:alm,Ma:rec} by $\theta'$ of the more trivial $2n$-dimensional calculus on $A$ where we set $\theta'=0$. 

Next, we turn to the geodesic velocity equation
$\doublenabla(\sigma_E )=0$ which depends on the choice linear connection $\nabla$ acting on $\Omega^1_A$. By  Proposition~\ref{propiii},  this is equivalent to  the auxiliary equation (1) in Proposition~\ref{propii} and an autoparallel equation,
\begin{equation}\label{auto12}  
(X\tens X)(\sigma-\id)=0,\quad (X\tens X)\nabla\extd a={1\over \mathrm{i}\hbar} [ X(\extd a),h],\quad \forall a\in A. \end{equation}
where we prefer to write $X(\id\tens X)$ as $X\tens X$ with the product of the result in $A$ understood, given that $X$ is a bimodule map.

\begin{proposition} \label{heisnabla} On the above $\Omega^1_A$, we have a natural right bimodule connection obeying $\doublenabla(\sigma_E)=0$, namely  $\nabla(\theta')=0$ and
\begin{align*} 
\nabla(\extd x^i)=\frac1{m}\theta'\tens\extd p_i,\quad \nabla(\extd p_i)=- \frac{\partial^2 V}{\partial x^i\partial x^j}\, \theta'\tens\extd x^j + \frac{\mathrm{i}\hbar}{2m}\, \frac{\partial  \del^2 V}{\partial x^i}  \,\theta'\tens\theta',
\end{align*} 
\[
\sigma ( \extd x^i \tens \extd x^j )=\extd x^j \tens \extd x^i,\quad 
\sigma ( \extd p_i \tens \extd p_j ) =  \extd p_j\tens\extd p_i, 
\]
\[ \sigma ( \extd x^i \tens \extd p_j ) = \extd p_j \tens\extd x^i +\frac{\mathrm{i}\hbar}{m}
\frac{\partial^2 V}{\partial x^j\partial x^i}\, \theta'\tens\theta',\quad \sigma ( \extd p_i \tens \extd x^j ) =\extd x^j\tens \extd p_i  - \frac{\mathrm{i}\hbar}{m}\,\frac{\partial^2 V}{\partial x^i \,\partial x^j} \theta' \tens\theta'
\]
and $\sigma={\rm flip}$ when one factor is $\theta'$. \end{proposition}
\proof The second half of (\ref{auto12}), explicitly, is \begin{align*}
( X\tens  X)\nabla (\theta') &=   0,  \cr
( X\tens  X)\nabla (\extd p_i)&=   -\frac1m\, \frac{\partial V}{\partial x^i\partial x^j}\, p_j
+ \frac{\mathrm{i}\hbar}{2m}\, \frac{\partial \del^2 V}{\partial x^i},   
\cr
( X\tens  X)\nabla (\extd x^i)&=     -\frac1m\, \frac{\partial V}{\partial x^i}.
\end{align*}
The stated  $\nabla$   is  then easily seen to obey these.  The calculation of $\sigma $ is then routine. Thus
\begin{align*}
\sigma ( \extd x^i \tens \extd p_j ) &= \extd p_j \tens\extd x^i +[\nabla (\extd p_j ),x^i] \cr
&= \extd p_j \tens\extd x^i + \big[
-\frac{\partial^2 V}{\partial x^j\partial x^k}\, \theta'\tens\extd x^k + \frac{\mathrm{i}\hbar}{2m}\, \frac{\partial^3 V}{\partial x^j\partial x^k\partial x^k}  \,\theta'\tens\theta',x^i\big]\cr
&= \extd p_j \tens\extd x^i +\frac{\mathrm{i}\hbar}{m}
\frac{\partial^2 V}{\partial x^j\partial x^i}\, \theta'\tens\theta',
 \end{align*}
\begin{align*}
\sigma ( \extd p_i \tens \extd x^j ) &= \nabla ( p_i . \extd x^j )-p_i .\nabla (  \extd x^j )
= \nabla ( [p_i , \extd x^j] ) + \nabla (\extd x^j.p_i) - p_i .\nabla (  \extd x^j )   \cr
&= \extd x^j\tens \extd p_i + [\nabla (\extd x^j)  ,  p_i] =
\extd x^j\tens \extd p_i +  [      \frac1{m} \theta'\tens\extd p_j     ,  p_i] \cr
&=
\extd x^j\tens \extd p_i  - \frac{\mathrm{i}\hbar}{m}\,\frac{\partial^2 V}{\partial x^i \,\partial x^j} \theta' \tens\theta',
\end{align*}
\begin{align*}
\sigma &( \extd p_i \tens \extd p_j ) = \nabla ( p_i . \extd p_j )-p_i .\nabla (  \extd p_j )
= \nabla ( [p_i , \extd p_j] ) + \nabla (\extd p_j.p_i) - p_i .\nabla (  \extd p_j )   \cr
&=\extd p_j\tens\extd p_i +\nabla ( \mathrm{i}\hbar\frac{\partial^2 V}{\partial x^i \,\partial x^j}\,\theta' ) 
+ [ \nabla (\extd p_j),p_i] \cr
&= \extd p_j\tens\extd p_i +  \mathrm{i}\hbar\, \theta' \tens \extd(\frac{\partial^2 V}{\partial x^i \,\partial x^j})
+  \big[
- \frac{\partial^2 V}{\partial x^j\partial x^k}\, \theta'\tens\extd x^k + \frac{\mathrm{i}\hbar}{2m}\, \frac{\partial^3 V}{\partial x^j\partial x^k\partial x^k}  \,\theta'\tens\theta',p_i\big] \cr
&= \extd p_j\tens\extd p_i +  \mathrm{i}\hbar\, \theta' \tens \extd(\frac{\partial^2 V}{\partial x^i \,\partial x^j})
-   \big[
\frac{\partial^2 V}{\partial x^j\partial x^k},p_i\big]\, \theta'\tens\extd x^k
+  
\frac{\mathrm{i}\hbar}{2m}\,  \big[\frac{\partial^3 V}{\partial x^j\partial x^k\partial x^k} ,p_i\big]  \,\theta'\tens\theta'\cr
&= \extd p_j\tens\extd p_i +  \mathrm{i}\hbar\, \theta' \tens \extd(\frac{\partial^2 V}{\partial x^i \,\partial x^j})
-  \mathrm{i}\hbar 
\frac{\partial^3 V}{\partial x^i\partial x^j\partial x^k}\, \theta'\tens\extd x^k
+  
\frac{(\mathrm{i}\hbar)^2}{2m}\, \frac{\partial^4 V}{\partial x^i\partial x^j\partial x^k\partial x^k}   \,\theta'\tens\theta'. 
\end{align*}
Substituting (\ref{dfx}) gives the result stated that $\sigma $ on the generators is just the flip. 

Finally, we have to check the first of (\ref{auto12}).  We have
\begin{align*}
&(X\tens X)(\sigma-\id)(\extd x^i\tens\extd x^j)=[ X(\extd x^j),X(\extd x^i)]=0  \cr
&(X\tens X)(\sigma-\id)(\extd p_i\tens\extd p_j)=[ X(\extd p_j),X(\extd p_i)]=0  \cr
&(X\tens X)(\sigma-\id)(\extd x^i\tens\extd p_j)=[ X(\extd p_j),X(\extd x^i)] +\frac{\mathrm{i}\hbar}{m}
\frac{\partial^2 V}{\partial x^j\partial x^i} =0\cr
&(X\tens X)(\sigma-\id)(\extd p_i\tens\extd x^j)=[ X(\extd x^j),X(\extd p_i)] -\frac{\mathrm{i}\hbar}{m}
\frac{\partial^2 V}{\partial x^j\partial x^i} =0, 
\end{align*}
which can be checked to hold for the form of $\nabla$. For example, for the last equation
\begin{align*}
 [ X(\extd x^j),X(\extd p_i)]  &= [\frac{p_j}{m},-\frac{\partial V}{\partial x^i}]=\frac{\mathrm{i}\hbar}{m}
\frac{\partial^2 V}{\partial x^j\partial x^i} 
\end{align*}
and similarly for the others.
\endproof

We are not asserting that $\nabla$ is unique, although we are not aware of any other solutions at least for generic $V(x)$. 
 It is natural in the sense of  playing well with central 1-forms in $\Omega^1_A$ and uniquely characterised by this as follows. 

\begin{proposition}\label{Gcov} (1) $\Omega^1_A$ has $2n$ central 1-forms  
\[ \omega_i=\extd p_i+\del_i V\theta',\quad \eta^i=\extd x^i-{p_i\over m}\theta'\]
such that $X(\omega_i)=X(\eta^i)=0$ for all $i$. Moreover, $\nabla$ is the unique right connection such that
\[ \nabla\theta'=\nabla \omega_i=\nabla\eta^i=0,\quad \forall i.\] 
(2) $\Omega^1_A\tens_A\Omega^1_A$ has a central element  
\[ G=\extd p_i\tens\extd x^i-\extd x^i\tens\extd p_i+ \theta'\tens\extd h -\extd h\tens\theta'+{\imath\hbar\over m}\del^2 V\theta'\tens\theta';\quad \extd h=(\extd x^i)\del_i V+{p_i\over m}\extd p_i\]
such that 
\[  ( X\tens\id)G=(\id\tens X)G=0,\quad \nabla  G=0.\]
\end{proposition}
\proof For (1), the commutation relations of Proposition~\ref{heiscalc} immediately give that $\omega_i,\eta^i$ are central, clearly annihilated by $X$ as stated there. That these are covariantly constant requires  $\nabla(\extd x^i)= \nabla(\theta'{p_i\over m})=\theta'\tens{\extd p_i\over m}+(\nabla\theta'){p_i\over m}$ and $\nabla(\extd p_i)=-\nabla(\theta'\del_i V)=-\theta'\tens\extd \del_i V-(\nabla\theta')\del_iV$ if $\nabla$ is a right connection. We then assume $\nabla\theta'=0$ and use (\ref{dfx}). 

Part (2) is immediate from part (1) once we compute that
\begin{align*}\omega_i\tens\eta^i-\eta^i\tens\omega_i&=(\extd p_i+\del_i V\theta')\tens(\extd x^i-{p_i\over m}\theta')-(\extd x^i-{p_i\over m}\theta')\tens(\extd p_i+\del_i V\theta')\\
&=\extd p_i\tens\extd x^i-\extd p_i{p_i\over m}\tens\theta'+\del_i V\theta'\tens\extd x^i-\del_iV{p_i\over m}\theta'\tens\theta'\\
&\quad-\extd x^i\tens\extd p_i-\extd x^i\tens\del_iV\theta'+{p_i\over m}\theta'\tens\extd p_i+{p_i\over m}\del_iV\theta'\tens\theta'\\
&=\extd p_i\tens\extd x^i-\extd x^i\tens\extd p_i+ \theta'\tens( \del_i V\extd x^i+{p_i\over m}\extd p_i) -(  \del_i V\extd x^i+{p_i\over m}\extd p_i )\tens\theta'\\
&\quad+{\imath\hbar\over m}\del^2V\theta'\tens\theta'\\
&=G
\end{align*}
where we used the commutation relations of the calculus to reorder $(\extd p_i) p_i$ and $(\extd x^i)\del_iV$ (the two corrections cancel) and the usual Heisenberg relations for the $\theta'\tens\theta'$ term for the 3rd equality. We then recognise the answer in terms of $\extd h$ (where a reorder of $\del_iV\extd x^i$ to match $\extd h$ cancels between the two terms). The formula for $\extd h$ follows from (\ref{dfx}) and the commutation relations of the calculus. \endproof

To discuss the quantum geometry further, we now need to specify $\Omega^2_A$. For every $\Omega^1_A$ there is a canonical `maximal prolongation' obtained by applying $\extd$ to the degree 1 relations, and other choices are a quotient. In our case, we impose $\extd\theta'=0$ and $\theta'^2=0$ and then the rest of the exterior algebra relations, obtained by applying $\extd$ to the degree 1 relations, are
\[ \{\extd x^i,\extd x^j\}=\{\extd x^i,\extd p_j\}=0,\quad \{\extd p_i,\extd p_j\}=\imath\hbar V_{,ijk}\extd x^k\theta'.\]
This has the expected dimension in degree 2 as for a $2n+1$-dimensional manifold and is therefore a reasonable
quotient of the maximal prolongation, where we imposed the $\theta'$ behaviour.  Since $\theta'^2=0$, corrections from the commutation relations vanish and one similarly has
\[ \{\eta^i,\eta^j\}=\{\eta^i,\omega_j\}=0,\quad \{\omega_i,\omega_j\}=\imath\hbar V_{,ijk}\eta^k\wedge\theta'.\]

\begin{corollary} The linear connection $\nabla$ in Proposition~\ref{heisnabla} is flat and has torsion 
\[ T_{\nabla }(\theta')=0  ,\quad T_{\nabla }(\extd x^i)=- {1\over m}\extd p_i\theta',\quad T_{\nabla }(\extd p_i)={\del^2 V\over\del x^j\del x^i}\extd x^j\theta'. \]
Moreover, the 2-form
\[ \tilde\omega=\wedge(G)=-2( \extd x^i\wedge\extd p_i+\extd h\wedge \theta')\]
is closed and covariantly constant under $\nabla$. 
\end{corollary}
\proof The torsion of a right connection is $T_\nabla=\wedge\nabla+\extd$ and comes out as shown. The formula for $\wedge (G)$ is immediate from the form of $G$ stated in Proposition~\ref{Gcov} given that $\theta'$ anticommutes with 1-forms. 
Note that the torsion and curvature for a right connection are right module maps but not necessarily  bimodule ones, which is indeed not the case for $T_\nabla$ here. It follows in our case that $R_\nabla=(\id\tens\extd +\nabla\wedge\id)\nabla=0$ as clearly $R_\nabla(\omega_i)=R_\nabla(\eta^i)=R_\nabla(\theta')=0$, since $\nabla$ itself vanishes on these.  \endproof

We see that $G$ is not quantum-symmetric in the sense of $\wedge(G)=0$ as needed for a strict quantum metric\cite{BegMa}; it is  `generalised quantum metric' in the notation there (and is, moreover, degenerate). Likewise, the torsion tensor does not vanish, so $\nabla$ is not a `quantum Levi-Civita connection' in the sense of quantum Riemannian geometry either.  Certainly, one can quotient out $\theta'=0$ to work in the unextended calculus on $A$, in which case $\tilde\omega$ has the same form as the canonical symplectic 2-form in the classical case, and $G$ becomes its lift. The geodesic velocity field $X$, however, does not descend to this quotient while  $\nabla=0$ at this quotient level, at least in the Heisenberg case studied here. Thus, the geometric picture is not exactly a quantum version of symplectic geometry either. We return to this in Section~\ref{secsym}.

 It should also be clear that the remarks at the end of Section~\ref{seciii} apply since the quantum vector field used above is explicitly constructed in Proposition~\ref{heiscalc} as the action of a  bimodule map $X:\Omega^1_A\to A$. In our case it is time independent and the Schr\"odinger representation used above is also time independent and faithful. It follows that $X$ obeys the geodesic velocity equations in Proposition~\ref{propii}, but we can also verify this directly since 
 \[ X(\extd X(\extd x^i))=X({\extd p_i\over m})=-{1\over m}\del_i V,\quad X(\extd X(\extd p_i))=-X(\extd \del_i V)=-\del_i\del_j V{ p_j\over m}+{\imath\hbar\over 2m}\del^2\del_iV\]
are the same expressions as computed for $(X\tens X)\nabla\extd x^i$ and $(X\tens X)\nabla \extd p_i$ in the proof of Proposition~\ref{heisnabla}. Since $X$ is a bimodule map, $[X,\kappa]=0$ for any $\kappa$. For the probabilistic interpretation in Proposition~\ref{presip} to apply with $\kappa=0$, we need a positive linear functional $\phi_0$ such that (\ref{phi0trace}) holds and, as explained there, the natural way to do this is to take a convex linear combination $\phi_0(a)=\sum_i \rho_i\<\psi_i|a|\psi_i\>$ of pure states associated to normalised energy eigenvectors $|\psi_i\>$ in the Schr\"odinger representation. Here $\rho_i\in[0,1]$, $\sum_i\rho_i=1$.   We can, for example, just take the ground state, $\phi_0(a)=\<0|a|0\>$. We also noted a non-standard quantum geodesic flow with $\kappa=\imath h/\hbar$.

\subsection{Example of the harmonic oscillator}

We look briefly at the case of $V={1\over 2}m\nu^2 \sum_i x_i^2$. For the calculus, geodesic velocity field and connection, we have
\[ [x^i,p_j]=\imath\hbar\delta_{ij},\quad [\extd x^i,x^j]=-{\imath\hbar\over m}\delta_{ij}\theta',\quad [\extd p_i,p_j]=-\imath\hbar m\nu^2\delta_{ij}\theta',\quad 
 [\extd x^i,p_j]=[\extd p_j,x^i]=0,\]
 \[  X(\extd x^i)={p^i\over m},\quad  X(\extd p_i)=-m\nu^2 x^i,\quad \nabla\extd x^i= \theta'\tens{\extd p^i\over m}, \quad \nabla\extd p_i=-m\nu^2\theta'\tens\extd x^i, \]
 \[ \sigma(\extd x^i\tens\extd p_j)=\extd p_j\tens\extd x^i+\imath\hbar\nu^2\delta_{ij}\theta'\tens\theta',\quad \sigma(\extd p_i\tens\extd x^j)=\extd x^j\tens\extd p_i-\imath\hbar\nu^2\delta_{ij}\theta'\tens\theta'\]
 and from Proposition~\ref{Gcov} we have the central element and 1-forms
 \[ G=\extd p_i\tens\extd x^i-\extd x^i\tens\extd p_i+\theta'\tens \extd h- \extd h\tens\theta'+\imath\hbar \nu^2\theta'\tens\theta'  \]
  \[ \extd h=m\nu^2 (\extd x^i)x^i +{p_i\over m}\extd p_i,\quad \omega_i=\extd p_i+m\nu^2 x^i\theta',\quad \eta_i=\extd x^i-{p_i\over m}\theta'\]
(sum over repeated indices) showing the expected symmetry between $x^i,p_i$. This quantum geometry quantises an extended phase space geometry as we discuss further in Section~\ref{secsym}, and underlies our interpretation of the usual Schr\"odinger evolution as a quantum geodesic flow. It means, for example, that
\begin{align*}
{\extd \over \extd t}\< \psi \big| p_i{}^2 \big| \psi\> = -m\nu^2 \,  \< \psi \big| p_i\,x^i+x^i\,p_i\big| \psi\>\ .
\end{align*}
according to (\ref{dexpdt}) whenever $|\psi\>$ is a solution of the  time-dependent Schr\"odinger equation for a quantum harmonic oscillator (this is also easy enough to see directly). 

We also have an underlying `anti-Heisenberg' flow (\ref{antiheisflow}) using $X:\Omega^1_A\to A$ and the machinery of Section~\ref{secii} with $\kappa=0$. This is $\dot a_t=-[H_t,a_t]$ which in our case is 
\[ \dot a_t=[{h\over\imath\hbar},a_t];\quad a_t=e^{t h\over\imath\hbar}a_0e^{-{t h\over\imath\hbar}}\]
as usual but with a reversed sign. We can make this concrete by restricting to the form
\begin{equation}\label{lina} a_t=\chi_i(t)x^i+\psi^i(t)p_i\end{equation}
for some evolving complex coefficients $\chi_i(t),\psi^i(t)$. Then 
\[ \dot\chi_i=\psi^i m\nu^2,\quad \dot\psi^i=-{ \chi_i\over m}\]
which is the expected simple harmonic motion among the coefficients. We can also introduce any $\kappa$ as far as Proposition~\ref{propii} is concerned, e.g. a constant $\kappa$ modifies the above with a damping term such that 
\[ \ddot \chi_i=-(\kappa^2+\nu^2) \chi_i-2\kappa \dot \chi_i.\]
On the other hand, such a flow is no longer unitary and indeed the natural choices in our analysis are either  (i) $\kappa=0$ and a suitable $\phi_0$ or (ii) $\kappa=H_t$ and any positive linear functional $\phi_0$ if we want a unitary flow according to Proposition~\ref{presip}. 

For case (i), the natural way to construct $\phi_0$ is as a convex linear combination of pure states obtained from energy eigenvectors. The latter are of course $|n\>$ labelled by $n=0,1,2,...$ and given by Hermite functions with eigenvalue $E_n=(n+{1\over 2})\hbar\nu$ of $h$. For example, the vacuum expectation value $\phi_0(a)=\<0|a|0\>$ provides a natural Hermitian inner product on $A$ with respect to which the anti-Heisenberg evolution is unitary. In the classical limit, the elements of $A$ become functions on phase space and the evolution becomes that of the classical harmonic oscillator. Case (ii), by contrast, is rather non-standard with
\[ \dot a_t={h\over \imath\hbar} a_t;\quad a_t=e^{t h\over \imath\hbar} a_0,\]
which is no longer closed for the linear ansatz (\ref{lina}).  Unlike Case (i), the expectation values in an energy eigenstate now evolve by phases, e.g. 
\[ \<n|a_t|n\>=e^{\imath (n+{1\over 2})\nu t}\<n|a_0|n\>.\]
The evolution is nevertheless $*$-preserving in the sense of Proposition~\ref{presip} and is best understood in terms of the associated positive linear functionals $\phi^t$ defined by $a_t$ according to (\ref{phia}), namely 
\[ \phi^t(b)=\phi_0(a_t^* b a_t)=\phi_0(a_0^*e^{-{th\over\imath\hbar}}be^{th\over\imath\hbar}a_0)=\phi^0(b_t^{heis})\]
where $b_t^{heis}=e^{-{th\over\imath\hbar}}be^{th\over\imath\hbar}$ is the usual Heisenberg flow of an initial $b\in A$ and $\phi^0=\phi_0(a_0^*(\ )a_0)$ is the initial value of $\phi^t$. If $\phi_0$ happens to be a convex linear combination of pure states given by energy eigenvectors then $\phi^t$ comes out the same  as the positive linear functional corresponding to the anti-Heisenberg flow of Case (i). In this case, if $a_0$ commutes with $h$ then $\phi^t=\phi^0$ does not evolve. These remarks are not specific to the harmonic oscillator but features of our formalism.

\section{Electromagnetic Klein Gordon equation as geodesic flow } \label{secKG}

Our goal in this section is to extend the geodesic picture of quantum mechanical evolution to a relativistic setting with flat spacetime metric $\eta={\rm diag}(-1,1,1,1)$ and an electromagnetic background with gauge potential $A_a$ in place of a Hamiltonian potential. Motion in such a background is not geodesic motion in a usual sense, so this will be a novel application of our formalism. Let $D_a= \frac{\partial }{\partial x^a}- \imath{q\over\hbar} A_a$,  where $q$ is the particle charge and we use a physical normalisation so that background electromagnetic fields will appear in the classical limit without extraneous factors of $\hbar$.  We introduce an external time parameter  $s$ for the geodesic flow so now $B=C^\infty(\R)$ for this parameter and we set $x^0=ct$ in terms of the usual time coordinate $t$. Our first guess might be 
\[ {\del\over\del s}\phi- \imath c \sqrt{\eta^{ab}D_aD_b}\phi=0\]
motivated by the formula for proper time in GR, but it is unpleasant to work with square roots and, as when working out geodesics in GR (where it is easier to extremise the integral of the proper velocity squared), we prefer to consider
\[  \frac{\partial\phi}{\partial s} -\frac{\imath\hbar}{2m}\eta^{ab} D_a D_b\phi =0\]
In effect, we write $\sqrt{\eta^{ab}D_aD_b}=\eta^{ab}D_aD_b/\sqrt{\eta^{cd}D_cD_d}$ and replace the denominator by its on-shell value $m c\over\hbar$ where $m$ is the particle rest mass. The half is to allow for the idea that any kind of variation of $\sqrt{\eta^{ab}D_aD_b}$ brings down a $1/2$ in comparison to that of
$\eta^{ab}D_aD_b$. Although somewhat novel, we will see that this ansatz lends itself to a reasonable quantum geodesic formulation. That this works out will appear as a minor miracle in terms of the amount of algebra,  which in itself lends credibility to our hypothesis. 

\subsection{Electromagnetic Heisenberg differential calculus} 

Motivated as above, we consider $\CH=L^2(\R^{1,3})$ with its 4D Schr\"odinger representation of the electromagnetic Heisenberg algebra $A$ with commutation relations
\[ [x^a,p_b]=\imath\hbar\delta^a{}_b,\quad [x^a,x^b]=0,\quad [p_a,p_b]=\imath\hbar q F_{ab},\]
given that  $[D_a,D_b]=-\imath{q\over\hbar} F_{ab}$ and $p_a$ is represented by $-\imath\hbar D_a$. Here $F_{ab}=A_{b,a}-A_{a,b}$ and the algebra is associative due to $\extd F=0$.

We set $E=\CH\tens C^\infty(\R)$, or more precisely $E=C^\infty(\R,\CH)$, and 
\[
\nabla_E\phi=\big(\frac{\partial \phi}{\partial s} - \frac{\imath\hbar}{2m}\eta^{ab} D_a D_b\phi  \big)\tens\extd s,\quad \sigma_E(\extd x^a\tens \phi)=-\frac{\imath\hbar}{m}\eta^{ab} D_b\phi\tens\extd s. 
\]
We can also write $\sigma_E(\extd a\tens \phi)=X(\extd a).\phi\tens \extd s$, where
\begin{equation}\label{Xh}
X(\extd x^a)= \frac1m \,\eta^{ab}p_b  \ ,\quad X(\extd p_c)= \frac{ q}{2m}\,\eta^{ab}(2 F_{ca}p_b-\imath \hbar F_{cb,a}) 
\end{equation}
maps to the algebra and its output then acts on $\CH$ in the 4D Schr\"odinger representation. For $j=1,2,3$, we have $X(\extd x^i)={p_i\over m}$ so in some sense  $\extd x^i\over\extd s$ is being identified with the value $p_i/m$  which is consistent with Special Relativity if $s$ were to be proper time.  We also define the Hamiltonian
\[ h={\eta^{ab}\over 2m\hbar} p_a p_b\] 
as  the operator in $\nabla_E$ relevant to our formalism. We now provide a suitable calculus for the above. 

\begin{proposition} \label{KGcalculus} The spacetime Heisenberg algebra $A$ has a first order differential calculus with an extra central direction $\theta'$, given by
\[
[\extd x^a,x^b]=-\frac{\imath\hbar}m \,\eta^{ab}\theta'   \ ,\quad [\extd x^a,p_c]=\frac{\imath \hbar q}{m}\,\eta^{ab} F_{bc}\theta' = [\extd p_c,x^a], 
\]
\[
[\extd p_c,p_d] = -  \imath\hbar q\, F_{ac,d}\,\extd x^a 
 - \frac{\hbar q}{2m}\,\eta^{ab}( \hbar F_{bc,ad}  +2\imath  q F_{ac}F_{bd}  )\theta'
\]
such that $X$ extended by $X(\theta')=1$ is a bimodule map $\Omega^1_A\to A$. 
\end{proposition}
\proof We give the two most difficult checks, first $\extd$ applied to the commutator of two $p$s:
\begin{align*}
[\extd p_c,p_d] &+[p_c,\extd p_d] = [\extd p_c,p_d] -[\extd p_d,p_c] \cr
&= - \imath\hbar q\, (F_{ac,d}-F_{ad,c})\,\extd x^a 
 - \frac{\hbar q}{2m}\,\eta^{ab}( \hbar F_{bc,ad}  +2\imath  q F_{ac}F_{bd} -\hbar F_{bd,ac}  -2\imath  q F_{ad}F_{bc}   )\theta' \cr
 &=  -\imath\hbar q\, F_{dc,a} \extd x^a 
 - \frac{\hbar^2 q}{2m}\,\eta^{ab}  F_{dc,ab}  \theta' =-  \imath\hbar q\, (F_{dc,a} \extd x^a 
 - \frac{\imath\hbar}{2m}\,\eta^{ab}  F_{dc,ab}  \theta' )
\end{align*}
which we compare to
\begin{align*}
\extd([p_c,p_d])=-(-\imath \hbar)^2\imath {q\over\hbar} \extd F_{cd}= \imath\hbar q\, \extd F_{cd}=- \imath\hbar q\, \extd F_{dc}
\end{align*}
and which agree when we remember to use the centrally extended formula for $\extd$.
Now we check the three $p$s Jacobi identity: 
\begin{align*}
[[\extd p_c,&p_d],p_e]  =  \big[ -\imath\hbar q\, F_{ac,d}\,\extd x^a 
 - \frac{\hbar q}{2m}\,\eta^{ab}( \hbar F_{bc,ad}  +2\imath  q F_{ac}F_{bd}  )\theta',p_e\big] \cr
  &=   \hbar^2 q\, F_{ac,de}\,\extd x^a  -  \imath\hbar q\, F_{ac,d}\,[\extd x^a ,p_e]
 -\frac{\hbar q}{2m}\,\eta^{ab}\big[( \hbar F_{bc,ad}  +2\imath  q F_{ac}F_{bd}  ),p_e\big] \theta'\cr
   &=   \hbar^2 q\, F_{ac,de}\,\extd x^a  -  \imath\hbar q\, F_{ac,d}\,\frac{\imath \hbar q}{m}\,\eta^{ab} F_{be}\theta'
 - \frac{\imath\hbar^2 q}{2m}\,\eta^{ab} ( \hbar F_{bc,ade}  +2\imath  q F_{ac,e}F_{bd}  +2\imath  q F_{ac}F_{bd,e}  ) \theta'\cr
    &=   \hbar^2 q\, F_{ac,de}\,\extd x^a  
 - \frac{\imath\hbar^2 q}{2m}\,\eta^{ab} ( \hbar F_{bc,ade}  +2\imath  q F_{ac,e}F_{bd}  +2\imath  q F_{ad,e} F_{bc}  +2\imath  q F_{ac,d} F_{be} ) \theta'
\end{align*}
and so
\begin{align*}
[[\extd p_c,p_d],p_e] -[[\extd p_c,p_e],p_d]  &= 
 - \frac{\imath\hbar^2 q}{2m}\,\eta^{ab} (- 2\imath  q F_{ae,d} F_{bc}   +2\imath  q F_{ad,e} F_{bc} ) \theta'\cr
  &= 
 - \frac{\hbar^2 q^2}{m}\,\eta^{ab}    F_{de,a} F_{bc}   \theta'
\end{align*}
and as
\[
[[ p_d,p_e],\extd p_c] =(-\imath\hbar)^2\imath {q\over\hbar} [\extd p_c,F_{de}] =-\imath\hbar q \frac{\imath \hbar q}{m}\,\eta^{ab} F_{bc}\theta' F_{de,a},  
\]
we see that the three $p$s Jacobi identity is satisfied.
\endproof

 We next want to choose $\nabla$ on $\Omega^1_A$ such that the conditions (\ref{auto12}) for $\doublenabla(\sigma_E)$ in Proposition~\ref{propiii} hold. 

\begin{theorem} \label{bigKG}
There is a right bimodule connection $\nabla$ on $\Omega^1_A$ given by $\nabla\theta'=0$ and
\begin{align*}
\nabla(\extd x^d) &=-  \frac{q}{m}\,\eta^{cd} F_{ac}\, \theta'\tens \extd x^a
+ \theta'\tens \frac{\imath\hbar q}{2m^2}\,\eta^{ab}\eta^{cd}F_{bc,a} \theta'    \ ,\cr
\nabla(\extd p_c) &=  -q\, F_{dc,e}\extd x^d\tens\extd x^e-\xi_c\tens\theta'-\theta'\tens\eta_c+N_c\theta'\tens\theta'
\end{align*}
such that  $\doublenabla(\sigma_E)=0$, where
\begin{align*}
  N_{c}
&=   
  -  \frac{\imath\hbar q^2}{2m^2}\, \eta^{nm} \,\eta^{ab}\big(2 F_{an}\,F_{mc,b}  +F_{bn,a}\,F_{mc} 
 \big) 
 + \frac{\hbar^2 q}{4m^2}\, \eta^{nm} \eta^{ab}   F_{bc,anm}   \ ,\cr
 \xi_c &=-\frac{\imath\hbar q}{2m}\,\eta^{nm}  F_{ac,nm}\extd x^a \ ,\quad 
 \eta_c = - \frac{\imath\hbar q}{2m}\,\eta^{nm} F_{nc,ma} \extd x^a   
  - \frac{q^2}{m}\,\eta^{eb} F_{ec}   F_{ab}\extd x^a \ .
\end{align*} 
Here $\sigma$ is the flip map when one factor is $\theta'$ and
\begin{align*} 
\sigma&(\extd x^e\tens\extd x^d) 
= \extd x^d\tens\extd x^e + \frac{\imath\hbar q}{m^2}\,\eta^{cd} F_{ac}\,\eta^{ae} \theta'\tens \theta', \cr
\sigma&(\extd p_e\tens\extd x^d) =  \extd x^d\tens\extd p_e +  \frac{\imath \hbar q}{m}\,\eta^{dc} \theta' \tens\big(-
 F_{ae,c}\,\extd x^a   - \frac{q}{m}\, F_{ac}\,\eta^{ab} F_{be}\theta' 
 +  \frac{\imath\hbar}{2m}\,\eta^{ab} F_{ae,cb} \theta' \big), \cr
 \sigma&(\extd x^a\tens\extd p_c) = \extd p_c\tens\extd x^a   +  \frac{\imath\hbar q}{m} \,\eta^{ea} F_{dc,e}  \extd x^d\tens\theta'   \cr
 & \qquad\qquad 
     - (   \frac{\imath\hbar q^2}{m^2}\,\eta^{bn}\,\eta^{ra}  F_{nc}  F_{rb}  
 - \frac{\hbar^2 q}{2m^2}\,\eta^{ar}\,\eta^{nb}  F_{bc,nr} ) \theta' \tens \theta',  \cr
 \sigma&(\extd p_e\tens\extd p_d )  =  \extd p_d\tens\extd p_e
+ \frac{ \imath\hbar q^2}{m}\,\eta^{rp} \big(  F_{rd}\, F_{ae,p}   \theta' \tens \extd x^a   -    F_{re}\,  F_{ad,p}    \extd x^a \tens \theta'  \big)   \cr
 & \qquad\qquad  +  \frac{\imath\hbar q^3}{  m^2 }\,  \eta^{rp}\,\eta^{ba}   F_{pe}  F_{ad}  F_{rb}   \theta' \tens \theta' 
    +  \frac{\hbar^2 q^2}{  2m^2 }\, \eta^{rp}\,\eta^{ab}      (  F_{pd}\,    F_{be,ar}    
   -  F_{pd,ar}     F_{be}  )  \theta' \tens \theta'. 
\end{align*}
\end{theorem}
\proof  First we have (on using the commutation relations for the $p_a$s),
\begin{align*}
[{h\over\imath\hbar},X(\extd x^d)] &= - \frac{q}{2m^2}\,\eta^{ab}\eta^{cd}(2 F_{ac}p_b-\imath \hbar F_{bc,a}) \ ,\cr
[{h\over\imath\hbar}, X(\extd p_c)] 
  &=     - \frac{q}{m^2}\,\eta^{ab}\eta^{de} F_{ac,e} p_b p_d   -    \frac{\imath \hbar q^2}{m^2}\,\eta^{ab}\eta^{de} F_{ac,e} F_{db}  
  + \frac{\imath\hbar q}{2m}\,\eta^{de} ( F_{ac,ed}+  F_{dc,ea}) X(\extd x^a)    \cr
&\quad   + \frac{\hbar^2 q}{4m^2}\,\eta^{ab}\eta^{de}     F_{bc,aed}    
  + \frac{q^2}{m}\,\eta^{eb} F_{ec}   F_{ab}X(\extd x^a)   -   \frac{\imath\hbar q^2}{2m^2}\,\eta^{ar} \eta^{eb} F_{ec}   F_{rb,a}
\end{align*}
and we can check that this obeys the autoparallel equation in (\ref{auto12}), in particular that $-X(\eta_c)-X(\xi_c)+N_c=[{h\over\imath\hbar}, X(\extd p_c)] $. 

From the value of $\nabla(\extd x^d)$,  
 we calculate
\begin{align}  \label{yodl}
\sigma(\extd x^e\tens\extd x^d) &= \extd x^d\tens\extd x^e -\nabla([\extd x^d,x^e])+[\nabla(\extd x^d),x^e]  \cr
&= \extd x^d\tens\extd x^e- \theta'\tens \frac{ q}{m}\,\eta^{cd} F_{ac}\,[\extd x^a,x^e]   \cr
&= \extd x^d\tens\extd x^e + \frac{\imath\hbar q}{m^2}\,\eta^{cd} F_{ac}\,\eta^{ae} \theta'\tens \theta'
\end{align}
and we check that this is a bimodule map, the difficult case being
\begin{align*}
[\extd x^e\tens\extd x^d,p_c] &=\frac{\imath \hbar q}{m}\,\eta^{eb} F_{bc}\theta' \tens\extd x^d+\extd x^e\tens \frac{\imath \hbar q}{m}\,\eta^{db} F_{bc}\theta' \cr
&= \frac{\imath \hbar q}{m}\big(\eta^{eb} F_{bc}\theta' \tens\extd x^d+\eta^{db} F_{bc} \extd x^e\tens  \theta' 
- \frac{\imath\hbar}m \,\eta^{ea}  \eta^{db} F_{bc,a}  \theta'  \tens \theta' \big)  \cr
\sigma([\extd x^e\tens\extd x^d,p_c]) &=  \frac{\imath \hbar q}{m}\big(\eta^{eb} F_{bc}\extd x^d \tens\theta'+\eta^{db} F_{bc} \theta'\tens \extd x^e
- \frac{\imath\hbar}m \,\eta^{ea}  \eta^{db} F_{bc,a}  \theta'  \tens \theta' 
\big) \cr
[\sigma(\extd x^e\tens\extd x^d) ,p_c] &= [  \extd x^d\tens\extd x^e   ,p_c]+ \frac{\imath \hbar q}{m^2}\,\eta^{bd} [F_{ab} ,p_c]\,\eta^{ae} \theta'\tens \theta'   \cr
&= \frac{\imath \hbar q}{m}\,\eta^{db} F_{bc}\theta'  \tens\extd x^e +  \extd x^d\tens \frac{\imath \hbar q}{m}\,\eta^{eb} F_{bc}\theta'  + \frac{\imath^2\hbar^2 q}{m^2}\,\eta^{bd} F_{ab,c} \eta^{ae} \theta'\tens \theta'   \cr
&= \frac{\imath \hbar q}{m}\big(\eta^{db} F_{bc}\theta'  \tens\extd x^e + \eta^{eb} F_{bc} \extd x^d\tens \theta' 
-  \frac{\imath\hbar }{m} \eta^{da}\eta^{eb} F_{bc,a} \theta'\tens \theta'  + \frac{\imath\hbar }{m}\,\eta^{bd} F_{ab,c} \eta^{ae} \theta'\tens \theta'  \big)
\end{align*}
as required.
From our value of $\nabla(\extd x^d)$, we also calculate
\begin{align*} 
\sigma&(\extd p_e\tens\extd x^d) = \extd x^d\tens\extd p_e -\nabla([\extd x^d,p_e])+[\nabla(\extd x^d),p_e]  \cr
&= \extd x^d\tens\extd p_e -\nabla(\frac{\imath \hbar q}{m}\,\eta^{db} F_{be}\theta' ) - \theta'\tens \frac{ q}{m}\,\eta^{cd} [F_{ac}\,\extd x^a,p_e] + \theta'\tens \frac{\imath\hbar q}{2m^2}\,\eta^{ab}\eta^{cd}[F_{bc,a},p_e] \theta'  \cr
&= \extd x^d\tens\extd p_e -  \frac{\imath \hbar q}{m}\,\eta^{dc} \theta' \tens \extd F_{ce}
 - \theta'\tens \frac{\imath\hbar q}{m}\,\eta^{cd} F_{ac,e}\,\extd x^a  \cr
 &\quad - \theta'\tens \frac{q}{m}\,\eta^{cd} F_{ac}\,\frac{\imath \hbar q}{m}\,\eta^{ab} F_{be}\theta' 
 - \theta'\tens \frac{\hbar^2 q}{2m^2}\,\eta^{ab}\eta^{cd}F_{bc,ae} \theta'  \cr
 &= \extd x^d\tens\extd p_e +  \frac{\imath \hbar q}{m}\,\eta^{dc} \theta' \tens\big(-
  \extd F_{ce}
- F_{ac,e}\,\extd x^a   - \frac{q}{m}\, F_{ac}\,\eta^{ab} F_{be}\theta' 
 +  \frac{\imath\hbar}{2m}\,\eta^{ab}F_{bc,ae} \theta' \big) \cr
  &= \extd x^d\tens\extd p_e +  \frac{\imath \hbar q}{m}\,\eta^{dc} \theta' \tens\big(
- (F_{ce,a}+F_{ac,e})\,\extd x^a   - \frac{q}{m}\, F_{ac}\,\eta^{ab} F_{be}\theta' 
 +  \frac{\imath\hbar}{2m}\,\eta^{ab}(F_{ce,ab}+F_{ac,be}) \theta' \big) \cr
   &= \extd x^d\tens\extd p_e +  \frac{\imath \hbar q}{m}\,\eta^{dc} \theta' \tens\big(
- F_{ae,c}\,\extd x^a   - \frac{q}{m}\, F_{ac}\,\eta^{ab} F_{be}\theta' 
 +  \frac{\imath\hbar}{2m}\,\eta^{ab} F_{ae,cb} \theta' \big) 
\end{align*}
as given. We check that this is commutes with commutators, 
\begin{align*}
\sigma([\extd p_e\tens\extd x^d,x^c]) &= \sigma( \frac{\imath \hbar q}{m}\,\eta^{cb} F_{be}\theta'  \tens\extd x^d - \extd p_e\tens \frac{\imath\hbar}m \,\eta^{dc}\theta')=\frac{\imath \hbar q}{m}\,\eta^{cb} F_{be}\extd x^d\tens\theta' -  \frac{\imath\hbar}m \,\eta^{dc}\theta'  \tens \extd p_e \cr
[\sigma(\extd p_e\tens\extd x^d),x^c] &=
[ \extd x^d\tens\extd p_e,x^c] -  \frac{\imath \hbar q}{m}\,\eta^{dc} \theta' \tens F_{ae,c}\,[\extd x^a   ,x^c]  \cr
&=
-\frac{\imath\hbar}m \,\eta^{dc}\theta' \tens\extd p_e + \extd x^d\tens \frac{\imath \hbar q}{m}\,\eta^{ca} F_{ae}\theta'  +  \frac{\imath \hbar q}{m}\,\eta^{dc} \theta' \tens F_{ae,c}\,\frac{\imath\hbar}m \,\eta^{ac}\theta'  
\end{align*}
as required, and 
\begin{align*}
[\extd p_e\tens\extd x^d,p_c] &= \big( -\imath\hbar q\, F_{ae,c}\,\extd x^a 
 - \frac{\hbar q}{2m}\,\eta^{ab}( \hbar F_{be,ac}  +2\imath  q F_{ae}F_{bc}  )\theta'\big)\tens \extd x^d 
 +\extd p_e\tens \frac{\imath \hbar q}{m}\,\eta^{db} F_{bc}\theta'  \cr
 &= - \imath\hbar q\, F_{ae,c}\,\extd x^a \tens \extd x^d 
 -\frac{\hbar q}{2m}\,\eta^{ab}( \hbar F_{be,ac}  +2\imath  q F_{ae}F_{bc}  )\theta' \tens \extd x^d \cr
&\quad  + \frac{\imath \hbar q}{m}\,\eta^{db} F_{bc} \extd p_e\tens \theta' 
 + \frac{\imath \hbar q}{m}\,\eta^{da} F_{ac,r} \frac{\imath \hbar q}{m}\,\eta^{rb} F_{be }\theta'  \tens \theta'  \cr
\sigma([\extd p_e\tens\extd x^d,p_c]) &=   - \imath\hbar q\, F_{ae,c}\big(     \extd x^d\tens\extd x^a + \frac{\imath\hbar q}{m^2}\,\eta^{bd} F_{rb}\,\eta^{ra} \theta'\tens \theta'  \big)
\\ &\quad - \frac{\hbar q}{2m}\,\eta^{ab}( \hbar F_{be,ac}  +2\imath  q F_{ae}F_{bc}  ) \extd x^d \tens\theta' \cr
&\quad  + \frac{\imath \hbar q}{m}\,\eta^{db} F_{bc} \theta'\tens \extd p_e
 + \frac{\imath \hbar q}{m}\,\eta^{da} F_{ac,r} \frac{\imath \hbar q}{m}\,\eta^{rb} F_{be }\theta'  \tens \theta'  
\end{align*}
versus
\begin{align*}
[\sigma&(\extd p_e\tens\extd x^d),p_c] = [\extd x^d\tens\extd p_e,p_c]  +  \frac{\imath \hbar q}{m}\,\eta^{dr} \theta' \tens [
 -F_{ae,r}\,\extd x^a   - \frac{ q}{m}\, F_{ar}\,\eta^{ab} F_{be}\theta' 
 +  \frac{\imath\hbar}{2m}\,\eta^{ab} F_{ae,rb} \theta' ,p_c]  \cr
 &= \frac{\imath \hbar q}{m}\,\eta^{db} F_{bc}\theta'  \tens\extd p_e + \extd x^d\tens\big( - \imath\hbar q\, F_{ae,c}\,\extd x^a 
 - \frac{\hbar q}{2m}\,\eta^{ab}( \hbar F_{be,ac}  +2\imath  q F_{ae}F_{bc}  )\theta'\big) \cr
 & \quad  -  \frac{ \hbar^2 q}{m}\,\eta^{dr} \theta' \tens \big(- 
 F_{ae,rc}\,\extd x^a   - \frac{q}{m}\, F_{ar,c}\,\eta^{ab} F_{be}\theta'   - \frac{q}{m}\, F_{ar}\,\eta^{ab} F_{be,c}\theta' 
 +  \frac{\imath\hbar}{2m}\,\eta^{ab} F_{ae,rbc} \theta' \big)  \cr
&\quad  -  \frac{\imath \hbar q}{m}\,\eta^{dr} \theta' \tens 
 F_{ae,r}\,\frac{\imath \hbar q}{m}\,\eta^{ab} F_{bc}\theta'   \cr
    &= \frac{\imath \hbar q}{m}\,\eta^{db} F_{bc}\theta'  \tens\extd p_e -  \imath\hbar q\,\extd x^d\tens F_{ae,c}\,\extd x^a 
 - \frac{\hbar q}{2m}\,\eta^{ab} ( \hbar F_{be,ac}  +2\imath  q F_{ae}F_{bc}  ) \extd x^d\tens \theta' \cr
 & \quad   + \frac{\imath\hbar^2 q}{2m^2}\,   \eta^{ab}  (\hbar  F_{be,acr}  +2\imath  q F_{ae,r}F_{bc}  +2\imath  q F_{ae}F_{bc,r} )  \,\eta^{dr}\theta' \tens  \theta'
  +  \frac{ \hbar^2 q}{m}\,\eta^{dr} F_{ae,rc}\, \theta' \tens  \extd x^a     \cr
&\quad  
  +  \frac{ \hbar^2 q^2}{m^2}\,\eta^{dr}\big( F_{ae,r}\,\eta^{ab} F_{bc}+ F_{ar,c}\,\eta^{ab} F_{be}   +  F_{ar}\,\eta^{ab} F_{be,c} \big)  \theta' \tens \theta'
   -  \frac{ \imath \hbar^3 q}{2m^2}\,\eta^{dr} \eta^{ab} F_{ae,rbc}   \theta' \tens \theta' \cr
        &= \frac{\imath \hbar q}{m}\,\eta^{db} F_{bc}\theta'  \tens\extd p_e -  \imath\hbar q\,F_{ae,c}\,\extd x^d\tens \extd x^a 
 - \frac{\hbar q}{2m}\,\eta^{ab} ( \hbar F_{be,ac}  +2\imath  q F_{ae}F_{bc}  ) \extd x^d\tens \theta' \cr
 & \quad  
  +  \frac{ \hbar^2 q^2}{m^2}\,\eta^{dr} \eta^{ab} \big(  F_{ar,c} F_{be}   +  F_{ar} F_{be,c} -  F_{ae}F_{bc,r} \big)  \theta' \tens \theta' 
\end{align*}
as required. Now we consider those values of $\sigma$ depending on $\nabla(\extd p_c)$. 
In particular, remembering that $\extd$ involves $\theta'$
\begin{align*}
\sigma&(\extd x^a\tens\extd p_c) = \extd p_c\tens\extd x^a - \frac{\imath \hbar q}{m}\,\eta^{ab}\nabla( F_{bc}\theta' ) -q\, F_{dc,e}[\extd x^d\tens\extd x^e,x^a]  \cr
&\quad -  [\xi_c,x^a]  \tens\theta'-\theta'\tens[\eta_c,x^a]   + [N_c,x^a] \theta'\tens\theta' \cr
&= \extd p_c\tens\extd x^a   + \frac{\imath\hbar q}{m} \,\eta^{ea} F_{dc,e}  \extd x^d\tens\theta'   
     - \frac{\hbar^2 q}{2m^2}\,\eta^{ab} \eta^{de}F_{bc,de} \theta' \tens \theta'
   \cr
&\quad -  [\xi_c,x^a]  \tens\theta'-\theta'\tens[\eta_c,x^a]   + [N_c,x^a] \theta'\tens\theta'   
\end{align*} 
as required. Later,  it will be convenient to set (defining $M^a{}_c$)
\begin{align} \label{mdidcal}
\sigma&(\extd x^a\tens\extd p_c) = \extd p_c\tens\extd x^a   +  \frac{\imath\hbar q}{m} \,\eta^{ea} F_{dc,e}  \extd x^d\tens\theta'   
     + M^a{}_c \theta' \tens \theta'
\end{align} 
where we can calculate
\[
 M^d{}_{e}  =  -  \frac{\imath\hbar q^2}{m^2}\,\eta^{ba}\,\eta^{rd}  F_{ae}  F_{rb}  
 + \frac{\hbar^2 q}{2m^2}\,\eta^{dr}\,\eta^{ab}  F_{be,ar}  \ .
\]
To calculate $\sigma  (\extd p_d\tens\extd p_c)$, we use
\begin{align*}
\sigma & (\extd p_d\tens\extd p_c) = \extd p_c\tens\extd p_d -\nabla([\extd p_c,p_d]) - q  [ F_{dc,e}\extd x^d\tens\extd x^e,p_d]  \cr
&\quad -  [\xi_c,p_d]  \tens\theta'-\theta'\tens[\eta_c,p_d]   + [N_c,p_d] \theta'\tens\theta'  \cr
&= \extd p_c\tens\extd p_d -\nabla\big( - \imath\hbar q\, F_{ac,d}\,\extd x^a 
 - \frac{\hbar q}{2m}\,\eta^{ab}( \hbar F_{bc,ad}  +2\imath  q F_{ac}F_{bd}  )\theta'  \big) - \imath\hbar q\,  F_{ac,ed}\extd x^a\tens\extd x^e \cr
&\quad -    \frac{\imath \hbar q^2}{m}\,F_{ac,e}  ( \eta^{ab} F_{bd}\theta' \tens\extd x^e +\eta^{eb} \extd x^a\tens  F_{bd}\theta'   )
 -  [\xi_c,p_d]  \tens\theta'-\theta'\tens[\eta_c,p_d]   + [N_c,p_d] \theta'\tens\theta'  
\end{align*} 
and using
\begin{align*}
\nabla & (  F_{ac,d}\,\extd x^a ) = \nabla(  \extd x^a\,F_{ac,d} + \frac{\imath\hbar}{m} \,\eta^{ab}\theta' F_{ac,db} ) \cr
&= \nabla(  \extd x^a)\,F_{ac,d}+ \extd x^a\tens\extd F_{ac,d}  + \frac{\imath\hbar}{m} \,\eta^{ab}\theta' \tens\extd F_{ac,db} \cr
&=  - \frac{q}{m}\,\eta^{ra} F_{er}\, \theta'\tens \extd x^e\,F_{ac,d}
+   \frac{\imath\hbar q}{2m^2}\,\eta^{eb}\eta^{ra}F_{br,e}\,F_{ac,d} \theta'\tens \theta'
+ \extd x^a\tens\extd F_{ac,d}  + \frac{\imath\hbar}{m} \,\eta^{ab}\theta' \tens\extd F_{ac,db}   \cr
&=  - \frac{q}{m}\,\eta^{ra} F_{er}F_{ac,d}\, \theta'\tens \extd x^e 
+ \frac{\imath\hbar q}{m^2}\, \eta^{ra} \,\eta^{eb} F_{er}\,F_{ac,db} \theta'\tens  \theta' 
+   \frac{\imath\hbar q}{2m^2}\,\eta^{eb}\eta^{ra}F_{br,e}\,F_{ac,d} \theta'\tens \theta'  \cr
&\quad    + \extd x^a\tens F_{ac,de}\extd x^e -\frac{\imath\hbar}{2m}\eta^{nm} \extd x^a\tens F_{ac,dnm}\theta' 
 + \frac{\imath\hbar}{m} \,\eta^{ab}F_{ac,dbe}  \theta' \tens  \extd x^e
\\ & \quad  - \frac{\imath\hbar}{m} \,\frac{\imath\hbar}{2m}\eta^{nm} \eta^{ab} F_{ac,dbnm} \theta' \tens  \theta'  \cr
  &= - \frac{q}{m}\,\eta^{ra} F_{er}F_{ac,d}\, \theta'\tens \extd x^e 
+ \frac{\imath\hbar q}{m^2}\, \eta^{ra} \,\eta^{eb} F_{er}\,F_{ac,db} \theta'\tens  \theta' 
+   \frac{\imath\hbar q}{2m^2}\,\eta^{eb}\eta^{ra}F_{br,e}\,F_{ac,d} \theta'\tens \theta'  \cr
&\quad    + F_{ac,de} \extd x^a\tens \extd x^e 
-\frac{\imath\hbar}{2m}\eta^{nm} F_{ac,dnm} \extd x^a\tens \theta' 
\end{align*}
and also
\begin{align*}
\eta^{ab}\nabla&\big( (\hbar  F_{bc,ad}  + 2\imath  q F_{ac}F_{bd}  )\theta'  \big) = \eta^{ab} \theta'\tens\extd (\hbar  F_{bc,ad}  + 2\imath  q F_{ac}F_{bd}  )\cr
&= \eta^{ab} (\hbar  F_{bc,ade}  + 2\imath  q F_{ac,e}F_{bd}  + 2\imath  q F_{ac}F_{bd,e}  ) \theta'\tens \extd x^e \cr
&\quad -\frac{\imath\hbar}{2m}\eta^{nm} \eta^{ab} ( \hbar F_{bc,adnm}  + 2\imath  q F_{ac,nm}F_{bd} + 4\imath  q F_{ac,n}F_{bd,m}  + 2\imath  q F_{ac}F_{bd,nm}  ) \theta'\tens \theta'
\end{align*}
we have 
\begin{align*}
\sigma & (\extd p_d\tens\extd p_c) = 
 \extd p_c\tens\extd p_d  +\imath\hbar q\big( -  \frac{q}{m}\,\eta^{ra} F_{er}F_{ac,d}\, \theta'\tens \extd x^e 
+ \frac{\imath\hbar q}{m^2}\, \eta^{ra} \,\eta^{eb} F_{er}\,F_{ac,db} \theta'\tens  \theta'  \cr
& \quad +   \frac{\imath\hbar q}{2m^2}\,\eta^{eb}\eta^{ra}F_{br,e}\,F_{ac,d} \theta'\tens \theta'  
-\frac{\imath\hbar}{2m}\eta^{nm} F_{ac,dnm} \extd x^a\tens \theta' 
 \big)
 \cr &\quad + \frac{\hbar q}{2m}\,  \eta^{ab} (  \hbar F_{bc,ade}  + 2\imath  q F_{ac,e}F_{bd}  + 2\imath  q F_{ac}F_{bd,e}  ) \theta'\tens \extd x^e
  \cr &\quad - \frac{\hbar q}{2m}\, \frac{\imath\hbar}{2m}\eta^{nm} \eta^{ab} (\hbar  F_{bc,adnm}  +2\imath  q F_{ac,nm}F_{bd} +4\imath  q F_{ac,n}F_{bd,m}  +2\imath  q F_{ac}F_{bd,nm}  ) \theta'\tens \theta'
 \cr
&\quad -    \frac{\imath \hbar q^2}{m}\,F_{ac,e}  ( \eta^{ab} F_{bd}\theta' \tens\extd x^e +\eta^{eb} \extd x^a\tens  F_{bd}\theta'   )
 -  [\xi_c,p_d]  \tens\theta'-\theta'\tens[\eta_c,p_d]   + [N_c,p_d] \theta'\tens\theta'  \cr
  &= 
 \extd p_c\tens\extd p_d    
  -  \frac{\hbar^2 q^2}{2m^2}\, \eta^{ra} \,\eta^{eb}\big(2 F_{er}\,F_{ac,db}  +F_{br,e}\,F_{ac,d} 
 \big)  \theta'\tens  \theta'  \cr
 &\quad  +   \frac{\hbar^2 q}{2m}\eta^{nm} F_{ac,dnm} \extd x^a\tens \theta' 
 + \frac{\hbar q}{2m}\,  \eta^{ab} (\hbar  F_{bc,ade}  + 2\imath  q F_{ac}F_{bd,e} -  2\imath  q F_{eb}F_{ac,d} ) \theta'\tens \extd x^e
  \cr &\quad - \frac{\hbar q}{2m}\, \frac{\imath\hbar}{2m}\eta^{nm} \eta^{ab} ( \hbar F_{bc,adnm}  +2\imath  q F_{ac,nm}F_{bd} + 4\imath  q F_{ac,n}F_{bd,m}  + 2\imath  q F_{ac}F_{bd,nm}  ) \theta'\tens \theta'
 \cr
&\quad -    \frac{\imath \hbar q^2}{m}\, \eta^{eb} F_{ac,e} F_{bd}  \extd x^a\tens   \theta'   
-    \frac{ \hbar^2 q^2}{m^2}\,  \eta^{eb}  \,\eta^{ar} F_{ac,e}   F_{bd,r}  \theta' \tens \theta'   
\\ &\quad -  [\xi_c,p_d]  \tens\theta'-\theta'\tens[\eta_c,p_d]   + [N_c,p_d] \theta'\tens\theta',  
\end{align*} 
giving the stated value.
The proof that these formulae for $\sigma$ are consistent with this extending as a bimodule map is extremely tedious and relegated to the appendix. 
Finally, we have to check the condition that $(X\tens X)(\sigma-\id)=0$. From the form of $\sigma$ in the statement, this means checking the following equations:
\begin{align*} 
[X(\extd x^e),&X(\extd x^d) ]
=   \frac{\imath\hbar q}{m^2}\,\eta^{cd} F_{ac}\,\eta^{ae}  \cr
[X(\extd p_e),&X(\extd x^d)] =     \frac{\imath \hbar q}{m}\,\eta^{dc} \big(
 - F_{ae,c}\,X(\extd x^a)   - \frac{ q}{m}\, F_{ac}\,\eta^{ab} F_{be}
 +  \frac{\imath\hbar}{2m}\,\eta^{ab} F_{ae,cb}  \big) \cr
[X(\extd x^a),&X(\extd p_c)] = \frac{\imath\hbar q}{m} \,\eta^{ea} F_{dc,e} X( \extd x^d ) 
     -   \frac{\imath\hbar q^2}{m^2}\,\eta^{bn}\,\eta^{ra}  F_{nc}  F_{rb}  
 + \frac{\hbar^2 q}{2m^2}\,\eta^{ar}\,\eta^{nb}  F_{bc,nr}   \cr
[X(\extd p_e),&X(\extd p_d )]  =  
 \frac{ \imath \hbar q^2}{m}\,\eta^{rp} \big(  F_{rd}\, F_{ae,p}    -    F_{re}\,  F_{ad,p}     \big)  X(\extd x^a)  \cr
 & \qquad  +  \frac{\imath\hbar q^3}{  m^2 }\,  \eta^{rp}\,\eta^{ba}   F_{pe}  F_{ad}  F_{rb}  
    +  \frac{\hbar^2 q^2}{  2m^2 }\, \eta^{rp}\,\eta^{ab}      (  F_{pd}\,    F_{be,ar}    
   -  F_{pd,ar}     F_{be}  ) \ .
\end{align*}
For example, we check the last and hardest case, computing
\begin{align*}
[X(\extd p_e),X(\extd p_d )]  &= \frac{ q^2}{4 m^2}\eta^{ab}\eta^{rp}
[(2F_{ae}p_b-\imath \hbar F_{be,a}), (2F_{rd}p_p-\imath \hbar F_{pd,r})]\\
 &= \frac{ q^2}{4 m^2}\eta^{ab}\eta^{rp}\Big(
\big(2F_{ae}[p_b,2F_{rd}]p_p - 2F_{rd}[p_p,2F_{ae}]p_b \big) \cr
& \quad + 2F_{ae}2F_{rd} [p_b,p_p] - \imath\hbar \big(  2F_{rd} [ F_{be,a} ,p_p] + 2F_{ae}  [p_b, F_{pd,r}  ]   \big)
\Big)
\end{align*}
using the fact that a commutator with a fixed element is a derivation. Expanding the commutators in the 
 three parts of the RHS gives the expression claimed.  \endproof

\begin{proposition} The above calculus on the Heisenberg algebra has a quotient $\Omega^1_{\rm red}$ with relations
\[ \extd x^0=-{p_0\over m}\theta',\quad \extd p_0=q F_{0i}\extd x^i -{\imath\hbar q\over 2m}F_{0i,i}\theta'\]
whereby the commutation relations of $\extd x^i,\extd p_i$ imply those required for $\extd x^0,\extd p_0$. Moreover, $X$ and $\nabla$ descend to this quotient. 
\end{proposition} 
\proof For the calculus we just check the hardest case
\begin{align*}
[q F_{0i}\extd x^i-{\imath\hbar q\over 2m}F_{0i,i}\theta',p_a]&=q[F_{0i},p_a]\extd x^i+q F_{0i}[\extd x^i,p_a]-{\imath\hbar q\over 2m}[F_{0i,i},p_a]\theta'\\
&=\imath\hbar q F_{0i,a}\extd x^i+{\imath \hbar q^2\over m}F_{0i}F_{ia}\theta'+{\hbar^2 q\over 2m}F_{0i,ia}\theta'
\end{align*}
which  agrees with $[\extd p_0,p_a]$.  That $X$ descends is immediate
and for $\nabla$ descending, the hardest case is showing that $\nabla(\extd p_0)$ is the same as
\begin{align*}
\nabla(& q F_{0i}\extd x^i -{\imath\hbar q\over 2m}F_{0i,i}\theta') = 
\nabla(q \extd x^i F_{0i}  +  {\imath\hbar q\over 2m}\theta' F_{0i,i}) \cr
&= q \nabla(\extd x^i) F_{0i}  +  q \extd x^i \tens  \extd F_{0i} +   {\imath\hbar q\over 2m}\theta' \tens  \extd F_{0i,i}   \cr
&= -\frac{ q^2}{m}\,\eta^{ci} F_{ac}\, \theta'\tens \extd x^a F_{0i} 
+ \frac{\imath\hbar q^2}{2m^2}\,\eta^{ab}\eta^{ci}  \theta'\tens F_{bc,a} \theta'  F_{0i} 
+ q \extd x^i \tens   F_{0i,a} \extd x^a    \cr
&\quad  -\frac{\imath\hbar q}{2m}  \extd x^i \tens  \Delta( F_{0i} )\theta'
+   {\imath\hbar q\over 2m}\theta' \tens   F_{0i,ia}  \extd x^a
-   {\imath\hbar q\over 2m}\theta' \tens   \frac{\imath\hbar}{2m} \Delta (F_{0i,i})\theta', 
\end{align*}
where we have used $\extd$ for functions $f$ of the $x^a$ in the analogous form to 
 (\ref{lapextd}), 
\begin{align} \label{lapextd4}
\extd f= f_{,a}\extd x^a - \frac{\imath\hbar}{2m} \Delta (f)\theta' 
\end{align}
with $\Delta( f)=\eta^{ab}f_{,ab}$. Ordering functions to the left and amalgamating terms gives the above as 
\begin{align*}
&= - \frac{ q^2}{m}\,\eta^{ci} F_{ac}F_{0i}  \theta'\tens \extd x^a 
+ \frac{\imath\hbar q^2}{m^2}\,\eta^{ci} F_{ac}\eta^{ab}F_{0i,b} \, \theta'\tens \theta' 
+ \frac{\imath\hbar q^2}{2m^2}\,\eta^{ab}\eta^{ci}  F_{bc,a}   F_{0i}  \theta'\tens \theta' \cr
&\quad   +  q  F_{0i,a} \extd x^i \tens   \extd x^a     - \frac{\imath\hbar q}{2m}  \Delta( F_{0i} ) \extd x^i \tens \theta'
-   {\imath\hbar q\over 2m}  F_{0i,ia}    \theta' \tens   \extd x^a
-   {\hbar^2 q\over 4m^2}  \Delta (F_{0i,i})   \theta' \tens   \theta' 
\end{align*}
which is $\nabla(\extd p_0)$ as required.
 \endproof

The $\extd x^0$ relation says that in this theory it is natural to identify $\theta'$ with the proper time interval $\extd\tau$  given that in Special Relativity ${\extd x^0\over\extd \tau}=-{p_0\over m}$ for our metric. With this in mind, the other relation  {\em roughly speaking} can be interpreted as the quantum analogue of 
\[ {\extd p_0 \over \extd\tau}=q F_{0i}{\extd x^i \over\extd\tau}\mp {\imath\hbar q\over 2m}F_{0i,i} \]
depending on which side we place the $F_{0i}$ before making our interpretation (with the 2nd term vanishing if we average the two). Here $F_{0i}=-{E_i\over c}$ so the first term here is the expected rate of change of energy $-c p_0$ due to the work done by the electric field $E_i$, while the `quantum correction' term is the divergence  $F_{0i,i}=-{\del\cdot E \over c}$ proportional to the charge density of the external source.  

In our formalism, we can also see this more precisely in terms of expectation values in an evolving $s$-dependent vector $\phi(s)=|\phi\>\in \CH$, where we have noted in general that (\ref{dexpdt}) holds, except that the geodesic time $t$ there is now being denoted by $s$, i.e. 
\begin{equation}\label{dsexpect} 
\frac{\extd}{\extd s} \<\phi | a | \phi \>= \<\phi  | X(\extd a) | \phi \>
\end{equation}
for all $a\in A$. Working in the full algebra (we do not need the above quotient), we have from (\ref{Xh}) that
\[ {\extd \<\phi |x^a| \phi \>\over \extd s}={\eta^{ab}\over m}\<\phi | p_b|\phi \>,\]
which says that in any state the expected momentum is $m$ times the proper velocity as classically, and if the  $F_{ab}$ are constant, 
\[ {\extd \<\phi |p_c| \phi \>\over \extd s}=\frac{ q}{m}\,\eta^{ab}F_{ca} \<\phi |p_b| \phi \>=  q F_{ca} {\extd \<\phi |x^a| \phi\>\over \extd s},
\]
which says that the expected proper acceleration is governed by the Lorentz force law again as classically. The $c=0$ instance of this is the relation discussed at the operator level in the quotient, now at the level of expectation values. If the $F_{ab}$ are not constant then we will have order $\hbar$ corrections due to the form of (\ref{Xh}). We next turn to the static case where $A_a$ are time independent. 

\begin{lemma}\label{ucentral} If $A_a$ is time independent then $\cu:=-p_0- q A_0$ is central in the Heisenberg algebra and $[\cu,x^0]=\imath\hbar$. Moreover, there is a subalgebra $\CA$  with subcalculus $\Omega^1_\CA$ of $\Omega^1_{red}$  generated by $x^i,p_i,\extd x^i,\extd p_i,\theta',\cu$ where $\cu$ is central in $\Omega^1_\CA$ and $\extd \cu=0$.  Moreover, $\nabla$ restricts on the generators to a connection on $\Omega^1_\CA$ and $h\in \CA$. 
\end{lemma}
\proof Clearly, $\cu$ is always canonically conjugate to $x^0$ as $p_0$ was. Also $[\cu,x^i]=0$ and when $A_a$ are time independent then $-[\cu, p_i]=[p_0,p_i]+[ q A_0,p_i]=\imath\hbar  qF_{0i}+ \imath\hbar q A_{0,i}=0$. For the differentials working in $\Omega^1_{\rm red}$ in the time independent case,
\[ -[\cu,\extd x^i]=[p_0+q A_0,\extd x^i]={\imath q \hbar\over m}(F_{i0}-A_{0,i})\theta'=0\] 
\[-[\extd p_i,\cu]= [\extd p_i,p_0+qA_0]=-{\imath\hbar q^2\over m}F_{ji}F_{j0}\theta'-\hbar q{\imath q \over m}F_{ij}A_{0,j}\theta'=0\]
We also have, using $x^0$ invariance of the $A_a$ and (\ref{lapextd4}) with $\Delta$ defined by $\eta^{ab}$, 
\begin{align*}
-\extd \cu &= \extd p_0+q\extd A_0 = \extd p_0+ q A_{0,i}\extd x^i-\frac{\imath \hbar q}{2m}
\Delta(A_0)\theta' =0
\end{align*}
 as $A_{0,i}=-F_{0i}$ and $\Delta A_0=A_{0,ii}=-F_{0i,i}$ by the relations in $\Omega^1_{\rm red}$. 

Next, we omit $x^0$ from our algebra as under our assumptions it does not appear in $F_{ab}$ or on the right hand side of any of the commutation relations other than as $\extd x^0=-p_0\theta'/m$. The remaining generators and relations are (\ref{Aured1})-(\ref{Aured3}) as listed below albeit $\cu$ a closed central generator. Further, $\nabla$ restricts to this subcalculus as any $\extd x^0$ terms given by $\nabla$ can be rewritten in terms of $u$ by the relations.  
 \endproof

We can clearly restrict $X$ as well, and obtain the equations for a quantum geodesic flow on $\CA$ with this calculus. Moreover as $\cu$ is central and closed,  we can consider it instead as a fixed real parameter. We  denote this quotient by $\CA_\cu$, with calculus $\Omega^1_{\CA_\cu}$  given by 
\begin{gather}\label{Aured1}[x^i,x^j]=0,\quad [x^i,p_j]=\imath\hbar\delta^i{}_j,\quad [p_i,p_j]= \imath\hbar q F_{ij},  \\ \label{Aured2} [\extd x^i, x^j]=-\imath{\hbar\over m}\delta^{ij}\theta',\quad [\extd x^i,p_j]=\imath{\hbar q\over m}F_{ij}\theta'=[\extd p_j,x^i],\\ \label{Aured3}
 [\extd p_i, p_j]=-\imath\hbar q F_{ki,j}\extd x^k-{\hbar^2 q\over 2m}F_{ki,kj}\theta'+{\imath\hbar q^2\over m}( F_{0i}F_{0j}-  F_{ki}F_{kj})\theta'- {\imath \hbar q\over m}F_{0i,j}(\cu+ q A_0)\theta'.\end{gather}
 This then deforms the Heisenberg algebra on spatial $\R^3$ in Section~\ref{secheis} by the background electrostatic and magnetostatic potentials $A_i,A_0$ with $\cu$ regarded as a real parameter.   This suggests to decompose our representation $\CH$ into fields where $\cu$ has constant value and this is what we will do in the next section. Thus we take $E=L^2(\R^3)\tens C^\infty(\R)$ and $\nabla_E$ given by the same Hamiltonian as above, now viewed as the representation of an element of $\CA_u$ on functions on $\R^3$ with fixed value of $\cu$. In this case,  we have quantum geodesic motion for the reduced Heisenberg algebra $\CA_u$ with calculus $\Omega^1_{\CA_u}$ as in Section~\ref{secheis} but now with Hamiltonian that contains magnetic potentials in the $p_i$ and an electric potential in the form of $V$.

\subsection{Relativistic amplitudes and hydrogen atom} 
 
 We now consider the probabilistic interpretation of the quantum geodesic evolution constructed above. We recall that  $x^0=ct$ where $c$ is the speed of light and time $t$ is in usual units in an inertial frame. So far, we considered the relativistic Heisenberg algebra acting by multiplication and $D_a$ on $\phi\in C^\infty(\R^{1,3})$ at each $s\in \R$. However, $\bar\phi\phi$ on spacetime is not suitable for a probabilistic interpretation in any laboratory as it involves probabilities spread over past and future in the laboratory  frame time. To address this, we work with fields $\psi(u,x^i)$ Fourier transformed from $t$ to a Fourier conjugate variable $u$ say, so
 \[ \psi(u,x^i)=\int\extd t e^{{\imath \over\hbar}ut} \phi(t,x^i),\quad \phi(t,x^i)=\int\extd u e^{-{\imath\over\hbar}tu}\psi(u,x^i).\]
 In physical terms, we can think of amplitudes $\psi$  with a probability distribution of energies and spatial positions. The Heisenberg algebra (as well as the Lorentz group) acts unitarily on this new space of fields completed to $L^2(\R^4)$ in these variables, just because it did before and Fourier transform in one variable can be viewed as an isometry (if also completed to $L^2(\R^4)$ on the spacetime side.)
 
In this form, we can fix one well-known problem with the Klein Gordon equations, namely we are only interested in fields with positive energy, i.e. we restrict support to $u\ge 0$. 
 Another way of looking at this is that the Klein-Gordon equation is second order in time, so to solve it we need additional information, e.g.\ the time derivative of $\phi$ at a fixed $t$. So far, $u$ stands for the classical Fourier conjugate variable to  $t$ but we also would like to identify it with the eigenvalues of an operator in the Heisenberg algebra. We chose this to be  $c\cu$ defined by $-p_0=\cu+ q A_0$ which then acts by multiplication by $u$ on our fields. This choice of $c\cu$ is adapted to the time-independent case but we can use it more generally also. The $c$ is needed since $\cu$ is conjugate to $x^0=ct$. The minus sign is needed due to the $-+++$ signature as classically $p^0=-p_0$ is positive for a future pointing time-like geodesic. The action of the electromagnetic Heisenberg algebra on $\psi(u,x^i)$ is by $x^i$ and $p_i=-\imath\hbar D_i$ as before,  and
\[ t=-\imath\hbar{\del\over\del u},\quad \cu={u\over c},\quad p_0=-{u\over c}-q A_0(t,x^i).\]
Moreover, $\nabla_E$ as before now appears as 
\begin{equation}\label{nablapsi}
\nabla_E\psi=\big(\frac{\partial \psi}{\partial s} -\frac{\imath\hbar}{2m}D_i D_i\psi -\frac{\imath}{2m\hbar}({u\over c}+q A_0)^2\psi \big)\tens\extd s.  \end{equation}
This is clear from the Fourier transform, but if one wants to check it directly,  $p_i,x^i$ are already represented as before and as $A_0(-\imath\hbar{\del\over\del u},x^i)$ does not depend on $u$, we still have $[p_i,x^0]=0$. Meanwhile, 
\begin{align*}
[p_0,p_i] &= \imath\hbar [ {u\over c}+ q A_0 ,D_i]= q [ {u\over c} ,A_i]
+\imath\hbar q [ A_0 ,\frac{\partial}{\partial x^i}] = \imath\hbar q F_{0i},  \cr
[p_0,x^0] &=[ {u\over c}+q A_0 ,  c\imath\hbar\,\frac{\partial}{\partial u}   ]=-\imath\hbar\ . 
\end{align*}

Now suppose that $A_a$ are indeed independent of $t$. Then by Lemma~\ref{ucentral} we can write $\psi(u,x^i)=\Psi(x^i)$ and regard $u$ as a fixed parameter energy since it is central in the Heisenberg algebra.  In this case the $\nabla_E$ on $\Psi$ is governed by a similar operator  as in Section~\ref{secheis} but with $p_i=-\imath\hbar D_i$ for a particle minimally coupled to the $A_i$ as a magnetostatic gauge potential and with potential energy
\[
V(x)=- \frac{1}{2m}({u\over c}+ q A_0(x))^2=- \frac{1}{2m c^2}(u- q \Phi(x))^2,\quad \Phi=cA^0=-c A_0, 
\]
where the upper index potential connects to usual conventions.  Thus,  $\nabla_E\Psi=0$ looks very much like Schr\"odinger's equation except that the geodesic time parameter is not $t$ but proper time $s$. Moreover, we have maintained Lorenz invariance (we could change our laboratory frame) in (\ref{nablapsi}) before we  fixed the energy $u$ in our laboratory frame.  Spacetime is still present and a mode concentrated at a specific $u$ appears in our original KG field $\phi(t,x^i)$  as $e^{-{\imath\over\hbar}ut}\Psi(x^i)$  with $\Psi(x^i)$ the amplitude distribution for such modes at different positions in space. Such a mode will not appear as concentrated at a fixed energy in another frame; this is our choice in the laboratory frame but the geodesic evolution is not dependent on this. We have suppressed that both $\phi$ and $\Psi$ are evolving and depend on the geodesic time parameter $s$.

\begin{example} (Free particle proper time relativistic wave packet.)\label{freeflow} We  consider the simplest case of a scalar field for mass $m$ in 1+1 Minkowski spacetime with zero electromagnetic potential. Then $\nabla_E\Psi=0$ is
\[ \imath\hbar{\del\over\del s}\Psi= h_u\Psi,\quad h_u={(-\imath\hbar)^2\over 2m}\del_x^2- {u^2\over 2m c^2}\]
with $h_u\Psi=-\CE_k \Psi$ with $\CE_k= (k^2-{u^2\over c^2})/ 2m $ for  $\psi_k(x)=e^{\imath {k x\over\hbar}}$. The on-shell fields (i.e. solving the KG equation) just evolve by an unobservable phase $e^{\imath {s m c^2\over 2\hbar}}$ but, as in GR,  we need to also look at nearby off-shell ones (in the case of GR to see that we are at maximum proper time). More precisely, we look at a wave packet which in spatial momentum space is centred on the positive on-shell value corresponding to energy $u$ but includes a Gaussian spread about this. We evolve this from $s=0$ to general $s$:
\[ \Psi(0,x)=\int\extd k e^{-{(c k - \sqrt{u^2-m^2 c^4})^2\over \beta}}e^{\imath k x\over\hbar}, \quad \Psi(s,x)=\int\extd k e^{\imath s {u^2-k^2c^2\over\hbar 2mc^2}}  e^{-{(c k- \sqrt{u^2-m^2 c^4})^2\over \beta}} e^{\imath k x\over\hbar}  \]
as plotted in Figure~1. It is easy enough to check the expectation values 
\[ {\<\Psi|p|\Psi\>\over \<\Psi|\Psi\>}={\int\extd k e^{-2{(c k- \sqrt{u^2-m^2 c^4})^2\over \beta}} k \over \int\extd k e^{-2{( c k- \sqrt{u^2-m^2 c^4})^2\over \beta}} }={1\over c}\sqrt{u^2-m^2 c^4},\quad {\<\Psi|p_0|\Psi\>\over \<\Psi|\Psi\>}=-{u\over c}\]
as expected. We also find  using $x=-\imath\hbar{\del\over\del k}$ on $\psi_k$ inside the upper of the ratio of integrals that 
\[{ \<\Psi|x|\Psi\>\over \<\Psi|\Psi\>}={s\over m c}  \sqrt{u^2-m^2 c^4}={s\over m}{\<\Psi|p|\Psi\>\over \<\Psi|\Psi\>}, \]
which verifies our identity (\ref{dsexpect}) and shows that our quantum geodesic evolves with proper velocity given by  the average spatial momentum/ $m$. We can also compute  $t=-\imath\hbar{\del\over\del u}$ applied in the upper of the ratio of integrals to find
\[ {\<\Psi|t|\Psi\>\over \<\Psi|\Psi\>}=s { u \over m c^2}={s\over \sqrt{1-{v^2\over c^2}}},\quad v:={\<\Psi|x|\Psi\>\over  \<\Psi|t|\Psi\>  }=c  \sqrt{1-\big({mc^2\over u}\big)^2} \]
as expected respectively for the proper velocity in the time direction in Special Relativity and the lab velocity $v$ in our case. Note that the Gaussian parameter $\beta>0$ does not enter into these expectation values but is visible in $\Psi$ as it sets the initial spread (which then increases during the motion). 
\end{example}
\begin{figure}
\includegraphics[scale=.35]{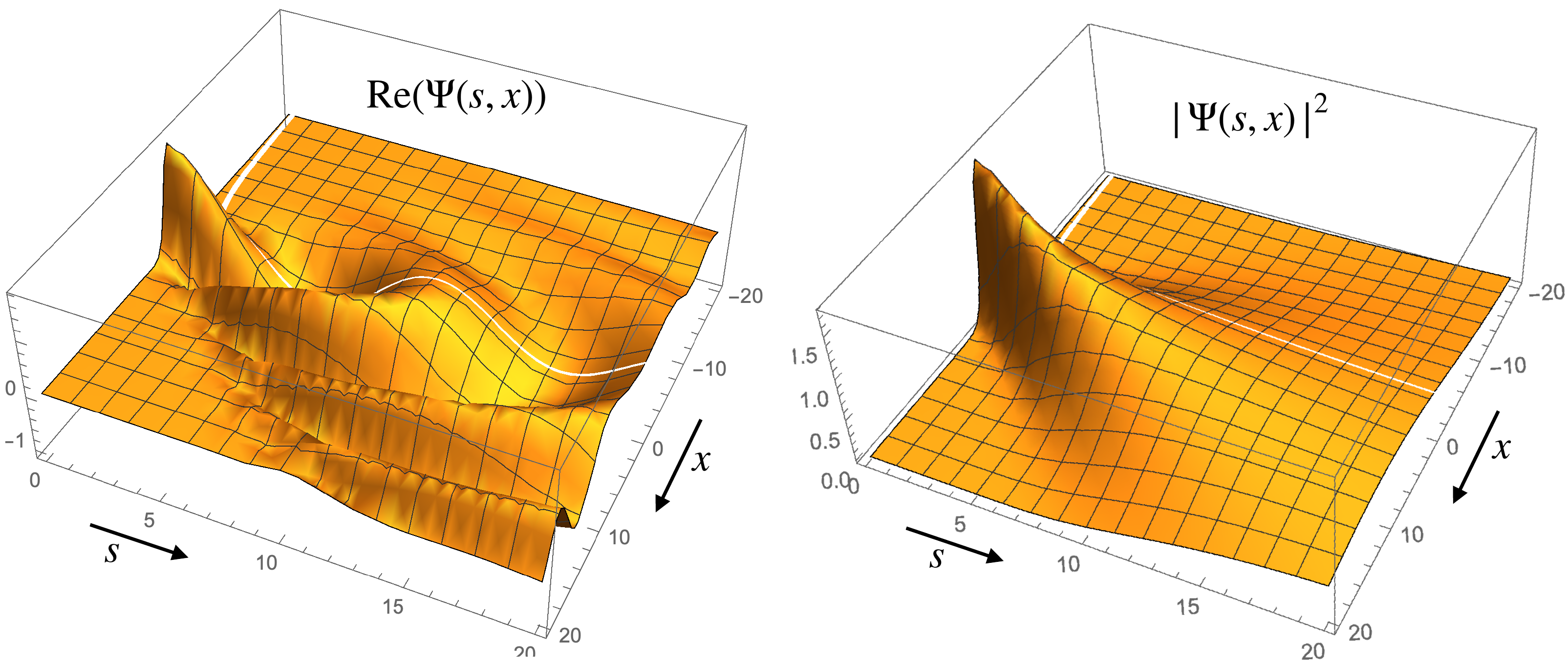}
\caption{Proper time $s$ relativistic wave packet dispersing as it moves down and to the right. Shown are the real and absolute values at $c=\hbar=m=1$ and $u=1.1$. Images produced by Mathematica. } \end{figure}

Although our quantum geodesic flow equation $\nabla_E \Psi=0$ is not Schr\"odingers equation, its close similarity means that we can use all the tools and methods of quantum mechanics with $s$ in place of time there and $u$ as a parameter in the Hamiltonian, as in the preceding example. This approach is somewhat different from the usual derivation of Schr\"odingers equation as a limit of the KG equation,  which involves writing $\phi(x^0,x^i)=e^{-\imath {m c^2 \over\hbar} t}\Psi_{KG}(t,x^i)$ where $ct=x^0$ and $\Psi_{KG}(t,x^i)$ is slowly varying to recover Schr\"odinger's equation for $\Psi_{KG}$ with corrections. The minus sign is due to the $-+++$ conventions. We do not need to make such slow variation assumptions and in fact we proceed relativistically. Our approach means that our differences from Schr\"odingers equation are now of a different nature from the usual ones coming from the KG equation, although they share some terms in common.

\begin{example}(Hydrogen atom revisited.) We consider a hydrogen-like atom or more precisely an electron of charge $q=-e$ in around a point source nucleus of atomic number $Z$ or charge $Ze$, so 
\[ \Phi(r)={Ze\over 4\pi\eps_0 r}. \]
Then the geodesic flow equation $\nabla_E\Psi=0$ at fixed $u$ is
\begin{align} \label{tuvo}
\imath\hbar\frac{\partial \Psi}{\partial s}=h_u\Psi,\quad h_u=\frac{(-\imath \hbar)^2}{2m} \del^2-\frac{1}{2mc^2}(u-q\Phi)^2
\end{align}
since there is no magnetic potential in the $D_i$ (we write $\del^2=\sum\del_i^2$ for the spatial Laplacian).  We are effectively in the Schr\"odinger equation setting of Section~\ref{secheis} with $h={p^2\over 2m}+V$  except that the geodesic parameter is now proper time while $u$ corresponds to a plane wave in laboratory time direction as explained above. We can still use the methods of ordinary quantum mechanics, with
\[ V(r)= -{1\over 2 m c^2}(u-q\Phi)^2=-{u^2\over 2m c^2}-  {u\hbar \over m c} {Z \alpha\over r}- {\hbar^2 \over 2 m }{Z^2\alpha^2\over r^2}, \]
where
\[
q\Phi=-\frac{Ze^2}{4\pi\epsilon_0r}=-\hbar c\frac{Z\alpha}{r}
\]
in terms of the fine struture constant $\alpha\sim 1/137$.  We solve this for eigenmodes 
\[ h_u\Psi=-\CE\Psi\]
where $\CE$ is positive due to a large negative offset in $h_u$ further minus a binding energy. This is solved by the same methods
as the usual hydrogen atom by separation of variables, namely set $\Psi=R(r)\chi_l$ where $\chi_l$ has only angular dependence and is given by an integer $l$ and a quantum number $m=-l,\cdots,l$ which does not change the energy. The remaining radial equation is then 
\[
\frac{\partial }{\partial r} \Big(r^2 \frac{\partial R}{\partial r} \Big)
-l(l+1)R   +  \frac{ m r^2}{\hbar^2} \Big(- 2\CE+ \frac{1}{mc^2} \Big(u+\hbar c\frac{Z\alpha}{r} \Big)^2 \Big)R =0
\]
which has the same form as for a usual atom but with shifted angular momentum $l'=l-\delta_l$ defined as in \cite[Chapter~2.3]{Itz} by
\[
l'(l'+1)=l(l+1) - Z^2\alpha^2;\quad \delta_l=l+{1\over 2}-\sqrt{(l+{1\over 2})^2- Z^2\alpha^2}.
\]
For every $n$ such that $n-(l+1)=d$ is a positive integer i.e. $l=0,...,n-1$, one has 
\[ R(r)=r^{l'}\,\mathrm{e}^{-kr}\CL^{  1 + 2 l'  }_d(2 k r);\quad k^2 =  \frac{ m}{\hbar^2} \Big( 2\CE- \frac{u^2}{mc^2}  \Big)\]
in terms of a generalised Laguerre polynomial of degree $d$. This gives eigenvalues
\begin{equation}\label{Enlu}
\CE_{n,l}=\frac{u^2}{2 mc^2}\Big(1+  \big(\frac{Z\alpha   }{n-\delta_l  }\big)^2\Big)
\end{equation}
for our Schr\"odinger-like geodesic flow equation. From our point of view,  we first consider when modes are on-shell, meaning the associated KG field obeys the Klein-Gordon equation. Given the way $\nabla_E$ was defined, this means to find the spectrum of $u$ such that the eigenvalue $\CE$ as above is ${mc^2\over 2}$. From (\ref{Enlu}),  these are 
\[
u_{n,l}=mc^2\,\frac{  1   }{\sqrt{  1+({Z\alpha \over n-\delta_l})^2 }}
\]
in agreement with the allowed `Schr\"odinger-like' energy spectrum coming from directly solving the KG equation \cite[Chapter~2.3]{Itz}. More generally, we are not obliged to stick to on-shell states and indeed we should not as we saw in the preceding example. For example, we can solve for each fixed $u$ as above and then a general evolution would be
\[ \Psi(s,x^i)=\sum_{n,l,m} e^{\imath\CE_{n,l}(u)\, s\over \hbar}c_{n,l,m}(u)\psi^{(u)}_{n,l,m}(x^i)\]
with eigenvectors $\psi^{(u)}_{n,l,m}$ at fixed $u$ as sketched above, and initial values set by coefficients $c_{n,l,m}(u)$.  One could then compute expectation values along the
quantum geodesic flow in a similar manner to Example~\ref{freeflow}.  

Finally, although not our main purpose, it is tempting to actually think of quantum geodesic flow as a modification of an atomic system and see what the differences are. For this we set $u=mc^2$ so that we have the correct $1/r$ term for an atom at least if we ignore that one should use the reduced mass and that $s$ is proper time. In this case 
\[ \CE_{n,l}=\frac{mc^2}{2}\Big(1+  \big(\frac{Z\alpha   }{n-\delta_l  }\big)^2\Big)={mc^2\over 2}+ {m c^2 Z^2\alpha^2\over 2 (n-\delta_l)^2 },\]
where the first term should be ignored and the  second would be minus the Rydberg binding energy for atomic number $Z$ if $\delta_l$ had been zero. In terms of potentials, at $u=mc^2$ we have
\[ V=-{1\over 2 m c^2}(u-q \Phi)^2=-{mc^2\over 2}+ q\Phi - {q^2\over 2 m c^2}\Phi^2\]
of which we discard the constant term so that the effective potential is 
\[  V(r)=-\hbar c{Z\alpha\over r}-{\hbar^2\over 2 m} {Z^2\alpha^2\over r^2}.\]
For a hydrogen atom, the two terms are of equal size at $r_{\rm crit}={e^2\over 8\pi\eps_0 m c^2}={\alpha\over 4\pi}\lambda_{c}={1\over 2}a_0\alpha^2$,  where $\alpha$ is the fine structure constant, $\lambda_c$ is the Compton wavelength and $a_0$ is the Bohr radius. For one electron,  this critical radius is about $1.4\times 10^{-15}$ metres compared to $0.85\times 10^{-15}$ metres for the size of a proton. But for a large atomic number $Z$ the critical radius would be $Z$ times this, so well outside the nucleus itself. However, in the more careful analysis above,  we still need $Z< {1\over 2\alpha}$ to have solutions for $l'$ as known in the context of solving the KG equation for this background\cite[Chapter 2.3]{Itz}. We see it directly from the potential and without the complications from double time derivatives. This correction is also different from the usual $1/r^3$ spin-orbit correction from allowing for the spin of the electron. Since $s$ is more like proper time, there would also be a relativistic correction compared to coordinate time much as in the usual relativistic correction to the $p^2$ component of the Hamiltonian. \end{example}

\section{Extended phase space Poisson geometry} \label{secsym}

Traditionally in physics, one starts at the Poisson level and then `quantises'. In our case the situation was reversed with  the quantum geometry of the Heisenberg algebra in Section~\ref{secheis} dictated by the algebraic set up. We now semiclassicalise this and similar models to a Poisson level version and present that independently. The first thing we notice is that there is an extra dimension $\theta'$ in the calculus, which is not a problem when $\hbar\ne 0$ but which means that we do not have an actual differential calculus when $\hbar=0$ as $\theta'$ is still present and not generated by functions and differentials of them. This suggests that to have an honest geometric picture, we should work on $\widetilde M=M\times \R$ where $(M,\omega,\bar\nabla)$ is a symplectic manifold with symplectic connection $\bar\nabla$ and symplectic form $\omega_{\mu\nu}$ in local coordnates (we denote its inverse by $\omega^{\mu\nu}$ with upper indices for the associated Poisson bivector inverse to it) and $\R$ corresponds to an external time variable $t$ with $\theta'=\extd t$. The latter recognises that noncommutative systems can generate their own time in a way that is not explicable in the classical limit. By a symplectic connection $\bar\nabla$ we mean torsion free and preserving the symplectic form. (Such connections always exist but are not unique.) 

In Section~\ref{secsymext}, we obtain a self-contained formulation where we fix a Hamiltonian function $h$ with $\bar X$ the associated Hamiltonian vector field, which we extend to $X={\del\over\del t}+\bar X$. We likewise extend  $\omega_{\mu\nu}$ to a (0,2) tensor $G$ on the extended phase space with 
\[ G_{\mu\nu}=\omega_{\mu\nu},\quad G_{0\mu}=-G_{\mu0}=\del_\mu h\]
and we {\em also} extend
$\omega^{\mu\nu}$ to a Poisson bivector on the extended phase space with 
\[ \omega^{0\mu}=-\omega^{0\nu}=\tau^\mu\]
for a suitable vector field $\tau=\tau^\mu\del_\mu$ on $M$. Our convention is that Greek indices {\em exclude} zero.   Both extensions are degenerate and no longer mutually inverse. For simplicity, both $h$ and $\tau$ are taken as time independent, i.e. defined by data on $M$. The main result will be to extend $\bar\nabla$ to a linear connection $\nabla$ on the extended phase space compatible with $G$ such that autoparallel curves are solutions of the original Hamiltonian-Jacobi equations with velocity vector field $X$. 

If $\omega=\extd\theta$ then $\wedge(G)=\extd \Theta$ where $\Theta=\theta +2 h\theta'$ for the usual contact form $\Theta$ on extended phase space as in \cite{contact}. On the other hand,  our specific results in this section are not related as far as we can tell to metrics on phase space such as the Jacobi metric in \cite{Gibbons,Ong}. Nevertheless, we do make use of a natural (possibly degenerate) classical metric $g^{\mu\nu}$ on $M$ induced by the Hamiltonian and we do not exclude the possibility that different approaches to geometry on phase space could be linked in future work. 

Rather, the bigger picture from our point of view is that $\omega^{\mu\nu},\tau^\mu=\omega^{0\mu}$ provide the Poisson bracket and hence quantisation data for the extended phase space:
\begin{equation}\label{relomega}  [x^\mu,x^\nu]=\imath\hbar\omega^{\mu\nu},\quad [t,x^{\mu}]=\imath\hbar\tau^\mu\end{equation}
and the extended linear connection $\nabla$ similarly provides semiclassical data for the quantisation of the differential structure cf\cite{Haw,BegMa:sem,BegMa}
\[  [x^\nu,\extd x^\mu]=\imath\hbar\omega^{\nu\beta}\nabla_{\beta}\extd x^\mu,\quad  [t,\extd x^\mu]=\imath\hbar\tau^\beta\nabla_{\beta}\extd x^\mu,\quad  [x^\nu,\extd t]=0,\quad [t,\extd t]=0\]
to errors $O(\hbar^2)$, which for the natural $\nabla$ found in (\ref{nablaext}) is
\begin{equation}\label{calccon} [x^\nu,\extd x^\mu]=\imath\hbar\omega^{\nu\beta}\bar\nabla_{\beta}\extd x^\mu+\imath\hbar g^{\mu\nu}\extd t,\quad [t,\extd x^\mu]=\imath\hbar\tau^\beta\bar\nabla_{\beta}\extd x^\mu- \imath\hbar g^{\mu\gamma}\omega_{\gamma\beta}\tau^\beta\extd t. \end{equation}
We require the connection to be Poisson compatible in the sense of \cite{BegMa:sem} and, for an associative algebra and calculus at order $\hbar^2$, we require $\omega^{\mu\nu},\tau^\mu$ to obey the Jacobi identity and $\nabla$ to be flat. Neither of the last two conditions is needed at the semiclassical level, while the Poisson compatibility holds if $\bar\nabla$ is symplectic and $\tau$ obeys some conditions deferred to the end of Section~\ref{secsymext}. Section~\ref{secsymph} checks the semiclassical limit of Section~\ref{secheis} extended by central $t$ with $\theta'=\extd t$, and shows that we obtain an example with $\bar\nabla=\tau=0$. 

\subsection{Hamiltonian vector fields as autoparallel on extended phase space}\label{secsymext}

Let $M$ be a symplectic manifold with coordinates $x^\mu$, $\mu=1,\dots,2n$, symplectic form $\omega_{\mu\nu}$ with  inverse  $\omega^{\mu\nu}$, and let
 $\bar\nabla$ be a symplectic connection with Christoffel symbols $\bar\Gamma^\mu{}_{\nu\rho}$ defined by $\bar\nabla_\mu\extd x^\nu=-\bar\Gamma^\nu{}_{\mu\rho}\extd x^\rho$.  We fix a function $h\in C^\infty(M)$ with  Hamiltonian vector field $\bar X{}^\mu=\omega^{\mu\nu}h_{,\nu}$ (we use $h_{,\nu}$ for the partial derivative of $h$ with respect to $x^\nu$). It is easy to see that in general $\bar X$ is not autoparallel as
\[ \nabla_{\bar X}\bar X{}^\mu = \omega^{\alpha\beta}h_{,\beta} \omega^{\mu \nu}\bar\nabla_\alpha h_{,\nu}= \omega^{\alpha\beta}h_{,\beta} \omega^{\mu \nu}( h_{,\nu\alpha}-\bar\Gamma^\gamma{}_{\alpha \nu}h_{,\gamma})= - g^{\mu \nu} h_{,\nu}
\]
where we define the possibly degenerate metric inner product
\[ (\extd x^\mu,\extd x^\nu)=g^{\mu\nu}=\omega^{\mu \gamma}\omega^{\nu \rho} (h_{,\rho}
)_{;\gamma}
\]
(with semicolon the $\bar\nabla$  connection). This  is symmetric as $\bar\nabla$ is torsion free.

To resolve this obstruction to $\bar X$ being autoparallel, we work on the extend phase space with $x^0=t$ and $X={\del\over\del_t}+\bar X$, and write down an extension of the symplectic connection on forms
\[ \nabla_\mu \extd x^\nu= \bar\nabla_\mu\extd x^\nu - \Gamma^\nu{}_{\mu 0}  \extd t   \  ,\quad
 \nabla_0 \extd x^\nu=  - \Gamma^\nu{}_{0\mu}  \extd x^\mu  \  ,\quad
\nabla_\mu \extd t=\nabla_0 \extd t=0 \]
for some additional Christoffel symbol data as shown. 

\begin{lemma} For generic $h$, $X$ is autoparallel with respect to $\nabla$ if and only if  $\Gamma^\mu{}_{\alpha0} +\Gamma^\mu{}_{0\alpha} =g^{\mu\beta}\omega_{\beta\alpha} $. 
\end{lemma}
\proof We require
\begin{align*}
\nabla_{X}X &= \nabla_{\bar X} \bar X+\nabla_{0} \bar X +\nabla_{\bar X} (\tfrac{\del}{\del t}) = (- g^{\mu \nu} h_{,\nu} +\Gamma^\mu{}_{0\alpha} \bar X{}^\alpha +\Gamma^\mu{}_{\alpha0} \bar X{}^\alpha  )\frac{\partial}{\partial x^\mu} \\
&=  (- g^{\mu \nu}  +(\Gamma^\mu{}_{\alpha0} +\Gamma^\mu{}_{0\alpha} )\omega^{\alpha\nu}  ) h_{,\nu} \,\frac{\partial}{\partial x^\mu}=0
\end{align*}
which gives the result stated. \endproof

We now turn to the classical symplectic form $\omega_{\mu\nu}\extd x^\mu\wedge\extd x^\nu$ and its torsion free symplectic connection $\bar\nabla$. In our extended calculus, it would be reasonable to find a related 2-form 
which is preserved by the extended covariant derivative $\nabla$. To the symplectic form, it is reasonable to add something wedged with $\extd t$ in order to be closed.

\begin{lemma}\label{extnabla}
The extended covariant derivative $\nabla$ preserves a 2-form of the form
$\omega_{\alpha\beta }\extd x^\alpha\wedge\extd x^\beta +\extd f\wedge\extd t$ 
for generic $f$ (derivative not vanishing identically on any open region) and has $X$ autoparallel if and only if $\bar\nabla_\mu(\extd(f+2h))=0$,
$\Gamma^\mu{}_{\alpha0} =g^{\mu\beta}\omega_{\beta\alpha} $ and
$\Gamma^\mu{}_{0\alpha} =0$. 
\end{lemma}
\proof Begin by calculating
\begin{align*}
\nabla_0(\omega_{\alpha \beta }\extd x^\alpha \wedge\extd x^\beta ) &= 2\omega_{\alpha \beta }\Gamma^\beta {}_{0\mu }\extd x^\mu \wedge\extd x^\alpha \\
\nabla_\mu(\omega_{\alpha \beta }\extd x^\alpha \wedge\extd x^\beta ) &= -2\omega_{\alpha \beta }\Gamma^\beta {}_{\mu 0}\extd x^\alpha \wedge \extd t  \\
\nabla_0(f_{,\alpha }\extd x^\alpha \wedge\extd t ) &= - f_{,\alpha }\Gamma^\alpha {}_{0\beta }\extd x^\beta \wedge\extd t  \\
\nabla_\mu(f_{,\alpha }\extd x^\alpha \wedge\extd t ) &= (f_{,\alpha \mu }-f_{,\beta }\Gamma^\beta {}_{\mu \alpha })\extd x^\alpha \wedge\extd t \ .
\end{align*}
By comparing these, we see that preserving the given 2-form requires that $f_{,\alpha}\Gamma^\alpha{}_{0\beta }=0$, which for generic $f$ requires $\Gamma^\alpha {}_{0\beta }=0$. Now from the autoparallel condition we have $\Gamma^\mu{}_{\alpha0} =g^{\mu\beta}\omega_{\beta\alpha} $, and 
\begin{align*}
\tilde\nabla_\mu(\omega_{\alpha \beta }&\extd x^\alpha \wedge\extd x^\beta+\extd f\wedge\extd t ) = 
(f_{,\alpha \mu }-f_{,\beta }\Gamma^\beta {}_{\mu \alpha }-2\omega_{\alpha \beta }\Gamma^\beta {}_{\mu 0 })\extd x^\alpha \wedge\extd t  \cr
&=    (f_{,\alpha \mu }-f_{,\beta }\Gamma^\beta {}_{\mu \alpha }-2\omega_{\alpha \beta }g^{\beta\gamma}\omega_{\gamma\mu} )\extd x^\alpha \wedge\extd t  \cr
&=  (f_{,\alpha \mu }-f_{,\beta }\Gamma^\beta {}_{\mu \alpha }-2\omega_{\alpha \beta }\omega^{\beta \pi}\omega^{\gamma \rho} (h_{,\rho})_{;\pi} \omega_{\gamma\mu} )\extd x^\alpha \wedge\extd t   \cr
&=    (f_{,\alpha \mu }-f_{,\beta }\Gamma^\beta {}_{\mu \alpha }+2(h_{,\mu})_{;\alpha} )\extd x^\alpha \wedge\extd t 
\end{align*}
so we require $\nabla_\mu(\extd(f+2h))=0$. \endproof

The obvious choice in this lemma is $f=-2h$ and we hence forth make this choice. This means that the classical symplectic geometry has a natural extension 
\begin{equation}\label{omegaext}\tilde\omega=\omega-2\extd h\wedge\extd t  \end{equation}
\begin{equation} \label{nablaext}  
 \nabla_\alpha\extd x^\mu=\bar\nabla_\alpha\extd x^\mu- g^{\mu\beta}\omega_{\beta\alpha} \extd t  \ ,\quad
 \nabla_0\extd x^\mu=0,\quad\nabla\extd t=0\end{equation}
arranged so that $\nabla_\mu\tilde\omega=\nabla_0\tilde\omega=0$ and $X$ is autoparallel. Using interior product $i_{\bar X}$ with a vector field $\bar X$ (defined as a graded derivation extending the degree 1 pairing), we obtain
 $i_{\bar X}(\omega)=-2\extd h$, where $\omega=\omega_{\alpha\beta}\extd x^\alpha\wedge \extd x^\beta$, which now appears in the extended terms as a kernel condition
\[ i_{X}\tilde\omega=0.\]
Equivalently,  one can check that the antisymmetric rank (0,2) tensor
\[ G=\omega_{\mu\nu}\extd x^\mu\tens\extd x^\nu+\extd t \tens\extd h-\extd h\tens\extd t =\omega_{\mu\nu}\eta^\mu\tens\eta^\nu\]
is covariantly constant under $\nabla$, where
\begin{equation}\label{omegamu}\eta^\mu:=\extd x^\mu-\bar X^\mu\extd t.\end{equation}
In a local patch with coordinates such that $\omega^{\mu\nu}$ are constant and  if $\bar\Gamma^\mu{}_{\nu\rho}=0$, then these 1-forms are also killed by $\nabla$. One can view them along with $\extd t $ as a local parallelisation  of $\widetilde M$.  The quotient of the cotangent bundle  where we set $\eta^\mu=0$, is dual to a sub-bundle the tangent bundle of $\widetilde M$ spanned by $X$. This, at any point of $\widetilde M$, is the tangent to the  Hamilton-Jacobi equations of motion regarded as a flow in $\widetilde M$ through that point. 

On the other hand, this extended connection from Lemma~\ref{extnabla} necessarily has torsion in this extended direction, 
\[ T^\mu{}_{\nu0}=-T^\mu{}_{0\nu}=g^{\mu\beta}\omega_{\beta\nu}.\]
We recall that adding torsion to a connection does not change the autoparallel curves but causes the directions normal to the curves to rotate about them. If we had taken the symmetric form of extension where $\Gamma^\mu{}_{\alpha0}=\Gamma^\mu{}_{0\alpha} ={1\over 2}g^{\mu\beta}\omega_{\beta\alpha} $, we would have had zero torsion but not compatibility with $\tilde\omega$.

\begin{proposition} The extended  $\nabla$ in (\ref{nablaext}) with torsion has curvature
\[
R^\alpha {}_{\beta\gamma\delta}=\bar R^\alpha {}_{\beta\gamma\delta}\ ,\quad
R^\alpha {}_{0\gamma \delta }=\omega^{\alpha \tau}\,h_{,\kappa}\, \bar R^\kappa{}_{\tau\gamma\delta}
\]
and zero for $R$ with index $0$ in all other positions,
where $\bar R^\alpha {}_{\beta\gamma\delta}$ is the curvature of $\bar\nabla$. 
\end{proposition}
\proof 
The curvature can be computed from the usual Christoffel symbol formula
with Roman indices including the index 0, and the derivative in the 0 direction vanishing, thus
\begin{align*}
R^a{}_{bcd}&=\Gamma^a{}_{db,c}-\Gamma^a{}_{cb,d}+\Gamma^a{}_{cs}  \Gamma^s{}_{db}
-\Gamma^a{}_{ds}  \Gamma^s{}_{cb}
\end{align*}
Recalling that $\Gamma^0{}_{ab}=\Gamma^a{}_{0b}=0$ we observe that $R^a {}_{bcd}$ is zero if any of $a,c,d$ are zero. Using Greek indices which cannot be zero, we observe from the formula that 
 the only possible nonzero values apart from $R^\alpha {}_{\beta\gamma\delta}=\bar R^\alpha {}_{\beta\gamma\delta}$ are
\begin{align*}
R^\alpha {}_{0\gamma \delta }&=\Gamma^\alpha {}_{\delta 0,\gamma }-\Gamma^\alpha {}_{\gamma 0,\delta }+\bar\Gamma^\alpha {}_{\gamma \pi}  \Gamma^\pi{}_{\delta 0}
-\bar\Gamma^\alpha {}_{\delta \pi}  \Gamma^\pi{}_{\gamma 0}\ .
\end{align*}
If we use semicolon for covariant differentiation with respect to $\bar\nabla$,  then
\[
\Gamma^\alpha{}_{\delta0} =g^{\alpha\beta}\omega_{\beta\delta} =
\omega_{\beta\delta} \omega^{\alpha \gamma}\omega^{\beta \rho} (h_{,\rho})_{;\gamma}
= - \omega^{\alpha \gamma}(h_{,\delta})_{;\gamma}= - \omega^{\alpha \gamma}(h_{,\gamma})_{;\delta}
\]
as $\bar\nabla$ has zero torsion. Now, as $\omega$ has zero covariant derivative with respect to $\bar\nabla$,
\begin{align*}
\Gamma^\alpha {}_{\delta 0,\gamma }&=-(\omega^{\alpha \tau}(h_{,\tau})_{;\delta} )_{,\gamma}=
-(\omega^{\alpha \tau}(h_{,\tau})_{;\delta} )_{;\gamma}+\bar\Gamma^\alpha{}_{\gamma \pi} \omega^{\pi \tau}
(h_{,\tau})_{;\delta} - \bar\Gamma^\rho{}_{\gamma \delta} \omega^{\alpha \tau}(h_{,\tau})_{;\rho} \cr
&= -\omega^{\alpha \tau}(h_{,\tau})_{;\delta;\gamma}-\bar\Gamma^\alpha{}_{\gamma \pi} 
\Gamma^\pi{}_{\delta0} +\bar\Gamma^\rho{}_{\gamma \delta} \Gamma^\alpha{}_{\rho0} 
\end{align*}
and substituting this into the equation for the Riemann tensor gives 
\begin{align*}
R^\alpha {}_{0\gamma \delta }&= \omega^{\alpha \tau}(h_{,\tau})_{;\gamma;\delta} -\omega^{\alpha \tau}(h_{,\tau})_{;\delta;\gamma}=\omega^{\alpha \tau}\,h_{,\kappa}\, \bar R^\kappa{}_{\tau\gamma\delta}\ ,
\end{align*}
where we have used $\bar\nabla$ torsion free again.\endproof

Note that the connection (\ref{nablaext}), its torsion and curvature  can be written in our more algebraic way, as a right connection, 
\[ \nabla \extd x^\mu
=\bar\nabla\extd x^\mu  - g^{\mu\nu}\omega_{\nu\alpha} \extd t \tens \extd x^\alpha,\]
\[ T_{\nabla}(\extd x^\mu)=g^{\mu\nu}\omega_{\nu\alpha}  \extd x^\alpha\wedge\extd t,\quad  R_{\nabla}(\extd x^\mu)= 
\frac12 \bar R^\mu{}_{\nu\alpha\beta} \eta^\nu 
\tens \extd x^\beta\wedge\extd x^\alpha
\]
where we used symmetry in the first two indices of the curvature of $\bar\nabla$ when raised by $\omega$, see \cite{Bie}. 

 From the point of view of this extended phase space geometry, we can now write (\ref{relomega})-(\ref{calccon}) compactly as
\begin{equation}\label{calcconext}[x^a,x^b]=\imath\hbar\omega^{ab},\quad [x^a,\extd x^b]=-\imath\hbar\omega^{ac}\Gamma^b{}_{cd}x^d,\end{equation}
where $a,b,c.d$ include 0 with $x^0=t$ and $\omega^{0\mu}=-\omega^{\mu 0}=\tau^\mu$. The condition for this to be Poisson compatible follows easily from the characterisation of Poisson compatibility in \cite[Lem.~9.21]{BegMa} and comes out  as 
\begin{equation}\label{contau} \bar\nabla\tau=0,\quad T^\mu{}_{\nu 0}\tau^\nu=0,\end{equation}
given our assumptions  on $\bar\omega$ and the form of the extensions above.

\subsection{Quantum geometry of the extended Heisenberg algebra}\label{secsymph}

Here we extend the Heisenberg algebra $A$ to $\tilde A=A\tens B$ where $B=C^\infty(\R)$, in the form of an
additional time variable $t$, {\em but} with calculus 
\[ \Omega_{\tilde A}=(\Omega_A\tens\Omega_B)/\<\theta'-\extd t\>\]
which makes sense as $\theta'$ and $\extd t$ are both graded central in the tensor product and killed by $\extd$. We then show that its semiclassical limit is an example of Section~\ref{secsymext}.  As $t$ is central, we have $\tau^\mu=0$.  

The first step is to identify the underlying classical symplectic structure and Poisson tensor
\[ -{1\over 2}\omega=\extd x^i\wedge\extd p_i;\quad \omega^{i, j+n}=-\omega^{i+n,j}=\delta_{ij},\quad \omega_{i,j+n}=-\omega_{i+n,j}=-\delta_{ij}.\]
The induced (possibly degenerate) metric $g^{\mu\nu}=(\extd x^\mu,\extd x^\nu)=\omega^{\mu\alpha}\omega^{\nu\beta}\nabla_\alpha h_{,\beta}$ in Section~\ref{secsymext} is given by
\begin{equation}\label{heisinner} (\extd x^i,\extd x^j)={1\over m}\delta_{ij},\quad (\extd p_i,\extd p_j)=V_{,ij}\end{equation}
and other entries zero. Here $\extd x^i$ and $\extd x^{i+n}=\extd p_i$ define $\extd x^\mu$. Writing $\del^2=\delta^{ij}\del_i\del_j$, the associated classical 2nd order Laplace-Beltrami operator is
\begin{equation}\label{heislap} \Delta={1\over m}\del^2+ V_{,ij}{\del^2\over\del p_i\del p_j}\end{equation}
as characterised by Leibnizator $L_{\Delta}(f,g)=2(\extd f,\extd g)$ for all $f,g$ on phase space (this makes sense without assuming $g^{\mu\nu}$ invertible,  but in the invertible case it would be the usual Laplacian-Beltrami operator $\Delta f=g^{\mu\nu}\nabla_\mu\del_\nu f$). 

The extended classical Hamiltonian vector field for our chosen Hamiltonian  as defined  in Section~\ref{secsymext} is
\[ X(\extd x^i)={1\over m}p_i,\quad X(\extd p_i)=-{\del V\over\del x^i},\quad X(\extd t)=1.\]
We have already used these values in the quantum case without change. We also combine the covariant 1-forms $\eta^i$ and $\omega_i=\eta^{i+n}$ as 1-forms $\eta^\mu$. 

\begin{corollary} The relations of the differential calculus, the quantum linear connection and the invariant 1-forms $\eta^\mu$  appear in Section~\ref{secheis} appear in terms of the above phase space structures as
\begin{align*} [x^\mu,x^\nu]&=\imath\hbar \omega^{\mu\nu},\quad [x^\mu,\extd x^\nu]=\imath\hbar g^{\mu\nu}\theta'  ,\quad \{\extd x^\mu,\extd x^\nu\}_\cdot=\imath\hbar g^{\mu\nu}{}_{,\rho}\extd x^\rho \theta' ,\\
 \nabla (\extd x^\mu)&=-\theta'  \tens g^{\mu\nu}\omega_{\nu\alpha}\extd x^\alpha-{\imath\hbar\over 2}\Delta( X(\extd x^\mu))\theta'  \tens\theta'  =\theta' \tens\extd(X(\extd x^\mu)),\\
 \sigma (\extd x^\mu\tens\extd x^\nu)&=\extd x^\nu\tens\extd x^\mu -\imath\hbar g^{\mu\alpha}g^{\nu\beta}\omega_{\alpha\beta}\theta' \tens\theta'  \\
 \eta^\mu&=\extd x^\mu-X(\extd x^\mu)\theta'  \end{align*}
along with $\nabla\theta' =0$ and $\sigma={\rm flip}$ when one factor is $\theta' $. These expressions also apply to $\tilde A$ with $\theta'=\extd t$.
\end{corollary}
\proof Note that the last of the calculus relations is given by applying $\extd$ to the middle relations (these relations do not need the metric to be invertible). The middle form of $\nabla(\extd x^\mu)$ requires some explanation. In fact, the quantum differential calculus in Proposition~\ref{heiscalc} has the structure of a general `central extension'\cite{Ma:alm,Ma:rec} by a 1-form $\theta'$ of the extended calculus on the Heisenberg algebra where we set $\theta'=0$. In this way, with $\theta'=\extd t$  one has 
\begin{equation}\label{lapextd} \extd f(x,p)= {\del f\over\del x^i}\extd x^i +{\del f\over\del p_i}\extd p_i  - {\imath\hbar\over 2}(\Delta f)\theta' ,\end{equation}
for $f(x,p)$ normal ordered so that $x$ is to the left of $p$ and $\Delta$ a certain 2nd order operator on the Heisenberg algebra which reduces to (\ref{heislap}) when $f(x,p)$ is normal ordered {\em and at most quadratic in $p$}. Otherwise $\Delta$ is more complicated with $O(\hbar)$ terms arising from (\ref{heisinner}) not being a bimodule map for general $V(x)$ on the unextended calculus. We used this $\Delta$ for $\nabla(\extd x^\mu)$ and recognised the result in terms of $\extd(X(\extd x^\mu))$.  \endproof

Comparing with (\ref{calccon}), we see that the calculus corresponds to $\nabla$ at the semiclassical level with $\bar\nabla=0$ (as well as $\tau^\mu=0$). The first term of the first form of $\nabla(\extd x^\mu)$ also then agrees with $\nabla$ in (\ref{nablaext}), with a further quantum correction.  In this way, the formulae in Section~\ref{secheis} can be written more geometrically on the extended phase space and the meaning of the connection $\nabla$ with respect to which Schr\"odinger's equation is `quantum geodesic flow' emerges as the semiclassical data for the quantum differential calculus. This also suggests how Section~\ref{secheis} could potentially be extended to other quantisations of symplectic manifolds, though this remains to be done.   We have only considered the time-independent theory and it seems likely that the above will extend also to the time-dependent case. Again, this remains to be done.                                                                                                                                                                                                                                                                                                                                                                                                                                                                                                                                                                                                                                                                                                                                                                                                                                                                                                                                                                                                                                                                                                                                                                                                                                                                                                                                

\section{Concluding remarks} \label{seccon}

In Section~\ref{seciii}, we extended the formalism of `quantum geodesics' in noncommutative geometry as introduced in \cite{Beg:geo} using $A$-$B$-bimodule connections from \cite{BegMa} to geodesics in representation spaces. We then applied this to ordinary quantum mechanics and showed in Section~\ref{secheis} that the usual Schr\"odinger equation can be viewed as a quantum geodesic flow for a certain quantum differential calculus on the quantum algebra of observables (the Heisenberg algebra) acting on wave functions in the Schr\"odinger representation. The quantum differential calculus here encodes the Hamiltonian much as in GR the Riemannian manifold determines the geodesic flow.  This idea that physics has new degrees of freedom in the choice of quantum differential structure has been around for a while now and is particularly evident at the Poisson level\cite{BegMa5}. Such a freedom was already exploited to encode Newtonian gravity by putting the gravitational potential into the spacetime differential structure\cite{Ma:newt}; our now results in Section~\ref{secheis} are in the same spirit but now on phase space in ordinary quantum mechanics and not as part of Planck scale physics. 

We then proceeded in Section~\ref{secKG} to a relativistic treatment based on the Klein-Gordon operator minimally coupled to an external field. Even the simplest 1+1 dimensional case without external field in Example~\ref{freeflow} proved interesting, with relativistic proper time wave packets $\Psi$ quantum geodesically flowing with constant velocity $v=\<\Psi|x|\Psi\>/\<\Psi|t|\Psi\>$ in the laboratory frame. The example illustrates well that quantum geodesic flow looks beyond the Klein-Gordon equation itself. Just as an ant moving on an apple has feet on either side of the geodesic which keeps it on the geodesic path, the quantum wave packet spreads off-shell on either side of a Klein-Gordon solution but on average evolves as expected. In that respect our relativistic approach to quantum mechanics is very different from previous ones and more resembles elements of scalar quantum field theory (where one goes off-shell in the functional integral). We showed how time-independent background fields nevertheless amount to a proper time Schr\"odinger-like equation if we analyse the geodesic flow at fixed energy $u$, allowing the usual tools of quantum mechanics to be adapted to our case. We illustrated this with a hydrogen-like atom of atomic number $Z$. Section~\ref{secsym} concluded with a look at the extended phase space geometry that emerges from our constructions at the semiclassical level. 

Clearly,  many more examples could be computed and studied using the formalism in this paper, including general (non-static) electromagnetic backgrounds to which the theory already applies. Also, in Section~\ref{secheis} we focussed on time-independent Hamiltonians, but the general theory in Proposition~\ref{propiii} does not require this.  It would be interesting to look at the time dependent case and the construction of conserved currents.  The present formalism also allows the possibility of more general algebras $B$ in place of $C^\infty(\R)$ for the geodesic time variable.

On the theoretical side,  the formalism can be extended to study quantum geodesic deviation, where classically one can see the role of Ricci curvature entering. This is not relevant to the immediate setting of the present paper since, at least in Section~\ref{secheis}, the quantum connection on phase space was flat and preserved the extended quantum symplectic structure (rather than being a quantum Levi-Civita connection). It will be looked at elsewhere as more relevant to quantum spacetime and quantum gravity applications, but we don't exclude the possibility of quantum mechanical systems where curvature is needed, e.g. with a more general form of Hamiltonian. Another immediate direction for further work would be to extend Section~\ref{secKG} from an electromagnetic background on the representation space to a curved Riemannian background on the latter, i.e. to gravitational backgrounds such as the quantum-geometric black-hole models in \cite{ArgMa2}. It could also be of interest to consider quantum geodesic flows using a Dirac operator or spectral triple\cite{Con} as in Connes' approach instead of the Klein Gordon operator. 

Finally, on the technical side, the role of $\theta'$ needs to be more fully understood from the point of view of the  quantum extended phase space and its reductions. In our case, it arises as an obstruction to the Heisenberg algebra differential calculus, which forces an extra dimension, but we ultimately identified it with the geodesic time interval. However, a very different approach to handle this obstruction is to drop the bimodule associativity condition in the differential structure\cite{BegMa:sem,BegMa6}, which could also be of interest here.

\appendix
\section{Proof that $\sigma$ for the KG connection is a bimodule map}

Here we complete the proof of Theorem~\ref{bigKG} by checking the remaining cases that $\sigma$ is a bimodule map. Begin with
\begin{align*}
[\extd x^a\tens\extd p_c,x^e] &= - \frac{\imath\hbar}m \,\eta^{ae}\theta'  \tens\extd p_c
 + \frac{\imath \hbar q}{m}\, \eta^{eb}\extd x^a\tens  F_{bc}\theta'    \cr
 &= - \frac{\imath\hbar}m \,\eta^{ae}\theta'  \tens\extd p_c
 +  \frac{\imath \hbar q}{m}\, \eta^{eb}   F_{bc}  \extd x^a\tens   \theta'  
  + \frac{ \hbar^2 q}{m^2}\, \eta^{eb} \eta^{ar} F_{bc,r}    \theta'\tens  \theta'     \cr
  [\extd p_c\tens\extd x^a,x^e] &= - \frac{\imath\hbar}m \,\eta^{ae}\extd p_c\tens\theta'
 +  \frac{\imath \hbar q}{m}\, \eta^{eb}   F_{bc} \theta'\tens \extd x^a
\\
\sigma([\extd x^a\tens\extd p_c,x^e])&  -  [\extd p_c\tens\extd x^a,x^e] 
=   \frac{ \hbar^2 q}{m^2}\, \eta^{eb} \eta^{ar} F_{bc,r}    \theta'\tens  \theta'  \ .
\end{align*}
and from (\ref{mdidcal})
\begin{align*}
\sigma([\extd x^a\tens\extd p_c,x^e])&  -  [\sigma(\extd x^a\tens\extd p_c),x^e] 
=     \frac{ \hbar^2 q}{m^2}\, \eta^{eb} \eta^{ar} F_{bc,r}    \theta'\tens  \theta'  -  \frac{\imath\hbar q}{m} \,\eta^{ba} F_{dc,b} [ \extd x^d,x^e] \tens\theta'   \cr
&=    \frac{ \hbar^2 q}{m^2}\, \eta^{eb} \eta^{ar} F_{bc,r}    \theta'\tens  \theta'  +  \frac{\imath\hbar q}{m} \,\eta^{ba} F_{dc,b} \frac{\imath\hbar}{m} \,\eta^{de}\theta'  \tens\theta'   =0\ .
\end{align*}
Also
\begin{align*}
[\extd p_e&\tens\extd x^d,p_c] = -\big( \imath\hbar q\, F_{ae,c}\,\extd x^a 
 + \frac{\hbar q}{2m}\,\eta^{ab}( \hbar F_{be,ac}  +2\imath  q F_{ae}F_{bc}  )\theta'\big)\tens \extd x^d 
 + \extd p_e\tens \frac{\imath \hbar q}{m}\,\eta^{db} F_{bc}\theta'  \cr
 &=  -\imath\hbar q\, F_{ae,c}\,\extd x^a \tens \extd x^d 
 - \frac{\hbar q}{2m}\,\eta^{ab}( \hbar F_{be,ac}  +2\imath  q F_{ae}F_{bc}  )\theta' \tens \extd x^d \cr
&\quad  + \frac{\imath \hbar q}{m}\,\eta^{db} F_{bc} \extd p_e\tens \theta' 
 + \frac{\imath \hbar q}{m}\,\eta^{da} F_{ac,r} \frac{\imath \hbar q}{m}\,\eta^{rb} F_{be }\theta'  \tens \theta'  \ ,  \cr
[\extd x^d& \tens\extd p_e,p_c] =- \extd x^d\tens \big( \imath\hbar q\, F_{ae,c}\,\extd x^a 
 + \frac{\hbar q}{2m}\,\eta^{ab}(\hbar  F_{be,ac}  +2\imath  q F_{ae}F_{bc}  )\theta'\big) 
 +  \frac{\imath \hbar q}{m}\,\eta^{db} F_{bc}\theta'    \tens \extd p_e  \cr
 &= - \imath\hbar q\, F_{ae,c}\, \extd x^d\tens  \extd x^a 
 - \frac{\hbar q}{2m}\,\eta^{ab} ( \hbar F_{be,ac}  +2\imath  q F_{ae}F_{bc}  )  \extd x^d\tens   \theta'
   + \frac{\imath \hbar q}{m}\,\eta^{db} F_{bc}\theta'    \tens \extd p_e  \cr
 & \quad - \frac{\hbar^2 q}m \,\eta^{dr}\, F_{ae,cr}\, \theta'\tens  \extd x^a 
 + \frac{\imath\hbar^2 q}{2m^2}\,\eta^{dr}\,\eta^{ab} ( \hbar F_{be,acr}  +2\imath  q F_{ae,r}F_{bc} 
 +2\imath  q F_{ae}F_{bc,r} )  \theta'\tens   \theta'
\end{align*}
and as a result, using the formula for $\sigma(\extd x^d\tens\extd x^a), $ 
\begin{align*}
\sigma&([\extd x^d \tens\extd p_e,p_c])  - [\extd p_e\tens\extd x^d,p_c] 
=   - \imath\hbar q\, F_{ae,c}\, (\sigma(\extd x^d\tens  \extd x^a)-\extd x^a \tens \extd x^d )
  \cr
 & \quad - \frac{\hbar^2 q}m \,\eta^{dr}\, F_{ae,cr}\,   \extd x^a \tens \theta'
 + \frac{\imath\hbar^2 q}{2m^2}\,\eta^{dr}\,\eta^{ab} ( \hbar F_{be,acr}  +2\imath  q F_{ae,r}F_{bc} 
 +2\imath  q F_{ae}F_{bc,r} )  \theta'\tens   \theta'   \cr
 &\quad   -  \frac{\imath \hbar q}{m}\,\eta^{da} F_{ac,r} \frac{\imath \hbar q}{m}\,\eta^{rb} F_{be }\theta'  \tens \theta' \cr
 &=    \frac{\hbar^2 q^2}{m^2}\,\eta^{ra}\,\eta^{bd}  F_{ae,c}  F_{br}\theta'\tens \theta'    +  \frac{\hbar^2 q^2}{m^2}\,\eta^{da} \eta^{rb} F_{ac,r} \,F_{be }\theta'  \tens \theta'
  \cr
 & \quad - \frac{\hbar^2 q}m \,\eta^{dr}\, F_{ae,cr}\,   \extd x^a \tens \theta'
 + \frac{\hbar^2 q}{2m^2}\,\eta^{dr}\,\eta^{ab} ( \imath \hbar F_{be,acr}  - 2  q F_{ae,r}F_{bc} 
 - 2  q F_{ae}F_{bc,r} )  \theta'\tens   \theta'   \cr
  &=    \frac{\hbar^2 q^2}{m^2}\,\eta^{ba}\,\eta^{rd}  (F_{ae,c}  F_{rb}  +F_{rb,c} \,F_{ae}  -F_{ae,r}F_{bc} 
 ) \theta'\tens \theta'   
  \cr
 & \quad - \frac{\hbar^2 q}m \,\eta^{dr}\, F_{ae,cr}\,   \extd x^a \tens \theta'
 + \frac{\imath\hbar^3 q}{2m^2}\,\eta^{dr}\,\eta^{ab}  F_{be,acr}   \theta'\tens   \theta'   
\end{align*}
and using (\ref{mdidcal}), 
\begin{align*}
\sigma&([\extd x^d \tens\extd p_e,p_c])  - [\sigma(\extd x^d\tens\extd p_e),p_c] 
  =    \frac{\hbar^2 q^2}{m^2}\,\eta^{ba}\,\eta^{rd}  (F_{ae,c}  F_{rb}  +F_{rb,c} \,F_{ae}  -F_{ae,r}F_{bc} 
 ) \theta'\tens \theta'   
  \cr
 & \quad - \frac{\hbar^2 q}m \,\eta^{dr}\, F_{ae,cr}\,   \extd x^a \tens \theta'
 + \frac{\imath\hbar^3 q}{2m^2}\,\eta^{dr}\,\eta^{ab}  F_{be,acr}   \theta'\tens   \theta'   
 - \frac{\imath\hbar q}{m} \,\eta^{rd} [F_{ae,r}  \extd x^a  ,p_c]  \tens\theta'   
   \\ &\quad  - [M^d{}_e,p_c] \theta' \tens \theta'  \cr
     &  =    \frac{\hbar^2 q^2}{m^2}\,\eta^{ba}\,\eta^{rd}  (F_{ae,c}  F_{rb}  +F_{rb,c} \,F_{ae}  -F_{ae,r}F_{bc} 
 ) \theta'\tens \theta'   
  \cr
 & \quad 
 + \frac{\imath\hbar^3 q}{2m^2}\,\eta^{dr}\,\eta^{ab}  F_{be,acr}   \theta'\tens   \theta'   
 - \frac{\imath\hbar q}{m} \,\eta^{rd} F_{ae,r}  \frac{\imath \hbar q}{m}\,\eta^{ab} F_{bc}\theta'  \tens\theta'   
     - \imath \hbar M^d{}_{e,c}\theta' \tens \theta'
\end{align*}
so we deduce that $\sigma([\extd x^d \tens\extd p_e,p_c])  =[\sigma(\extd x^d\tens\extd p_e),p_c] $.
Also
\begin{align*}
[\extd p_a\tens\extd p_c,x^e] &= \frac{\imath \hbar q}{m}\, \eta^{eb} (  F_{ba}\theta'  \tens\extd p_c
 +   \extd p_a\tens  F_{bc}\theta'  )  \cr
 &=     \frac{\imath \hbar q}{m}\, \eta^{eb} (  F_{ba}\theta'  \tens\extd p_c
 +   F_{bc} \extd p_a\tens   \theta'  +  \frac{\imath \hbar q}{m}\,\eta^{rd} F_{da}  F_{bc,r} \theta' \tens   \theta' )
\end{align*}
and as a result, 
\begin{align*}
\sigma&([\extd p_a\tens\extd p_c,x^e])-[\extd p_c\tens\extd p_a,x^e]  = - \frac{\hbar^2q^2}{m^2}\,  \eta^{eb}   \eta^{rd} (F_{da}  F_{bc,r}   -   F_{dc}  F_{ba,r}   ) \theta' \tens   \theta' 
\end{align*}
which, with a little work, implies $\sigma([\extd p_a\tens\extd p_c,x^e])=[\sigma(\extd p_a\tens\extd p_c),x^e]$.
Finally, we look at the condition
$\sigma([\extd p_e\tens\extd p_d,p_c] )=[\sigma(\extd p_e\tens\extd p_d),p_c] $, beginning with
\begin{align*}
&[\extd p_e\tens\extd p_d,p_c] = - \imath\hbar q\, F_{ae,c}\,\extd x^a \tens \extd p_d
 - \frac{\hbar q}{2m}\,\eta^{ab}(\hbar  F_{be,ac}  + 2\imath  q F_{ae}F_{bc}  )\theta'\tens \extd p_d \cr
 &\quad\quad\quad\quad\quad\quad\quad - \imath\hbar q\, \extd p_e\tens F_{ad,c}\,\extd x^a 
 - \frac{\hbar q}{2m}\,\eta^{ab}\extd p_e\tens(\hbar  F_{bd,ac}  +2\imath  q F_{ad}F_{bc}  )\theta' \cr
 &= - \imath\hbar q\, F_{ae,c}\,\extd x^a \tens \extd p_d
 - \frac{\hbar q}{2m}\,\eta^{ab}(\hbar  F_{be,ac}  + 2\imath  q F_{ae}F_{bc}  )\theta'\tens \extd p_d \cr
 &\quad - \imath\hbar q\, F_{ad,c}\,\extd p_e\tens \extd x^a 
 - \frac{\hbar q}{2m}\,\eta^{ab}( \hbar F_{bd,ac}  +2\imath  q F_{ad}F_{bc}  )\extd p_e\tens \theta' +  \frac{\hbar^2 q^2}{m}\, F_{ad,cr} \,\eta^{rp} F_{pe}\theta' \tens \extd x^a\\  
&\quad - \frac{\imath \hbar^2 q^2}{2m^2}\,\eta^{ab}(\hbar F_{bd,acr}  +2\imath  q F_{ad,r}F_{bc}  +2\imath  q F_{ad}F_{bc,r}  )  \eta^{rp} F_{pe}\theta' \tens \theta' 
 \end{align*}
 we get
 \begin{align*}
\sigma&([\extd p_e\tens\extd p_d,p_c] ) - [\extd p_d\tens\extd p_e,p_c] 
\\
& = - \imath\hbar q\, F_{ae,c}\,\sigma(\extd x^a \tens \extd p_d)
 - \frac{\hbar q}{2m}\,\eta^{ab}(\hbar  F_{be,ac}  + 2\imath  q F_{ae}F_{bc}  )\extd p_d \tens\theta' \cr
 &\quad - \imath\hbar q\, F_{ad,c}\,\sigma(\extd p_e\tens \extd x^a )
 - \frac{\hbar q}{2m}\,\eta^{ab}(\hbar  F_{bd,ac}  + 2\imath  q F_{ad}F_{bc}  )\theta'  \tens\extd p_e \cr
  &\quad +  \frac{\hbar^2 q^2}{m}\, F_{ad,cr} \,\eta^{rp} F_{pe} \extd x^a \tens \theta' 
 - \frac{\imath \hbar^2 q^2}{2m^2}\,\eta^{ab}( \hbar F_{bd,acr}  +2\imath  q F_{ad,r}F_{bc}  +2\imath  q F_{ad}F_{bc,r}  )  \eta^{rp} F_{pe}\theta' \tens \theta' \cr
 &\quad +  \imath\hbar q\, F_{ad,c}\,\extd x^a \tens \extd p_e
 + \frac{\hbar q}{2m}\,\eta^{ab}( \hbar F_{bd,ac}  +2\imath  q F_{ad}F_{bc}  )\theta'\tens \extd p_e \cr
 &\quad + \imath\hbar q\, F_{ae,c}\,\extd p_d\tens \extd x^a 
 + \frac{\hbar q}{2m}\,\eta^{ab}(\hbar  F_{be,ac}  +2\imath  q F_{ae}F_{bc}  )\extd p_d\tens \theta' \cr
  &\quad -  \frac{\hbar^2 q^2}{m}\, F_{ae,cr} \,\eta^{rp} F_{pd}\theta' \tens \extd x^a 
 + \frac{\imath \hbar^2 q^2}{2m^2}\,\eta^{ab}(  \hbar F_{be,acr}  + 2\imath  q F_{ae,r}F_{bc}  + 2\imath  q F_{ae}F_{bc,r}  )  \eta^{rp} F_{pd}\theta' \tens \theta' \cr
 & =  -\imath\hbar q\, F_{ae,c}\,\big( \sigma(\extd x^a \tens \extd p_d) - \extd p_d\tens \extd x^a 
 \big)    - \imath\hbar q\, F_{ad,c}\,\big(   \sigma(\extd p_e\tens \extd x^a )  -  \extd x^a \tens \extd p_e
\big)    \cr
  &\quad +  \frac{\hbar^2 q^2}{m}\, F_{ad,cr} \,\eta^{rp} F_{pe} \extd x^a \tens \theta' 
 - \frac{\imath \hbar^2 q^2}{2m^2}\,\eta^{ab}(\hbar  F_{bd,acr}  +2\imath  q F_{ad,r}F_{bc}  +2\imath  q F_{ad}F_{bc,r}  )  \eta^{rp} F_{pe}\theta' \tens \theta' \cr
  &\quad -  \frac{\hbar^2 q^2}{m}\, F_{ae,cr} \,\eta^{rp} F_{pd}\theta' \tens \extd x^a 
 + \frac{\imath \hbar^2 q^2}{2m^2}\,\eta^{ab}( \hbar F_{be,acr}  +2\imath  q F_{ae,r}F_{bc}  +2\imath  q F_{ae}F_{bc,r}  )  \eta^{rp} F_{pd}\theta' \tens \theta' 
 \end{align*}
 so using (\ref{mdidcal}), 
  \begin{align*}
\sigma&([\extd p_e\tens\extd p_d,p_c] ) - [\extd p_d\tens\extd p_e,p_c] \cr
  & =- \frac{\hbar^2 q^2}{  m }\, F_{ae,c}\,\big( -  \eta^{pa} F_{sd,p}  \extd x^s \tens\theta'   
     - (  - \frac{q}{m}\,\eta^{bp}\,\eta^{ra}  F_{pd}  F_{rb}  
 - \frac{\imath\hbar }{2m}\,\eta^{ar}\,\eta^{pb}  F_{bd,pr} ) \theta' \tens \theta'
 \big)     \cr
 &\quad + \frac{ \hbar^2 q^2}{m}\, F_{ad,c}\,    \eta^{ap} \theta' \tens\big(-
 F_{re,p}\,\extd x^r   - \frac{q}{m}\, F_{rp}\,\eta^{rb} F_{be}\theta' 
 +  \frac{\imath\hbar}{2m}\,\eta^{rb} F_{re,pb} \theta' \big) 
   \cr
  &\quad +  \frac{\hbar^2 q^2}{m}\, F_{ad,cr} \,\eta^{rp} F_{pe} \extd x^a \tens \theta' 
 - \frac{\imath \hbar^2 q^2}{2m^2}\,\eta^{ab}( \hbar F_{bd,acr}  +2\imath  q F_{ad,r}F_{bc}  +2\imath  q F_{ad}F_{bc,r}  )  \eta^{rp} F_{pe}\theta' \tens \theta' \cr
  &\quad -  \frac{\hbar^2 q^2}{m}\, F_{ae,cr} \,\eta^{rp} F_{pd}\theta' \tens \extd x^a 
 + \frac{\imath \hbar^2 q^2}{2m^2}\,\eta^{ab}(\hbar  F_{be,acr}  +2\imath  q F_{ae,r}F_{bc}  +2\imath  q F_{ae}F_{bc,r}  )  \eta^{rp} F_{pd}\theta' \tens \theta' 
 \end{align*}
 and here the $\extd x$ containing terms are
   \begin{align*}
  & \frac{ \hbar^2 q^2}{m}\,\eta^{rp} \big( - (  F_{rd,c}\, F_{ae,p}  +   F_{ae,cr} \, F_{pd})  \theta' \tens \extd x^a   +   (F_{re,c}\,  F_{ad,p}    +  F_{ad,cr} \,F_{pe} )\extd x^a \tens \theta'   
  \big)  \cr
    &=  \frac{ \hbar^2 q^2}{m}\,\eta^{rp} \big(-  (  F_{rd}\, F_{ae,p}  )_{,c}  \theta' \tens \extd x^a   +    (F_{re}\,  F_{ad,p}  )_{,c}\extd x^a \tens \theta'   
  \big)  
 \end{align*}
 Next, 
    \begin{align*}
   [ \frac{ \imath \hbar q^2}{m}\,\eta^{rp} &\big(  F_{rd}\, F_{ae,p}   \theta' \tens \extd x^a   -    F_{re}\,  F_{ad,p}    \extd x^a \tens \theta'  \big)  ,p_c  ]    \cr
      &=  \frac{ \hbar^2 q^2}{m}\,\eta^{rp} \big(  -(  F_{rd}\, F_{ae,p}  )_{,c}  \theta' \tens \extd x^a   +   (F_{re}\,  F_{ad,p}  )_{,c}\extd x^a \tens \theta'    \big)  \cr
      &\quad - \frac{ \hbar^2  q^3}{m^2}\,\eta^{rp} \,\eta^{ab} \big(  F_{rd}\, F_{ae,p}   F_{bc}  -    F_{re}\,  F_{ad,p}     F_{bc} \big)\theta'  \tens \theta' 
 \end{align*}
 so that 
   \begin{align*}
\sigma([\extd p_e&\tens\extd p_d,p_c] ) - [\extd p_d\tens\extd p_e,p_c]
-  [ \frac{ \imath \hbar q^2}{m}\,\eta^{rp} \big(  F_{rd}\, F_{ae,p}   \theta' \tens \extd x^a   -    F_{re}\,  F_{ad,p}    \extd x^a \tens \theta'  \big)  ,p_c  ]  \cr
  & =  \frac{\hbar^2 q^2}{  m }\, F_{ae,c}\,   
      (  - \frac{q}{m}\,\eta^{bp}\,\eta^{ra}  F_{pd}  F_{rb}  
 - \frac{\imath\hbar }{2m}\,\eta^{ar}\,\eta^{pb}  F_{bd,pr} ) \theta' \tens \theta'
     \cr
 &\quad + \frac{ \hbar^2 q^2}{m}\, F_{ad,c}\,    \eta^{ap}\big(
- \frac{q}{m}\, F_{rp}\,\eta^{rb} F_{be}
 +  \frac{\imath\hbar}{2m}\,\eta^{rb} F_{re,pb}  \big)  \theta' \tens \theta'
   \cr
  &\quad 
 - \frac{\imath \hbar^2 q^2}{2m^2}\,\eta^{ab}(\hbar  F_{bd,acr}  +2\imath  q F_{ad,r}F_{bc}  +2\imath  q F_{ad}F_{bc,r}  )  \eta^{rp} F_{pe}\theta' \tens \theta' \cr
  &\quad 
 + \frac{\imath \hbar^2 q^2}{2m^2}\,\eta^{ab}( \hbar F_{be,acr}  +2\imath  q F_{ae,r}F_{bc}  +2\imath  q F_{ae}F_{bc,r}  )  \eta^{rp} F_{pd}\theta' \tens \theta' \cr
 &\quad +\frac{  \hbar^2 q^3}{m^2}\,\eta^{rp} \,\eta^{ab} \big(  F_{pd}\, F_{ae,r}   F_{bc}  -    F_{pe}\,  F_{ad,r}     F_{bc} \big)\theta'  \tens \theta'      \cr
   & =  \frac{\hbar^2 q^2}{  2m^2 }\, F_{ae,c}\,   \eta^{bp}\,\eta^{ra} 
      (  - 2 q\, F_{pd}  F_{rb}  
 - \imath \hbar F_{bd,pr} ) \theta' \tens \theta'
     \cr
 &\quad + \frac{\hbar^2  q^2 }{2m^2}\, F_{ad,c}\,   \eta^{rb}  \eta^{ap}\big(
- 2  q\, F_{rp}\,F_{be}
 +  \imath\hbar F_{re,pb}  \big)  \theta' \tens \theta'
   \cr
  &\quad 
 - \frac{ \hbar^2 q^2}{2m^2}\,\eta^{ab}( \imath \hbar F_{bd,acr}  -  2  q F_{ad}F_{bc,r}  )  \eta^{rp} F_{pe}\theta' \tens \theta' \cr
  &\quad 
 + \frac{ \hbar^2 q^2}{2m^2}\,\eta^{ab}( \imath\hbar  F_{be,acr}   -  2  q F_{ae}F_{bc,r}  )  \eta^{rp} F_{pd}\theta' \tens \theta'    \cr
    & = - \frac{\hbar^2 q^3}{  m^2 }\,  \eta^{rp}\,\eta^{ba}  \big( -F_{ae,c}  F_{pd}  F_{rb}  
 + F_{ad,c} F_{rb}\,F_{pe} 
 -  F_{ad}F_{bc,r}     F_{pe}
 +    F_{ae}F_{bc,r}     F_{pd}\big)\theta' \tens \theta'  \cr
    & \quad +  \frac{\imath\hbar^3 q^2}{  2m^2 }\,
    \big( - F_{ae,c}\,   \eta^{bp}\,\eta^{ra}      \, F_{bd,pr}     
 +  F_{ad,c}\,   \eta^{rb}  \eta^{ap}  F_{re,pb}    
   -  \eta^{ab} F_{bd,acr}     \eta^{rp} F_{pe} 
 \\&\quad+ \eta^{ab}  F_{be,acr}      \eta^{rp} F_{pd}    \big)   \theta' \tens \theta'   \cr
    & = - \frac{\hbar^2 q^3}{  m^2 }\,  \eta^{rp}\,\eta^{ba}  \big( -F_{pe,c}  F_{ad}  F_{br}  
 + F_{ad,c} F_{rb}\,F_{pe} 
 -  F_{ad}F_{bc,r}     F_{pe}
 +    F_{pe}F_{rc,b}     F_{ad}\big)\theta' \tens \theta'  \cr
   & \quad +  \frac{\imath\hbar^3 q^2}{  2m^2 }\, \eta^{rp}\,\eta^{ab}      \, 
    \big( - F_{be,c}   \,  F_{pd,ar}    
 +  F_{pd,c}\,    F_{be,ar}    
   -  F_{pd,acr}     F_{be} 
 +  F_{be,acr}   F_{pd}    \big)   \theta' \tens \theta'  \cr
   & = - \frac{\hbar^2 q^3}{  m^2 }\,  \eta^{rp}\,\eta^{ba}  ( F_{pe}  F_{ad}  F_{rb}  )_{,c}  \theta' \tens \theta' 
    +  \frac{\imath\hbar^3 q^2}{  2m^2 }\, \eta^{rp}\,\eta^{ab}      \, 
(    F_{pd}\,    F_{be,ar}    
   -  F_{pd,ar}     F_{be}    )_{,c}   \theta' \tens \theta'       \cr
    & =[     \frac{\imath\hbar q^3}{  m^2 }\,  \eta^{rp}\,\eta^{ba}   F_{pe}  F_{ad}  F_{rb}   \theta' \tens \theta' 
    +  \frac{\hbar^2 q^2}{  2m^2 }\, \eta^{rp}\,\eta^{ab}      (  F_{pd}\,    F_{be,ar}    
   -  F_{pd,ar}     F_{be}  )  \theta' \tens \theta' ,p_c]
 \end{align*}
 and this gives a value for $\sigma(\extd p_e\tens\extd p_d )$ which would imply the bimodule map condition. Subtracting the value from the last long calculation from the value calculated from $\nabla$, we get the condition
\begin{align*}
0&=   
  -  \frac{\hbar^2 q^2}{2m^2}\, \eta^{ra} \,\eta^{eb}\big(2 F_{er}\,F_{ac,db}  +F_{br,e}\,F_{ac,d} 
 \big)  \theta'\tens  \theta'   +   \frac{\hbar^2 q}{2m}\eta^{nm} F_{ac,dnm} \extd x^a\tens \theta' 
 \\ &\quad + \frac{\hbar q}{2m}\,  \eta^{ab} (\hbar  F_{bc,ade}  + 2\imath  q F_{ac}F_{bd,e} -  2\imath  q F_{eb}F_{ac,d} 
 - 2 \imath q F_{ac}\, F_{ed,b} ) \theta'\tens \extd x^e 
   \cr &\quad - \frac{\imath\hbar^2 q}{4m^2}\, \eta^{nm} \eta^{ab} ( \hbar F_{bc,adnm}  +2\imath  q F_{ac,nm}F_{bd} +4\imath  q F_{ac,n}F_{bd,m}  +2\imath  q F_{ac}F_{bd,nm}  ) \theta'\tens \theta'
 \cr
&\quad 
-    \frac{ \hbar^2 q^2}{m^2}\,  \eta^{eb}  \,\eta^{ar} F_{ac,e}   F_{bd,r}  \theta' \tens \theta'   
 +  [\xi_c,p_d]  \tens\theta'+\theta'\tens[\eta_c,p_d]   + [N_c,p_d] \theta'\tens\theta'  \cr
 & \quad  -  \frac{\imath\hbar q^3}{  m^2 }\,  \eta^{rp}\,\eta^{ba}   F_{pd}  F_{ac}  F_{rb}   \theta' \tens \theta' 
    -  \frac{\hbar^2 q^2}{  2m^2 }\, \eta^{rp}\,\eta^{ab}      (  F_{pc}\,    F_{bd,ar}    
   -  F_{pc,ar}     F_{bd}  )  \theta' \tens \theta' 
\end{align*} 
and substituting the values for $\xi_c,\eta_c$  and $N_c$ from the statement satisfies this.

\end{document}